\documentclass[review,10pt]{elsarticle}

\usepackage{setspace}
\usepackage[a4paper, portrait, margin=3cm]{geometry}
\usepackage[english]{babel}
\usepackage[utf8]{inputenc}
\usepackage[T1]{fontenc}
\usepackage{helvet}
\usepackage[retainorgcmds]{IEEEtrantools}
\usepackage{etoolbox}
\usepackage{graphicx}
\usepackage{titlesec}
\usepackage{caption}
\usepackage{csquotes}
\usepackage{tabularx} 
\usepackage{booktabs}
\usepackage{xcolor} 
\usepackage[most]{tcolorbox}
\usepackage{tikz}
\usepackage{accents}
\usetikzlibrary{calc,arrows,chains,shapes.geometric,decorations.markings}
\usetikzlibrary{3d}
\usetikzlibrary{external}   
\usetikzlibrary{calc, decorations.pathreplacing, arrows.meta}

\usepackage{pgfplots}
\usepgfplotslibrary{}
\pgfplotsset{compat=newest}
\usepgfplotslibrary{fillbetween}
\tcbuselibrary{listingsutf8} 

\usepackage{flafter} 
\usepackage[obeyFinal]{todonotes}
\usepackage{subcaption}
\usepackage{stmaryrd}
\usepackage{bm}
\usepackage{dsfont} 
\usepackage{amsmath,amssymb, amsthm,mathtools,mathrsfs,stmaryrd,titletoc}
\usepackage{wasysym}
\usepackage[norefs,nocites,ignoreunlbld]{refcheck} 
\usepackage[acronym]{glossaries}
\usepackage{xfrac} 
\usepackage{nicefrac} 
\usepackage{numprint} 
\usepackage{physics}
\usepackage{listings} 
\usepackage{algorithm} 
\usepackage{algpseudocode} 
\usepackage{siunitx} 
\usepackage{threeparttable}
\usepackage{xspace}

\usepackage{microtype} 

\usepackage[colorlinks, citecolor=cyan]{hyperref}
\hypersetup{
	colorlinks=true,
	linkcolor=blue,
	filecolor=magenta,      
	urlcolor=blue,
	citecolor=blue 
}

\usepackage[nameinlink]{cleveref}

 \AtBeginDocument{%
    \crefname{equation}{equation}{equations}%
    \crefname{chapter}{chapter}{chapters}%
    \crefname{section}{section}{sections}%
    \crefname{appendix}{appendix}{appendices}%
    \crefname{enumi}{item}{items}%
    \crefname{footnote}{footnote}{footnotes}%
    \crefname{figure}{figure}{figures}%
    \crefname{table}{table}{tables}%
    \crefname{theorem}{theorem}{theorems}%
    \crefname{lemma}{lemma}{lemmas}%
    \crefname{corollary}{corollary}{corollaries}%
    \crefname{proposition}{proposition}{propositions}%
    \crefname{definition}{definition}{definitions}%
    \crefname{result}{result}{results}%
    \crefname{example}{example}{examples}%
    \crefname{remark}{remark}{remarks}%
    \crefname{note}{note}{notes}%
}

\usepackage{caption}
\graphicspath{ {./images/} }
\graphicspath{{inkscape/}}
\usepackage{fancyhdr}
\usepackage{graphicx}

\fancypagestyle{plain}{
	\fancyhf{}

	\lhead{\color{cyan}\small \textbf{~}\\ \color{black}
	\textit{~}\\ }	
}

\usepackage{afterpage}
\usepackage{longtable} 
\usepackage{rotating}

\newcommand{\ie}{\textit{i.e.}~} 
\newcommand{\eg}{\textit{e.g.}~}

\newcommand{\mr}[1]{\mathrm{#1}}

\newcommand{\gradv}[1]{\overrightarrow{\grad}}

\newcommand{\hpdv}[2]{\ensuremath{\partial#1/\partial#2}}

\newcommand{\hfrac}[2]{\ensuremath{#1/#2}}
\newcommand{\I}{\textbf{I}}

\newcommand{\stress}{\ensuremath{\bm{\sigma}}}
\newcommand{\Mstress}{\ensuremath{\bm{\Sigma}}}

\newcommand{\D}{\ensuremath{\textbf{D}}}

\newcommand{\dt}{\ensuremath{\Delta t}}

\newcommand{\sigmaEq}{\ensuremath{\bar{\sigma}}}
\newcommand{\SigmaEqi}{\ensuremath{\bar{\Sigma}}_{i}}

\newcommand{\azero}{\ensuremath{\bm{a}_0}}
\newcommand{\ahat}{\ensuremath{\hat{\bm{a}}}}
\newcommand{\Ppind}{\ensuremath{\mathbb{P}}}
\newcommand{\Ifourth}{\ensuremath{\mathbb{I}}}

\newcommand{\F}{\ensuremath{\textbf{F}}}

\newcommand{\Fe}{\ensuremath{\textbf{F}_\mathrm{e}}}

\newcommand{\PKtwo}{\ensuremath{\textbf{S}}}
\newcommand{\C}{\ensuremath{\textbf{C}}}
\newcommand{\Lv}{\ensuremath{\textbf{L}}}

\newcommand{\W}{\ensuremath{\textbf{W}}}

\newcommand{\B}{\ensuremath{\textbf{B}}}

\newcommand{\A}{\ensuremath{\textbf{A}}}

\newcommand{\epsdot}{\ensuremath{\dot{\varepsilon}}}
  
\newcommand{\N}{\textbf{N}}

\newcommand{\transpose}[1]{\ensuremath{#1^{\mathrm{T}}}}

\newcommand{\purple}[1]{\textcolor{black}{#1}}

\definecolor{TUDblue}{RGB}{0,166,214} 
\definecolor{Donkerblauw}{RGB}{12,35,64}
\definecolor{Turkoois}{RGB}{0,184,200} 
\definecolor{Blauw}{RGB}{0,118,194} 
\definecolor{Paars}{RGB}{111,29,119} 
\definecolor{Roze}{RGB}{165,0,52} 
\definecolor{Rood}{RGB}{224,60,49} 
\definecolor{Oranje}{RGB}{237,104,66} 
\definecolor{Lichtgroen}{RGB}{108,194,74} 
\definecolor{Donkergroen}{RGB}{0,155,119}

\algblockdefx{Start}{End}[1]{\textbf{#1}}{\textbf{end}}

\newtcolorbox[auto counter,crefname={Box}{Boxes}]{MyBox}[2][]{%
  colback=gray!5!white,colframe=blue!75!black,fonttitle=\bfseries,before skip=20pt plus
  2pt,after skip=20pt,
  title=Box~\thetcbcounter: #2, #1}

\tikzstyle{codeblock} = [rectangle, draw=blue, fill=none, minimum width=4em, rounded corners,minimum height=2.3em,align=center,font=\scriptsize]
\tikzstyle{block} = [rectangle, draw=none, fill=Paars, minimum width=4em, rounded corners,minimum height=2.3em,align=center,font=\footnotesize\color{white}]
\tikzstyle{decision} = [diamond, draw=none, fill=blue, minimum width=4em, rounded corners,minimum height=2.3em,align=center,font=\footnotesize\color{white}]
\tikzstyle{blob} = [ellipse, draw=blue, fill=none, minimum width=4em, rounded corners,minimum height=2.3em,align=center,font=\footnotesize]

\usepackage{setspace}
\biboptions{sort&compress}

\journal{~}

\begin{document}

\begin{frontmatter}



  \title{A viscoplasticity model with an invariant-based non-Newtonian flow rule for unidirectional thermoplastic composites}


\author{P. Hofman}
\author{D. Kovačević}
\author{F. P. van der Meer}
\author{L. J. Sluys}

%
\affiliation[]{organization={Delft University of Technology},
             city={Delft},
             country={Netherlands}}

\affiliation[]{P.Hofman@tudelft.nl}

\begin{abstract}
  A three-dimensional mesoscopic viscoplasticity model for simulating rate-dependent plasticity and creep in unidirectional thermoplastic composites is presented.
The constitutive model is a transversely isotropic extension of an isotropic finite strain viscoplasticity model for neat polymers.
Rate-dependent plasticity and creep are described by a non-Newtonian flow rule where the viscosity of the material depends on an equivalent stress measure through an Eyring-type relation. 
In the present formulation, transverse isotropy is incorporated by defining the equivalent stress measure and flow rule as functions of transversely isotropic stress invariants. 
In addition, the Eyring-type viscosity function is extended with anisotropic pressure dependence.
As a result of the formulation, plastic flow in fiber direction is effectively excluded and pressure dependence of the polymer matrix is accounted for.
The re-orientation of the transversely isotropic plane during plastic deformations is incorporated in the constitutive equations, allowing for an accurate large deformation response.
The formulation is fully implicit and a consistent linearization of the algorithmic constitutive equations is performed to derive the consistent tangent modulus.
The performance of the mesoscopic constitutive model is assessed through a comparison with a micromechanical model for carbon/PEEK, with the original isotropic viscoplastic version for the polymer matrix and with hyperelastic fibers.
The micromodel is first used to determine the material parameters of the mesoscale model with a few stress-strain curves.
It is demonstrated that the mesoscale model gives a similar response to the micromodel under various loading conditions. 
Finally, the mesoscale model is validated against off-axis experiments on unidirectional thermoplastic composite plies.
\end{abstract}



%
%

\end{frontmatter}



\section{Introduction}
\label{sec:introduction}

Unidirectional fiber reinforced polymer composites are increasingly used in the aerospace and automotive industry because of their appealing properties. 
These materials, with superior stiffness and strength compared to more traditional metallic materials, allow for lighter structural components, resulting in significant weight-savings in airplanes and automobiles and therefore less fuel consumption and environmental impact \cite{timmisEnvironmental2015}. 

In recent years, there has been a growing interest in the use of thermoplastics in fiber reinforced polymer composites. Structural elements made of thermoplastic composites can be fusion bonded, without the need of additional materials such as adhesives or bolts, resulting in more weight-savings, faster processing cycles and the possibility to manufacture composite parts with more complex geometries. 
However, the mechanical performance of these fusion bonded thermoplastic composites strongly depends on the processing conditions \cite{valverdeEffect2018, valverdeInfluence2020,akkermanAnalysis2020, neveuManufacturing2022}.
At present, the understanding of processing effects on the mechanical response is not fully matured and the lack of sophisticated performance prediction tools forms an obstacle to the wide-spread use of fusion bonded thermoplastic composites.
To improve the prediction abilities, it is essential to develop accurate, efficient and robust constitutive models, capable of simulating the material response under short- and long-term loadings.

A constitutive model that unifies stain-rate dependent yielding and creep in glassy polymers is the Eindhoven Glassy Polymer (EGP) model \cite{tervoortMultimode1996, tervoortConstitutive1997, govaertInfluence2000, klompenModeling2005, vanbreemenExtending2011, lendersElasto2023}. 
This is an isotropic viscoplastic model and is part of a family of models for polymers without an explicit yield function \cite{hawardUse1968, boyceLarge1988, boyceEffects1992}. Instead of a separation in an elastic and plastic response, it is assumed that an applied stress always produces plastic flow and that the rate of plastic flow depends on the stress level. The rate of plastic deformation is then described with a non-Newtonian flow rule following an Eyring-type relation \cite{eyringViscosity1936}. 

The isotropic EGP model has been successfully applied to micromechanical analyses of polymer composites with representative volume elements \cite{kovacevicRateDependentFailure2022,lendersPeriodic2024,kovacevicMicromechanical2024}, where fibers and matrix are explicitly modeled. 
A representative volume element is sufficient for studying the composites' behavior under homogeneous deformations at the mesocale level---that is, the level at which the composite can be considered a homogeneous medium. For more complex structural analyses of composites with inhomogeneous deformations, a multiscale approach can be used. This requires a coupling between the microscale and mesoscale, where two finite element analyses are performed simultaneously and information is exchanged in between. However, such approaches remain computationally infeasible and are still subject of ongoing research in the case of localization \cite{oliverContinuum2015,keComputational2022}. To overcome the computational burden of multiscale analyses, either surrogate models \cite{maiaPhysically2023, maiaSurrogatebased2025}, homogenized micromechanics-based models \cite{larssonMicromechanically2020, singhMicromechanics2023} or mesoscopic phenomenological constitutive models are required.

Extensions of the EGP model for simulating anisotropic rate-dependent plasticity and creep have previously been proposed \cite{vanerpPrediction2009, sendenAnisotropic2013, amiri-radAnisotropic2019, amiri-radImproved2021}. The key element in these works is the incorporation of anisotropy in the (hyper-)elasticity and rate-dependent plasticity relations.
Van Erp et al. \cite{vanerpPrediction2009} proposed an anisotropic flow rule based on the classical Hill yield criterion \cite{hillr.Theory1948}. Senden et al. \cite{sendenAnisotropic2013} used this flow rule in the EGP model for predicting anisotropic yielding in injection molded polyethylene and Amiri-Rad et al.~further developed it for \emph{short} fiber \cite{amiri-radAnisotropic2019} and \emph{long} fiber reinforced polymer composites \cite{amiri-radImproved2021}. \purple{However, a suitable version for \emph{continuous} fiber reinforced polymer composites does not yet exist.}

In continuous fiber reinforced polymers, fibers behave elastically until fracture, while the polymer matrix is responsible for the viscoelastic/viscoplastic response. Combined in a composite, this results in a mostly elastic response when loaded in fiber direction and in a viscoplastic response under off-axis loads. In a constitutive model, strong transverse isotropy can be achieved through the use of transversely isotropic stress invariants \cite{spencerKinematic1987, eidelAnisotropic2004} for describing yield criteria, as previously done with Perzyna-type viscoplastic models \cite{koerberExperimental2018,gerbaudInvariant2019, rodrigueslopesInvariantbased2022}. These models have been successfully applied to the simulation of rate-dependent anisotropic plasticity in \emph{thermosetting} polymer composites under short term loadings.  
As opposed to \emph{thermosets}, \emph{thermoplastics} lack primary (chemical) bonds between polymer chains \cite{brinsonPolymer2015}. When subjected to stress, the polymer response transitions from solid-like to fluid-like, which is described in the EGP model with an Eyring-type non-Newtonian flow rule. With the non-Newtonian flow rule, creep and rate-dependent plasticity are treated in a unified manner. In addition, the effects of temperature can be taken into account through the Eyring relation, as well as the effects of pressure \cite{govaertInfluence2000, govaertMicromechanical2001} and aging \cite{klompenModeling2005}.

In this manuscript, we combine the use of transversely isotropic invariants and non-Newtonian flow, and propose an invariant-based mesoscopic extension of the EGP model for simulating rate-dependent plasticity and creep in continuous fiber reinforced thermoplastic composites. 
For assessing the accuracy of the mesoscopic constitutive model, a detailed micromodel of a carbon/PEEK composite \cite{kovacevicStrainRateArclength2022,  kovacevicRateDependentFailure2022} is used with fibers and matrix explicitly modeled. The micromodel first serves to identify the parameters of the mesoscopic constitutive model through numerical homogenization \cite{vandermeerMicromechanical2016, daghiaValidation2023, liuNumerical2020} with a parameter identification procedure based on a few stress-strain curves. Subsequently, the response of the mesoscale model under off-axis constant strain rates and creep loads is assessed. Finally, unidirectional plies subjected to off-axis strain rates are simulated and compared against experiments. 

Scalars are represented by italic symbols (\eg $a$), while vectors are denoted using italic bold lower case symbols (\eg $\bm{a}$). Second-order tensors are expressed with bold upper case Roman symbols (\eg $\textbf{A}$), and fourth-order tensors are indicated by bold blackboard symbols (\eg $\mathbb{A}$). 
  The symmetric and skew-symmetric parts of a second order tensor $\textbf{A}$ are given by $\textbf{A}^{\mr{sym}} = \hfrac{1}{2} \left(\textbf{A} + \textbf{A}^{\mr{T}}\right)$ and $\textbf{A}^{\mr{skw}} = \hfrac{1}{2} \left(\textbf{A} - \textbf{A}^{\mr{T}}\right)$.  
The product of two second-order tensors $\textbf{A}$ and $\textbf{B}$ is expressed as $\textbf{A} \cdot \textbf{B} = A_{ik} B_{kj}$, while the double contraction is given by $\textbf{A} : \textbf{B} = A_{ij} B_{ij}$.  
Finally, the dyadic product of two vectors $\bm{a}$ and $\bm{b}$ is written as $\bm{a} \otimes \bm{b} = a_i b_j$.

\begin{table*}
  \caption*{\textbf{Major Symbols}}
   \centering
     \setlength\fboxsep{0pt}
\vskip-\topsep%
\smallskip%
\colorbox{gray!10}{%
   \begin{tabularx}{1.0\textwidth}{llX}
   \toprule
   Variable & Type  & Meaning\\
  \midrule
  \textit{General} & &  \\
  $\theta_0$ & scalar & Initial off-axis angle\\
  $\theta$ & scalar & Off-axis angle\\
  $\varepsilon$ & scalar & True strain \\
  $\sigma$ & scalar & True stress \\
  $\varepsilon_\mr{eng}$ & scalar & Engineering strain \\
  $\sigma_\mr{eng}$ & scalar & Engineering stress \\
  $\PKtwo$ & 2$^\mr{nd}$ order tensor & 2$^\mr{nd}$ Piola Kirchoff-stress\\
  $\textbf{F}$ & 2$^\mr{nd}$ order tensor & Deformation gradient\\
  $\mathbf{C}$ & 2$^\mr{nd}$ order tensor & Right Cauchy-Green tensor\\
  $\mathbf{B}$ & 2$^\mr{nd}$ order tensor & Left Cauchy-Green tensor\\
               &  & \\
  \textit{All modes} & & \\
  $N$ & scalar & Number of modes\\
  $\mu_{\mr{p}}$ & scalar & Pressure dependency parameter\\
  $\sigma_0$ & scalar & Nonlinearity parameter\\
  $\eta_{0}$ & scalar & Maximum initial viscosity\\
  $\alpha_2$ & scalar & Anisotropy parameter\\
  $\sigmaEq$ & scalar & Total equivalent stress\\
  $a_\sigma$ & scalar & Stress shift factor\\
  $I_1, I_2, I_3$ & scalars & Transversely isotropic invariants \\
  $\bm{a}_0$ & vector & Fiber vector in initial configuration \\
  $\bm{a}$ & vector & Fiber vector in current configuration \\
  $\bar{\bm{a}}$ & vector & Normalized fiber vector in current configuration \\
  $\bar{\textbf{A}}$ & 2$^\mr{nd}$ order tensor & Structural tensor in current configuration\\
  $\stress$ & 2$^\mr{nd}$ order tensor & Cauchy stress \\
  $\stress^\mr{pind}$ & 2$^\mr{nd}$ order tensor & Plasticity inducing Cauchy stress\\
  $\Ppind$ & 4$^\mr{th}$ order tensor & Tensor that maps $\stress$ to $\stress^\mr{pind}$  \\
   & & \\
  \textit{Mode} $i$ & & \\
  $\bar{\Sigma}_i$ & scalar & Equivalent stress \\
  $\dot{\gamma}_{\mr{p}i}$ & scalar & Equivalent rate of plastic deformation \\
  $m_i$ & scalar & Ratio of elastic constants in relaxation spectrum\\
  $\eta_i$ & scalar & Stress-dependent viscosity \\
  $\eta_{0i}$ & scalar & Initial viscosity\\
  $\lambda_i, \mu_i, \alpha_i, \beta_i, \gamma_i$ & scalars & Hyperelastic model parameters \\
  $\hat{I}_{1i}, \hat{I}_{2i}, \hat{I}_{3i}$ & scalars & Transversely isotropic invariants \\
  $\hat{\bm{a}}_i$ & vector & Fiber vector in intermediate configuration \\
  $\textbf{F}_{\mr{p}i}$ & 2$^\mr{nd}$ order tensor & Plastic deformation gradient \\
  $\textbf{F}_{\mr{e}i}$ & 2$^\mr{nd}$ order tensor & Elastic deformation gradient \\
  $\hat{\textbf{A}}_i$ & 2$^\mr{nd}$ order tensor & Structural tensor in intermediate configuration \\
  $\hat{\mathbf{L}}_{\mr{p}i}$ & 2$^\mr{nd}$ order tensor & Plastic velocity gradient \\
  $\hat{\mathbf{D}}_{\mr{p}i}$ & 2$^\mr{nd}$ order tensor & Rate of plastic deformation \\
  $\hat{\mathbf{W}}_{\mr{p}i}$ & 2$^\mr{nd}$ order tensor & Plastic material spin \\
  $\hat{\mathbf{N}}_{\mr{p}i}$ & 2$^\mr{nd}$ order tensor & Plastic normal \\
  $\hat{\mathbf{B}}_{\mr{e}i}$ & 2$^\mr{nd}$ order tensor & Elastic left Cauchy-Green tensor  \\
  $\hat{\mathbf{C}}_{\mr{e}i}$ & 2$^\mr{nd}$ order tensor & Elastic right Cauchy-Green tensor \\
  $\stress_i$ & 2$^\mr{nd}$ order tensor & Cauchy stress tensor\\
  $\Mstress_i$ & 2$^\mr{nd}$ order tensor & Mandel-like stress tensor\\
  $\Mstress^\mr{sym}_i$ & 2$^\mr{nd}$ order tensor & Symmetric part of Mandel-like stress tensor \\
  $\Mstress^\mr{pind}_i$ & 2$^\mr{nd}$ order tensor & Plasticity inducing Mandel-like stress tensor  \\
  $\hat{\Ppind}_i$ & 4$^\mr{th}$ order tensor & Tensor that maps $\Mstress^\mr{sym}_i$ to $\Mstress^\mr{pind}_i$ \\
  \bottomrule
 \end{tabularx}%
 }
 \label{tab:notations}
\end{table*}

\section{Formulation of the constitutive model}
\label{sec:formulation}

\begin{figure}
  \centering
  \begin{tikzpicture}
        \node at (0,0)
  {
      {\def\svgwidth{0.4\columnwidth}{\scalebox{1.0}{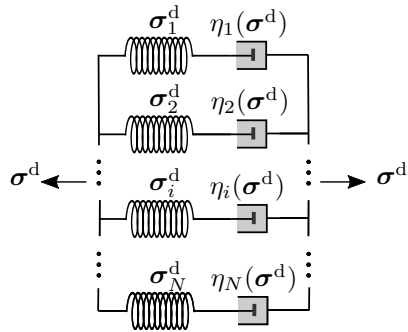}}}
  };
  \end{tikzpicture}
  \caption{Rheological model of the driving stress.}
  \label{fig:rheological-model}
\end{figure}

The mesoscopic constitutive model for the composite material is based on the EGP model for neat polymers \cite{tervoortMultimode1996,tervoortConstitutive1997, vanbreemenExtending2011}, which assumes two contributions to the stress: a driving stress $\bm{\sigma}^{\mr{d}}$ and a hardening stress $\bm{\sigma}^{\mr{h}}$ 
\begin{equation}
  \stress = \bm{\sigma}^\mr{d} + \bm{\sigma}^\mr{h} 
  \label{eq:total-stress-split}
\end{equation}
The driving stress is described by a spectrum of relaxation times, which is incorporated in the model by adding $N$ nonlinear spring-dashpots (denoted as \emph{modes}) in parallel. The driving stress is the sum of the driving stresses $\stress^{\mr{d}}_i$ in each mode $i$
\begin{equation}
  \stress^\mr{d} = \sum^N_i \stress^{\mr{d}}_i
\end{equation}
For thermorheologically simple materials, it can be assumed that the viscosity of each mode $\eta_i$ has the same functional dependence on the \emph{total} driving stress $\stress^\mr{d}$ \cite{tervoortMultimode1996}. The rheological model of the driving stress contribution is shown in \Cref{fig:rheological-model}. 

In this manuscript, the focus is on the driving stress contribution for describing anisotropic rate-dependent plasticity in the pre-yield and yield regime. Therefore, the  hardening contribution is not taken into account ($\bm{\sigma}^\mr{h}=0$). To improve readability, the superscript ($\mr{d}$) in the driving stress is dropped in the remainder of the text.

\subsection{Kinematics}
\label{sec:kinematics}

\begin{figure}
  \centering
  \begin{tikzpicture}
        \node at (0,0)
  {
      {\def\svgwidth{0.4\columnwidth}{\scalebox{1.0}{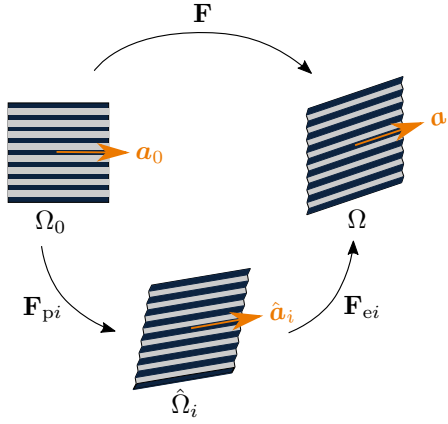}}}
  };
  \end{tikzpicture}
  \caption{Decomposition of total deformation in elastic and plastic deformation for \emph{each} mode $i$, with the corresponding \textit{initial} $\Omega_0$, \textit{intermediate} $\Omega_i$ and \textit{current} configuration $\Omega$.}
  \label{fig:configurations}
\end{figure}
In each mode $i$, a multiplicative decomposition of the total deformation gradient $\textbf{F}$ into an elastic $\textbf{F}_{\mr{e}i}$ and a plastic $\textbf{F}_{\mr{p}i}$ deformation gradient is assumed \cite{kroenerAllgemeine1959, leeElasticPlastic1969}
\begin{equation}
  \F = \F_{\mr{e}i} \cdot \F_{\mr{p}i}
  \label{eq:deformation-decomposition}
\end{equation}
The plastic deformation gradient maps the neighborhood of a mesoscopic material point from the \emph{initial} configuration $\Omega_0$ to a fictitious, locally stress-free, \emph{intermediate} configuration $\hat{\Omega}_i$. Subsequently, the elastic deformation maps it from the \emph{intermediate} configuration to the \emph{current} configuration $\Omega$ (see \Cref{fig:configurations}).
The plastic velocity gradient in the \emph{intermediate} configuration reads
\begin{equation}
  \hat{\Lv}_{\mr{p}i} = \dot{\F}_{\mr{p}i}  \cdot \F^{-1}_{\mr{p}i} = 
  {\underbrace{\left(\dot{\F}_{\mr{p}i}  \cdot \F^{-1}_{\mr{p}i}\right)}_{\hat{\D}_{\mr{p}i}}}^\mr{sym} + 
  {\underbrace{\left(\dot{\F}_{\mr{p}i} \ \cdot \F^{-1}_{\mr{p}i}\right)}_{\hat{\W}_{\mr{p}i}}}^\mr{skw}  \label{eq:velocity gradient}
\end{equation}
where $\hat{\D}_{\mr{p}i}$ is the \emph{rate of plastic deformation} and $\hat{\W}_{\mr{p}}$ is the \emph{plastic material spin} \cite{dafaliasPlastic1998}.
To overcome the non-uniqueness of the multiplicative decomposition with regards to the orientation of the \emph{intermediate} configuration, we choose $\hat{\W}_{\mr{p}i} = \bm{0}$ \cite{boyceKinematics1989, boyceEffects1992}.
Therefore, the evolution of $\F_{\mr{p}i}$ is described with the following differential equation
\begin{equation}
   \dot{\F}_{\mr{p}i} = \hat{\D}_{\mr{p}i} \cdot \F_{\mr{p}i}   
   \label{eq:plastic-evolution}
\end{equation}
The transverse isotropy that originates from the microstructure of the unidirectional polymer composite is characterized by fiber direction vectors $\bm{a}_0$, $\hat{\bm{a}}_i$ and $\bm{a}$ in the \emph{initial}, \emph{intermediate} and \emph{current} configurations, respectively. In the present mesoscopic constitutive model, the fiber vector represents \emph{continuous} fibers in the composite and is assumed to remain affinely attached to the material during deformation,\footnote{For short fiber composites, this assumption is debatable as pointed out by Ref. \cite{dafaliasPlastic1998}, where short fibers may evolve differently from the mesoscopic kinematics.} which is described by the following transformations using the multiplicative decomposition in \Cref{eq:deformation-decomposition}:
\begin{align}
  \bm{a} = \textbf{F}_{\mr{e}i} \cdot \hat{\bm{a}}_i = \F_{\mr{e}i} \cdot \F_{\mr{p}i} \cdot \bm{a}_0  = \F \cdot \bm{a}_0
  \label{eq:a}
\end{align}
Furthermore, plastic deformation is assumed to be isochoric:
\begin{equation}
   \det\left(\F_{\mr{p}i}\right) = 1
  \label{eq:isochoric-assumption}
\end{equation}

\subsection{Viscoplasticity relations}
\label{sec:viscoplasticity}
The rate of plastic deformation in the intermediate configuration in each mode $i$ follows a non-Newtonian flow rule 
\begin{equation}
  \hat{\D}_{\mr{p}i} = \dot{\gamma}_{\mr{p}i} \hat{\N}_{\mr{p}i}
\label{eq:flow-rule}
\end{equation}
where $\dot{\gamma}_{\mr{p}i}$ is the (scalar) equivalent plastic strain rate and $\hat{\N}_{\mr{p}i}$ is the direction of plastic flow. The equivalent plastic strain rate is given by
\begin{equation}
  \dot{\gamma}_{\mr{p}i} = \frac{\SigmaEqi}{\eta_i}
  \label{eq:equiv-strain}
\end{equation}
where $\SigmaEqi$ is the equivalent stress in mode $i$. The viscosity $\eta_i$ is determined as 
\begin{equation}
  \eta_i = \eta_{0i}\,a_\sigma
  \label{eq:viscosity}
\end{equation}
where $a_\sigma$ is the \emph{stress shift factor}\footnote{The name \textit{stress shift factor} refers to its effect of reducing the initial viscosity with increasing stress, resulting in horizontal shifts at different stress levels in creep-compliance curves on logarithmic time scales \cite{tervoortConstitutive1997}.} and $\eta_{0i}$ is the initial viscosity of mode $i$. 
The \emph{stress shift factor} follows an Eyring relation and is a function of the \emph{total} driving stress $\stress$ through a \emph{total} equivalent stress $\sigmaEq$ and may depend on the temperature, pressure and aging \cite{govaertInfluence2000, govaertMicromechanical2001, klompenNonlinear1999}.
Neglecting these influences, the \emph{stress shift factor} reads
\begin{equation}
  a_\sigma  = \frac{\nicefrac{\sigmaEq}{\sigma_0}}{\sinh\left(\nicefrac{\sigmaEq}{\sigma_0}\right)}
  \label{eq:shift-factor-old}
\end{equation}
where $\sigma_0$ is a parameter that controls the stress-induced exponential decrease of the viscosity. 
Note that the viscosity in each mode is different because of the different initial viscosities $\{\eta_{0i}\}$. However, $a_\sigma$ is the same across all modes, representing a thermorheologically simple material \cite{tervoortMultimode1996}. 

For describing plastic flow, a Mandel-like stress tensor \cite{mandel1972plasticite} is introduced as
\begin{equation}
  \bm{\Sigma}_{i} = \F^{\mr{T}}_{\mr{e}i} \cdot \stress_i \cdot \F^{\mr{-T}}_{\mr{e}i} 
  \label{eq:mandel-like}
\end{equation}
which is work-conjugate to $\hat{\D}_{\mr{p}i}$ and is in general not symmetric for anisotropic materials \cite{lublinerPlasticity2008}. To ensure a symmetric $\hat{\D}_{\mr{p}i}$ and to remain consistent with the choice of a vanishing $\hat{\W}_{\rm{p}i}$ (see \Cref{sec:kinematics}), it is assumed that \emph{only} the symmetric part of $\bm{\Sigma}_i$ determines the plastic flow direction \cite{rodrigueslopesInvariantbased2022,eidelAnisotropic2004,deanFinite2016}, \ie
\begin{equation}
  \hat{\N}_{\mr{p}i} = \pdv{\SigmaEqi}{\bm{\Sigma}^{\mr{sym}}_{i}}
  \label{eq:plastic-normal}
\end{equation} 
In the (original) isotropic EGP model, the equivalent stress(es) are proportional to the Von Mises stress \cite{tervoortMultimode1996, tervoortConstitutive1997, govaertInfluence2000, klompenModeling2005, vanbreemenExtending2011}. For \emph{short} and \emph{long} fiber reinforced polymer composites, they can be proportional to the Hill effective stress \cite{amiri-radAnisotropic2019,amiri-radImproved2021}. In this work, strong transverse isotropy of \emph{continuous} fiber reinforced polymer composites is taken into account by defining the equivalent stresses $\sigmaEq$ and $\SigmaEqi$ 
as functions of transversely isotropic stress invariants. In addition, anisotropic pressure dependency is incorporated by modifying the Eyring-type relation \Cref{eq:shift-factor-old}. 
The invariant-based formulation is presented in the next section.

\subsection{Invariant formulation}

Fiber reinforced polymer composites can be considered transversely isotropic at the mesoscale. The response of the mesoscopic constitutive model should therefore
be invariant with respect to the symmetry transformations for transverse isotropy \cite{boehlerApplications1987}.
For unidirectional fiber reinforced polymer composites with strong anisotropy, additional requirements can be specified: (\textit{i}) the material should not flow in the direction of the fiber, (\textit{ii}) the plastic deformation should be isochoric (as stated in \Cref{eq:isochoric-assumption}) and (\textit{iii}) the pressure dependence of the polymer matrix should be taken into account. 
These requirements can be satisfied by using transversely isotropic invariants \cite{spencer1972deformations, boehlerApplications1987} for defining the equivalent stresses $\sigmaEq$ and $\bar{\Sigma}_i$ and by extending the Eyring relation (\Cref{eq:shift-factor-old}) to account for anisotropic pressure dependence.

\subsubsection{Total equivalent stress}
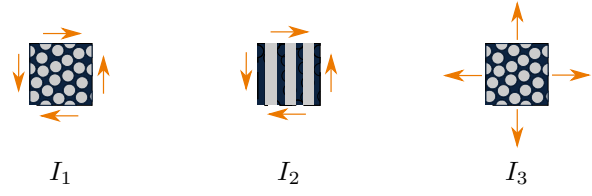
\begin{figure}
  \centering
\begin{tikzpicture}

  \node[]            at (-3.0,0)
    {
      {\def\svgwidth{0.13\columnwidth}{\scalebox{1.0}{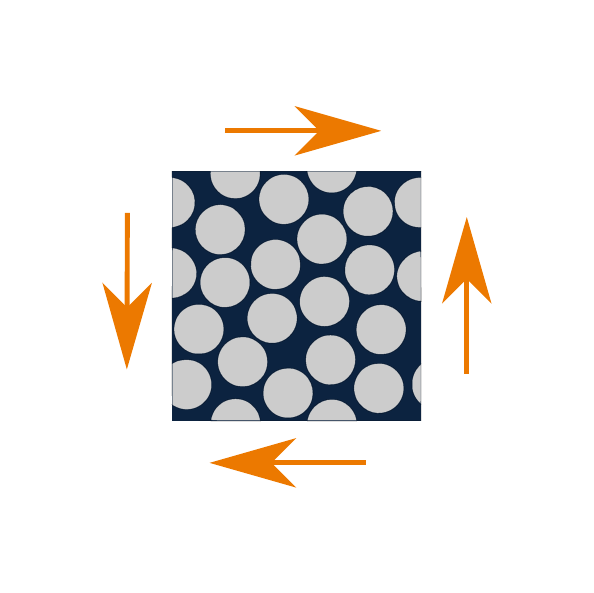}}}
    };

  \node[] at (-3.0,-1.3) {$I_1$};
  \node[rotate=0]            at (0.0,0)
    {
      {\def\svgwidth{0.13\columnwidth}{\scalebox{1.0}{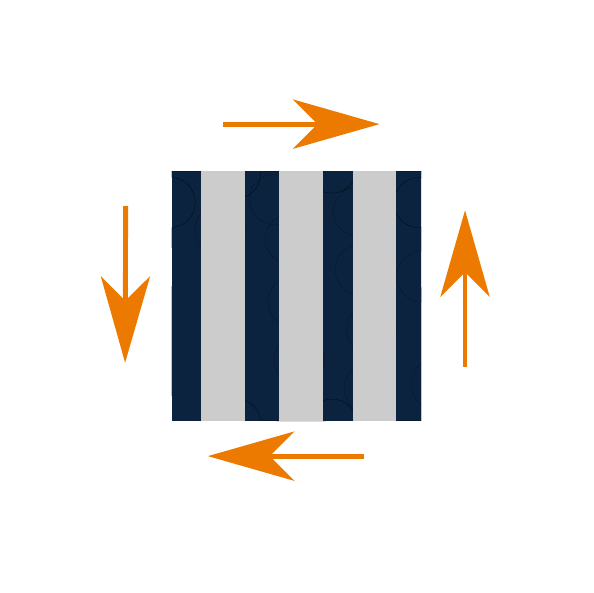}}}
    };
  \node[] at (0.0,-1.3) {$I_2$};

  \node[] at (3.0,-1.3) {$I_3$};
  \node[]            at (3.0,0.0)
    {
      {\def\svgwidth{0.13\columnwidth}{\scalebox{1.0}{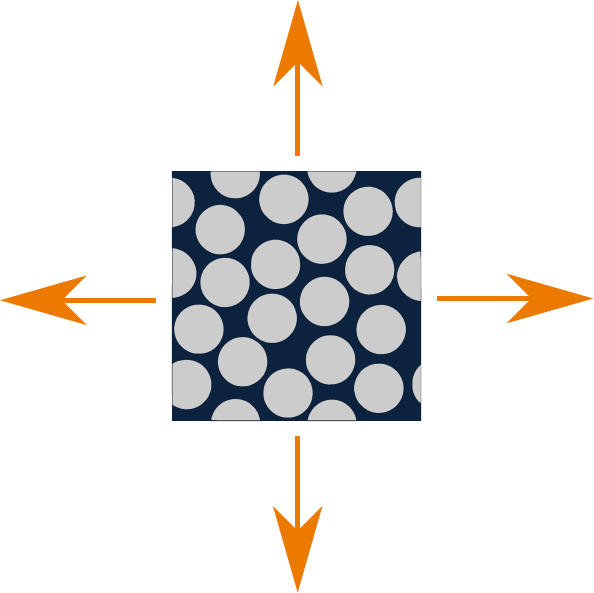}}}
    };
\end{tikzpicture}
\caption{The transversely isotropic stress invariants are related to transverse shear (\textit{left}), longitudinal shear (\textit{middle}) and biaxial tension or compression (\textit{right}).}
  \label{fig:invariants-diagrams}
\end{figure}

The material symmetries of the fiber reinforced polymer composite are represented with fiber direction (unit) vectors $\bm{a}_0$ and $\bar{\bm{a}}=\hfrac{\bm{a}}{\norm{\bm{a}}}$ in the \emph{initial} and \emph{current} configurations, respectively (see \Cref{fig:configurations}). Furthermore, the stress is first split into a plasticity inducing $\bm{\sigma}^{\mathrm{pind}}$ and a remaining (elastic) part \cite{spencer1972deformations, spencerKinematic1987}
\begin{equation}
  \bm{\sigma}^{\mathrm{pind}} = \bm{\sigma}  - (p~\textbf{I} + \sigma_\mathrm{f}~\bar{\A})
\end{equation}
where $p$ is the pressure, $\sigma_\mathrm{f}$ the part of the stress projection onto the fiber direction that exceeds the pressure and $\bar{\A} = \bar{\bm{a}} \otimes \bar{\bm{a}}$. The plasticity inducing stress can be determined from the total stress $\stress$ with the mapping
\begin{equation}
  \stress^\mr{pind} = \Ppind : \stress
  \label{eq:stress-pind}
\end{equation}
where $\Ppind$ is a fourth order tensor, given as 
\begin{align}
  \Ppind = \Ifourth - \frac{1}{2} \I \otimes \I - \frac{3}{2} \bar{\A} \otimes \bar{\A} + \frac{1}{2} \left( \bar{\A} \otimes \I - \I \otimes \bar{\A} \right)
  \label{eq:Ppind}
\end{align}
with $\Ifourth_{ijkl}=\delta_{ik} \delta_{jl}$. The following three transversely isotropic invariants are introduced
\cite{voglerModeling2013}
\begin{align}
  I_1 &= \frac{1}{2} \trace\left[\stress^{\mr{pind}}  
    \cdot \stress^{\mr{pind}} \right] - \bar{\bm{a}} \cdot \left[\stress^{\mr{pind}}  
  \cdot \stress^{\mr{pind}} \right]  \cdot \bar{\bm{a}} \\
    I_2 &= \bar{\bm{a}} \cdot \left[\stress^{\mr{pind}}  
    \cdot \stress^{\mr{pind}} \right] \cdot \bar{\bm{a}} \\
    I_3 &= \trace\left[\stress\right] - \bar{\bm{a}} \cdot \stress \cdot \bar{\bm{a}} \label{eq:matrix-pressure}
\end{align}
In local frame, where $\bm{e}_1$ is aligned with the fiber direction vector $\bm{a}$, the invariants read
\begin{align}
  I_1 &= \frac{1}{4} \left(\sigma_{22} - \sigma_{33} \right)^2 + \sigma^{2}_{23} \\ 
  I_2 &= \sigma^2_{12} + \sigma^2_{13}  \\
  I_3 &= \sigma_{22} + \sigma_{33}  
\end{align}
Note that $I_1$ is related to transverse shear, $I_2$ to longitudinal shear and $I_3$ to biaxial tension or compression in the transverse plane (see \Cref{fig:invariants-diagrams}). With these invariants, an equivalent stress can be constructed that does not induce yielding due to stress projections in the fiber direction. The \emph{total} equivalent stress $\sigmaEq$ that drives the evolution of the viscosity through \emph{stress shift factor} $a_\sigma$ is proposed as
\begin{equation}
  \sigmaEq = \sqrt{2\left(I_1+\alpha_2 I_2\right)} 
  \label{eq:total-equiv-stress}
\end{equation}
where $\alpha_2$ is a model parameter. 
The third invariant $I_3$ is used to describe pressure dependence of the polymer matrix, by extending the Eyring relation \Cref{eq:shift-factor-old} as
\begin{equation}
  a_\sigma  = \frac{\nicefrac{\sigmaEq}{\sigma_0}}{\sinh\left(\nicefrac{\sigmaEq}{\sigma_0}\right)}
  \exp\left(-\mu_{\mr{p}} \frac{I_3}{\sigma_0}\right)
  \label{eq:shift-factor}
\end{equation}
where $\mu_\mr{p}$ is a pressure dependency parameter. This relation is different from previous modifications of the Eyring relation for isotropic polymers \cite{govaertInfluence2000}, where the hydrostatic pressure was used.

\subsubsection{Equivalent stress of each mode}
\label{sec:seq-mode}
As mentioned in \Cref{sec:viscoplasticity}, the equivalent stress of each mode $\SigmaEqi$ is a function of the symmetric part of $\bm{\Sigma}_i$ and the fiber direction vector in the \emph{intermediate} configuration $\hat{\bm{a}}_i$. Replacing in \Cref{eq:stress-pind} and \Cref{eq:Ppind} quantities referring to the \emph{current} configuration $\{\stress,\bm{a}\}$ by quantities referring to the \emph{intermediate} configuration $\{\Mstress^{\mr{sym}}_i,\hat{\bm{a}}_i\}$, gives the plasticity inducing part of $\bm{\Sigma}^{\mr{sym}}_i$
\begin{equation}
  \Mstress^\mr{pind}_i = \hat{\Ppind}_i : \Mstress^{\mr{sym}}_i
\end{equation}
with corresponding fourth order tensor $\hat{\Ppind}_i$ 
\begin{align}
  \hat{\Ppind}_i = \Ifourth - \frac{1}{2} \I \otimes \I - \frac{3}{2} \hat{\A}_i \otimes \hat{\A}_i + \frac{1}{2} \left( \hat{\A}_i \otimes \I - \I \otimes \hat{\A}_i \right)
\end{align}
and invariants for each mode $i$
\begin{align}
  \hat{I}_{1i} &= \frac{1}{2} \trace\left[\Mstress^{\mr{pind}}_{i}
  \cdot \Mstress^{\mr{pind}}_{i} \right] - \hat{\bm{a}}_i \cdot \left[\Mstress^{\mr{pind}}_{i}
\cdot \Mstress^{\mr{pind}}_{i}\right] \cdot \hat{\bm{a}}_i \\
  \hat{I}_{2i} &= \hat{\bm{a}}_i \cdot \left[\Mstress^{\mr{pind}}_{i} 
\cdot \Mstress^{\mr{pind}}_{i} \right] \cdot \hat{\bm{a}}_i
  \label{eq:modal-invariants}
\end{align}
To prevent plastic flow in fiber direction and account for plastic incompressibility, only invariants $\hat{I}_{1i}$ and $\hat{I}_{2i}$, which are functions of $\Mstress^{\mr{pind}}_i$, are used to describe the direction of plastic flow through \Cref{eq:plastic-normal}. Similar to the \emph{total} equivalent stress $\sigmaEq$, the equivalent stress of mode $i$ is defined as
\begin{equation}
  \SigmaEqi = \sqrt{2\left(\hat{I}_{1i}+\alpha_2 \hat{I}_{2i}\right)} 
  \label{eq:modal-equiv-stress}
\end{equation}
with plastic normal direction
\begin{equation}
  \textbf{N}_{\mr{p}i} =  \pdv{\SigmaEqi}{\Mstress^\mr{sym}_i} = \frac{1}{\SigmaEqi} \left[\pdv{\hat{I}_{1i}}{\Mstress^{\mr{sym}}_i} + \alpha_2  \pdv{\hat{I}_{2i}}{\Mstress^\mr{sym}_i} \right]
  \label{eq:plastic-normal2}
\end{equation}
where $\alpha_2$ is the same model parameter as in \Cref{eq:total-equiv-stress}, to limit the number of parameters and aid their identification procedure. The derivatives of the invariants read
\begin{align}
  \pdv{\hat{I}_{1i}}{\Mstress^{\mr{sym}}_i} &= \purple{\left[\left(\I-\hat{\A}_i\right) \cdot \Mstress^{\mr{pind}}_i -\Mstress^{\mr{pind}}_i \cdot \hat{\A}_i \right]: \hat{\Ppind}_i \label{eq:dI1-dMs}} \\
  \pdv{\hat{I}_{2i}}{\Mstress^{\mr{sym}}_i} &= \left[\hat{\A}_i \cdot \Mstress^{\mr{pind}}_i + \Mstress^{\mr{pind}}_i \cdot \hat{\A}_i \right]: \hat{\Ppind}_i \label{eq:dI2-dMs} 
\end{align}

\subparagraph{Remark 1}
The \emph{total} equivalent stress $\sigmaEq$ is a function of $\stress$ and $\bar{\bm{a}}$, instead of $\bm{\Sigma}_i$ and $\hat{\bm{a}}_i$. The reason for this is that the latter quantities refer to an \emph{intermediate} configuration, which is different for each mode (see \Cref{fig:configurations}). Therefore, 'total versions' of $\bm{\Sigma}$ and $\hat{\bm{a}}$ do not exist.
\subparagraph{Remark 2}
In the present contribution, thermorheologically simple material behavior is assumed. The model can be extended to simulate thermorheologically complex behavior with several relaxation processes. A multiprocess model can be obtained by adding multiple driving stress contributions in parallel, where each contribution obeys an Eyring relation with a different parameter $\sigma_0$ \cite{klompenNonlinear1999} and a different relaxation spectrum.

\subparagraph{Remark 3}
As pointed out by \cite{vandermeerMicromechanical2016},
the difference between stress combinations $\sigma_{12}-\sigma_{22}$ and $\sigma_{12}-\sigma_{33}$ is not considered in the invariant formulation.
Furthermore, the effect on the yielding of a stress in fiber direction is removed. Although the material should not flow in the fiber direction, the stress in the fiber direction should contribute to the yielding of the polymer matrix under combined loading, for example longitudinal shear and stress in fiber direction. These assumptions remain limitations of the present mesoscale model.

\subparagraph{Remark 4}
In the equivalent stress definitions, only $\alpha_2$ is used as a coefficient of invariant $I_2$. The fact that $\alpha_2$ is cancelled in a transverse uniaxial tension and compression test simplifies the parameter identification procedure as will be shown in \Cref{sec:analytical-parameter-identification}.

\subsection{Embedded hyperelastic constitutive relations}
\label{sec:hyperelastic}
A hyperelastic transversely isotropic constitutive model \cite{bonetSimple1998} is used in this work to compute the stress in the composite material. The second Piola Kirhchhoff stress $\PKtwo$ is decomposed in an isotropic (iso) and a transversely isotropic part (trn) as
\begin{equation}
  \PKtwo = \PKtwo_{\mr{iso}} + \PKtwo_{\mr{trn}}
\end{equation}
Without plastic deformations, these contributions are given as
\begin{align}
  \begin{split}
  \PKtwo_{\mr{iso}} &= \mu (\I - \C^{-1} ) + \lambda J ( J - 1 ) \C^{-1} \\
  \PKtwo_{\mr{trn}} &= 2\beta (\xi_2 - 1)\I + 2\left[\alpha + \beta (\xi_1-3) + 2 \gamma (\xi_2-1)\right] \azero \otimes \azero - \alpha \left(\C \cdot \azero \otimes \azero + \azero \otimes \C \cdot \azero \right) 
\end{split}
\end{align}
where $\C=\transpose{\F}\cdot\F$ is the right Cauchy-Green deformation tensor and $J=\det\left(\F\right)$. The parameters $\lambda, \mu, \alpha, \beta$ and $\gamma$ are material constants that can be computed from the Young moduli and the Poisson ratios
\begin{equation}
\begin{aligned}
  n &= \frac{E_{22}}{E_{11}} \\
  m &= 1-\nu_{21} - 2n \nu^2_{21} \\
  \lambda &= E_{22} \frac{\nu_{21} + n\, \nu^2_{21}}{m (1+\nu_{21})} \\
  \mu &=  \frac{E_{22}}{2(1+\nu_{21})} \\
  \alpha &= \mu - G_{12}  \\
  \beta &= \frac{E_{22}\,\nu^2_{21}(1-n)}{4 m (1+\nu_{21})} \\
  \gamma &= \frac{E_{11}(1-\nu_{21})}{8 m } - \frac{\lambda + 2\mu}{8} + \frac{\alpha}{2} - \beta
\end{aligned}
  \label{eq:hyperelastic-params}
\end{equation}
Furthermore, $\xi_1$ and $\xi_2$ are defined as
\begin{align}
  \xi_1 &= \trace \left(\C\right) \\
  \xi_2 &= \bm{a} \cdot \bm{a}
\end{align}
In the present contribution, we use this hyperelastic transversely isotropic constitutive model to compute the stress in each mode $i$ when the material is mapped from its \emph{intermediate} configuration to the \emph{current} configuration (see \Cref{fig:configurations}). To this end, the following quantities are replaced by quantities that refer to the \emph{intermediate} configurations:
$\{\PKtwo, \azero, \C, \xi_1, J \} \rightarrow \{ \hat{\PKtwo}_i,  \ahat_i, \C_{\mr{e}i},  \xi_{1\mr{e}i}, J_{\mr{e}i} \}$. The relations for the hyperelastic model of each mode $i$ become
\begin{align}
  \begin{split}
  \hat{\PKtwo}_{\mr{iso},i} &= \mu_i (\I - \C^{-1}_{\mr{e}i} ) + \lambda_i J_{\mr{e}i} ( J_{\mr{e}i} - 1 ) \C^{-1}_{\mr{e}i} \\
  \hat{\PKtwo}_{\mr{trn},i} &= 2\beta_i (\xi_2 - 1)\I + 2\left[\alpha_i + \beta_i (\xi_{1\mr{e}i}-3) + 2 \gamma_i (\xi_{2}-1)\right] \ahat_i \otimes \ahat_i - \alpha_i \left(\C_{\mr{e}i} \cdot \ahat_i \otimes \ahat_i + \ahat_i \otimes \C_{\mr{e}i} \cdot \ahat_i \right) 
\end{split}
\label{eq:hyperelastic-intermediate}
\end{align}
Note that each mode has a different set of elastic constants. 
Furthermore, the vector $\ahat_i$ is a unit vector since plastic flow is excluded in fiber direction. 
Pushing forward \Cref{eq:hyperelastic-intermediate} from the \emph{intermediate} to the \emph{current} configuration gives the Cauchy stress contributions
\begin{align}
  \begin{split}
    \stress_{\mr{iso},i} &= \frac{\mu_i}{J_{\mr{e}i}} \left(\B_{\mr{e}i} - \I \right) + \lambda_i ( J_{\mr{e}i} - 1 ) \I  \\
  J_{\mr{e}i} \stress_{\mr{trn},i} &= 2\beta_i (\xi_2 - 1)\B_{\mr{e}i} + 2\left[\alpha_i + \beta_i (\xi_{1\mr{e}i}-3) + 2 \gamma_i (\xi_2-1)\right] \bm{a} \otimes \bm{a} - \alpha_i \left(\B_{\mr{e}i} \cdot \bm{a} \otimes \bm{a} + \bm{a} \otimes \B_{\mr{e}i} \cdot \bm{a} \right) 
\end{split}
  \label{eq:embedded-hyperelastic-relation}
\end{align}
where $\textbf{B}_{\mr{e}i} = \textbf{F}_{\mr{e}i} \cdot \textbf{F}^\mr{T}_{\mr{e}i}$ is the elastic right Cauchy-Green deformation tensor.
Note that the kinematics in \Cref{fig:configurations}, with re-orienting fiber direction vector(s) in the \emph{intermediate} configuration(s), are taken into account in the embedded hyperelastic model.

\subsection{Multimode model}
\label{sec:multimode-model}
\begin{figure}
  \centering
  \begin{tikzpicture}
        \node at (0,0)
  {
      {\def\svgwidth{0.6\columnwidth}{\scalebox{1.0}{
\begingroup%
  \makeatletter%
  \providecommand\color[2][]{%
    \errmessage{(Inkscape) Color is used for the text in Inkscape, but the package 'color.sty' is not loaded}%
    \renewcommand\color[2][]{}%
  }%
  \providecommand\transparent[1]{%
    \errmessage{(Inkscape) Transparency is used (non-zero) for the text in Inkscape, but the package 'transparent.sty' is not loaded}%
    \renewcommand\transparent[1]{}%
  }%
  \providecommand\rotatebox[2]{#2}%
  \newcommand*\fsize{\dimexpr\f@size pt\relax}%
  \newcommand*\lineheight[1]{\fontsize{\fsize}{#1\fsize}\selectfont}%
  \ifx\svgwidth\undefined%
    \setlength{\unitlength}{236.99352264bp}%
    \ifx\svgscale\undefined%
      \relax%
    \else%
      \setlength{\unitlength}{\unitlength * \real{\svgscale}}%
    \fi%
  \else%
    \setlength{\unitlength}{\svgwidth}%
  \fi%
  \global\let\svgwidth\undefined%
  \global\let\svgscale\undefined%
  \makeatother%
  \begin{picture}(1,0.51316768)%
    \lineheight{1}%
    \setlength\tabcolsep{0pt}%
    \put(0,0){\includegraphics[width=\unitlength,page=1]{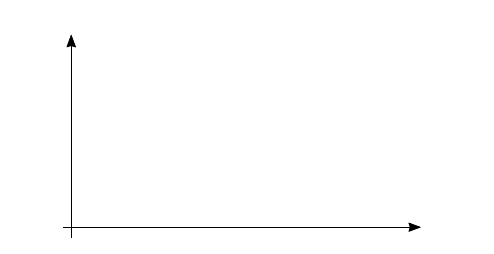}}%
    \put(1.13834859,-0.2650538){\color[rgb]{0,0,0}\makebox(0,0)[lt]{\begin{minipage}{0.05952906\unitlength}\raggedright \end{minipage}}}%
    \put(-0.30452208,0.44646){\color[rgb]{0,0,0}\makebox(0,0)[lt]{\begin{minipage}{1.01199376\unitlength}\raggedright \end{minipage}}}%
    \put(0.13288736,0.45830228){\color[rgb]{0,0,0}\makebox(0,0)[lt]{\lineheight{1.25}\smash{\begin{tabular}[t]{l}$\bm{\sigma}$\end{tabular}}}}%
    \put(0.90599035,0.04434827){\color[rgb]{0,0,0}\makebox(0,0)[lt]{\lineheight{1.25}\smash{\begin{tabular}[t]{l}$\bm{\varepsilon}$\end{tabular}}}}%
    \put(0,0){\includegraphics[width=\unitlength,page=2]{single-multi-mode-v3.pdf}}%
  \end{picture}%
\endgroup%
}}}
  };
  \end{tikzpicture}
  \caption{Stress-strain response with a single mode and with multiple modes.} 
  \label{fig:single-multi}
\end{figure}

Direction-, pressure- and rate-dependent yielding can be described by a single mode (see \Cref{fig:single-multi}), requiring four parameters: $\alpha_2$, $\mu_\mr{p}$, $\sigma_0$ and $\eta_0$. However, for polymers and polymer composites, a single viscosity is not sufficient to describe the nonlinear response prior to yielding \cite{tervoortMultimode1996, vanbreemenExtending2011}. A more accurate representation of the time-dependent pre-yield (and creep) response is obtained by including multiple modes (see \Cref{fig:single-multi}). With $N$ modes, the yield stress is then determined by the mode with the highest initial viscosity $\eta_0 = \max\{\eta_{0i}\}$.

A relaxation spectrum can be determined from a single stress-strain curve, obtained from a test under a constant strain rate as described in Ref. \cite{vanbreemenExtending2011}. This procedure was originally developed for isotropic polymers and recently extended to anisotropic yielding in \emph{short}- and \emph{long}-fiber composites \cite{amiri-radAnisotropic2019}. The same procedure is applied to the present model for \emph{continuous} fiber reinforced polymer composites and is briefly outlined here for completeness. For more details, the reader is referred to the Refs. \cite{vanbreemenExtending2011, amiri-radAnisotropic2019}.

The method makes use of a Boltzmann integral with $N$ unknown relaxation times to fit a 1D equivalent stress-strain curve from a constant strain rate test under off-axis angle $\theta$. The result of the procedure is a spectrum of moduli $\{E_{\theta i}\}$ and initial viscosities $\{\eta_{0i}\}$. It is then assumed that the ratio
\begin{equation}
  m_i = \frac{{E_\theta}_i}{\sum^N_i {E_\theta}_i}
\end{equation}
is the same for $E_{11}$, $E_{22}$ and $G_{12}$. With the set of ratios $\{m_i\}$, the elastic constants are obtained for each mode
\begin{align}
  \begin{split}
  E_{11i} &= m_i E_{11} \\
  E_{22i} &= m_i E_{22} \\
  G_{12i} &= m_i G_{12} \\
  \nu_{21i} &= \nu_{21} \\
\end{split}
  \label{eq:stiffness-distribution}
\end{align}
The hyperelastic parameters for each mode are obtained with \Cref{eq:hyperelastic-params}, replacing constants $\{E_{11}, E_{22}, G_{12}\}$ by
$\{E_{11i}, E_{22i}, G_{12i}\}$.

\subsection{Integration of the constitutive relations}
To compute the stress in each mode from the elastic deformation, the plastic deformation must be known, which in turn depends, through the non-Newtonian flow rule, on the stress in each mode and on the \emph{total} stress through the stress-dependent shift factor. 
This renders a nonlinear relation between the total stress and deformation gradient, that must be solved with an iterative scheme.

\subsubsection{Nested scheme}
 
\begin{figure}
  \hspace{-0.5cm}
  \begin{tikzpicture}
        \node at (0.0,0)
  {
      {\def\svgwidth{1.10\columnwidth}{\scalebox{1.0}{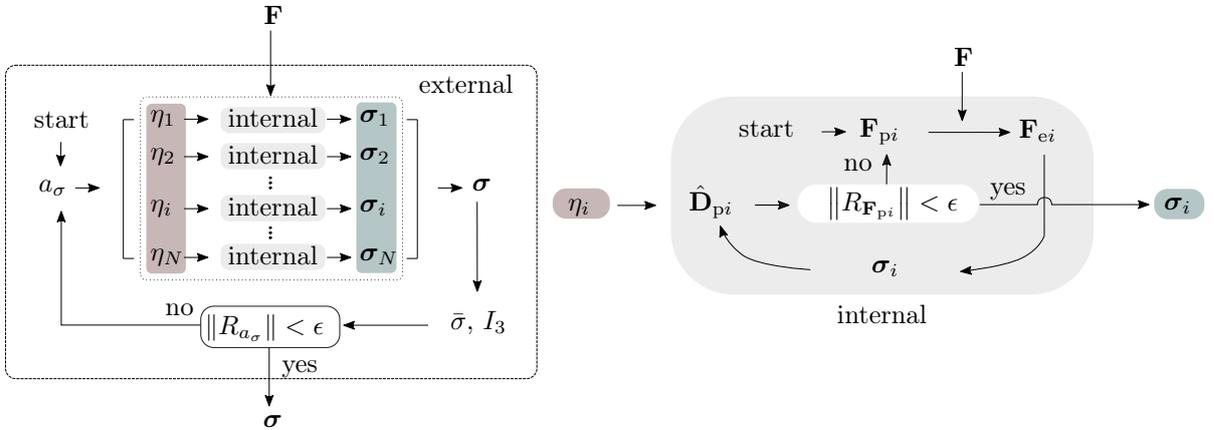}}}
  };
  \end{tikzpicture}
  \caption{Nested external-internal solution scheme. At every external iteration (\textit{left}), $N$ internal schemes are solved, one for each mode $i$ (\textit{right}).}
  \label{fig:nested}
\end{figure}

Following Ref. \cite{khaleghiThermodynamically2022}, a nested scheme with an \emph{external} and \emph{internal} solution process is used (see \Cref{fig:nested}). In the \emph{external} scheme, the \emph{stress shift factor} $a_\sigma$ is iteratively solved with Newton iterations. For every \emph{external} iteration, the viscosities $\{\eta_i\}$ of the modes are known, which allows for computing the stress in each mode $\stress_i$ separately with an \emph{internal} Newton-Raphson scheme. 
 
\subsubsection{External Newton-Raphson scheme}
For solving the \emph{stress shift factor} $a_\sigma$, \Cref{eq:shift-factor} is cast in residual form
\begin{align}
  R_{a_\sigma} = a_{\sigma} - \frac{\nicefrac{\bar{\sigma}}{\sigma_0}}{\sinh\left( \nicefrac{\bar{\sigma}}{\sigma_0}\right)
  } \exp\left(-\mu_\mr{p} \frac{I_3}{\sigma_0}\right)
  \label{eq:external-residual}
\end{align}
The root of this equation is found with Newton iterations $j=1\dots N_\mr{iter}$ by updating $a_\sigma$ as follows
\begin{align}
  a^{(j+1)}_\sigma = a^{(j)}_\sigma - \frac{R^{(j)}_{a_\sigma}}{{\left.\pdv{R_{a_\sigma}}{a_\sigma}\right|^{(j)}_{\F}}}
  \label{eq:external-iter-scheme}
\end{align}
where $\left.\hpdv{R_{a_\sigma}}{a_\sigma}\right|_{\F}$ is the Jacobian for the \emph{external} scheme, which is derived in \Cref{sec:jacobian-external}.
  For each \emph{external} Newton iteration $j$, the stress in each mode $\stress_i$ is found with the \emph{internal} Newton-Raphson scheme, with viscosity  $\eta^{(j)}_i = \eta_{0i} {a_\sigma}^{(j)}$. Subsequently, the \emph{total equivalent stress} $\sigmaEq$ is computed and the residual $R_{a_\sigma}$ and the Jacobian $\left.\hpdv{R_{a_\sigma}}{a_\sigma}\right|_{\F}$ are evaluated to update the stress shift factor $a^{(j+1)}_{\sigma}$ for the next iteration with \Cref{eq:external-iter-scheme}. 

\subsubsection{Internal Newton-Raphson scheme}
In the \emph{internal} scheme, the plastic deformation $\textbf{F}_{\mr{p}i}$ is chosen as primary unknown. The time integration of \Cref{eq:plastic-evolution} is performed with an implicit exponential map \cite{eterovicHyperelasticbased1990,weberFinite1990} to retain plastic incompressibility (\Cref{eq:isochoric-assumption}) \cite{sansourLarge1998}. 
The deformation gradient at the current time step $\F_{\mr{p}i}$ is computed from the deformation gradient at the previous time step $\F^0_{\mr{p}i}$ as
\begin{align}
  \F_{\mr{p}i} = \exp(\hat{\D}_{\mr{p}i} \dt) \cdot \F^0_{\mr{p}i}
  \label{eq:exponential-map}
\end{align}
where the tensor exponential function is replaced by a Pad\'{e} approximation
\cite{baaserPadeapproximation2004}
\begin{align}
  \exp\left(\hat{\D}_{\mr{p}i} \dt\right) \approx \bm{\Pi}\left(\hat{\D}_{\mr{p}i}, \dt\right) = \left(\I-\frac{\dt}{2} \hat{\D}_{\mr{p}i}\right)^{-1} \cdot \left(\I+\frac{\dt}{2} \hat{\D}_{\mr{p}i}\right)
\label{eq:pade-approx}
\end{align}
Casting this equation in residual form yields
\begin{align}
  \textbf{R}_{\F_{\mr{p},i}} &= \F_{\mr{p}i} - \bm{\Pi}\left(\dt, \hat{\D}_{\mr{p}i}\right)\cdot \textbf{F}^0_{\mr{p}i}
  \label{eq:internal-residual}
\end{align}
The root of this equation is solved by updating the plastic deformation for each \emph{internal} iteration $k=1 \dots N_\mr{iter}$ as follows
\begin{equation}
  {\F}^{(k+1)}_{\mr{p}i} = {\F}^{(k)}_{\mr{p}i} - \left[{\pdv{\textbf{R}_{\F_{\mr{p}i}}}{\F_{\mr{p}i}} }^{(k)}\right]^{-1} : \textbf{R}^{(k)}_{\F_{\mr{p}i}}
\label{eq:external-iters}
\end{equation}
where $\hpdv{\textbf{R}_{\textbf{F}_{\mr{p}i}}}{\textbf{F}_{\mr{p}i}}$, is the Jacobian for the \emph{internal} Newton-Raphson scheme, which is given in \Cref{sec:jacobian-internal}. With the plastic deformation $\F_{\mr{p}i}$, the elastic deformation in each mode $\F_{\mr{e}i}$ is computed with \Cref{eq:deformation-decomposition} and the stress $\stress_i$ with \Cref{eq:embedded-hyperelastic-relation}. Subsequently, the total equivalent stress $\sigmaEq$ is computed with \Cref{eq:total-equiv-stress}, after which the \emph{internal} residual $\textbf{R}_{\F_{\mr{p}i}}$ and Jacobian $\hpdv{{\textbf{R}}_{\textbf{F}_{\mr{p}i}}}{\textbf{F}_{\mr{p}i}}$, are evaluated to update the plastic deformation for the next iteration with \Cref{eq:external-iters}.

\subparagraph{Remark 5}
The time step dependence from the time integration scheme with Pad\'{e} approximation, \Cref{eq:internal-residual}, is assessed in \Cref{sec:results}. 
For a better approximation of the exponential map, a higher-order Pad\'{e} approximation \cite{baaserPadeapproximation2004} could be used.

\subsection{Jacobians}
\label{sec:jacobians}
The Jacobians for the \emph{internal} and \emph{external} Newton-Raphson schemes and the consistent tangent modulus for the \emph{global} implicit solution scheme are derived in this section.

\subsubsection{Jacobian of the internal scheme}
\label{sec:jacobian-internal}
The Jacobian of the \emph{internal} residual (\Cref{eq:internal-residual}) reads
\begin{align}
  \pdv{\textbf{R}_{\F_{\mr{p}i}}}{\textbf{F}_{\mr{p}i}} = 
  \Ifourth + \pdv{\textbf{R}_{\F_{\mr{p}i}}}{\bm{\Pi}_i} 
  : \pdv{\bm{\Pi}_i}{\hat{\textbf{D}}_{\mr{p}i}} :
  \left[
    \pdv{\hat{\textbf{D}}_{\mr{p}i}}{\bm{\Sigma}^\mr{sym}_i} :
    \pdv{\bm{\Sigma}^{\mr{sym}}_{i}}{\textbf{F}_{\mr{e}i}} :
    \pdv{\textbf{F}_{\mr{e}i}}{\textbf{F}_{\mr{p}i}}
    + \pdv{\hat{\textbf{D}}_{\mr{p}i}}{\hat{\bm{a}}_i} \cdot  \pdv{\hat{\bm{a}}_i}{\textbf{F}_{\mr{p}i}}
\right]
  \label{eq:internal-jacobian}
\end{align}
The derivatives in this expression are given in \ref{sec:internal-jac}. 

\subsubsection{Jacobian of the external scheme}
\label{sec:jacobian-external}
The Jacobian of the \emph{external} residual (\Cref{eq:external-residual}) reads
\begin{align}
  \pdv{R_{a_\sigma}}{a_{\sigma}} = 1 + \left[\pdv{R_{a_\sigma}}{\sigmaEq}
  \pdv{\bar{\sigma}}{\stress}  +
\pdv{R_{a_\sigma}}{I_3} \pdv{I_3}{\stress} \right]: \left[\sum^{N}_{i=1} \pdv{\stress_i}{\F_{\mr{e}i}} : \pdv{\F_{\mr{e}i}}{\F_{\mr{p}i}} : \pdv{\F_{\mr{p}i}}{a_\sigma} \right]
 \label{eq:external-jac}
\end{align}
where $\hpdv{\bar{\sigma}}{\stress}$ follows from \Cref{eq:plastic-normal2} by replacing \emph{intermediate} quantities $\{\bar{\Sigma}_i,\bm{\Sigma}^\mr{sym},\hat{I}_{1i}, \hat{I}_{2i}\}$ with \emph{current} quantities $\{\sigmaEq,\stress,I_1, I_2\}$.
The first and second terms in the sum on the RHS are given by \Cref{eq:dsigma-dFe} and \Cref{eq:dFe-dFp}, respectively. The other terms are given in \ref{sec:derivatives-external}. The terms $\{\hpdv{{\F}_{\mr{p}i}}{a_\sigma}\}$ are obtained as follows.
The \emph{internal} residual for mode $i$ is a function of independent variables $a_\sigma$ and $\textbf{F}_{\mr{p}i}$. Therefore, the variation of the residual reads
\begin{align}
  \delta \textbf{R}_{\F_{\mr{p}i}} = \pdv{ \textbf{R}_{\F_{\mr{p}i}}  }{a_\sigma} \delta a_\sigma + \pdv{\textbf{R}_{\F_{\mr{p}i}}}{\textbf{F}_{\mr{p}i}} : \delta \textbf{F}_{\mr{p}i} 
  \label{eq:consistency-2}
\end{align}
Since we solve iteratively for the root of $\textbf{R}_{\F_{\mr{p}i}}$ with the \emph{internal} scheme, its variation between \emph{external} iterations $j$ vanishes, \ie $\delta \textbf{R}_{\F_{\mr{p}i}} = \bm{0}$. This is a \emph{consistency condition} that can be used for finding $\hpdv{\textbf{F}_{\mr{p}i}}{a_\sigma}$, similar to what is done in deriving consistent tangent moduli in classical plasticity models with return mapping schemes.

The \emph{consistency condition} $\delta \textbf{R}_{\F_{\mr{p}i}} = \bm{0}$ gives, after rewriting, the sought-after derivative $\hpdv{\textbf{F}_{\mr{p}i}}{a_\sigma}$ 
\begin{align}
  \delta \F_{\mr{p}i} = \underbrace{-\left[\pdv{  \textbf{R}_{\F_{\mr{p}i}}}{\textbf{F}_{\mr{p}i}}\right]^{-1} : \pdv{\textbf{R}_{\F_{\mr{p}i}} }{a_\sigma}}_{\pdv{\textbf{F}_{\mr{p}i}}{a_\sigma}} \delta a_\sigma
  \label{eq:dFp-da}
\end{align}
where the first term on the RHS  is the Jacobian for the \emph{internal} scheme (\Cref{eq:internal-jacobian}). The second term on the RHS is given in \ref{sec:derivatives-external}.

\subsubsection{Consistent tangent modulus}
\label{sec:global-tangent}

The derivative of the Cauchy stress with respect to the deformation gradient reads
\begin{equation}
  \pdv{\stress}{\F} = \sum^{N}_i \left[\pdv{\stress_i}{\F_{\mr{e}i}} : \left(\pdv{\F_{\mr{e}i}}{\F} +  \pdv{\F_{\mr{e}i}}{{\F_{\mr{p}i}}} : \pdv{\F_{\mr{p}i}}{\F} \right) + \pdv{\stress_i}{\bm{a}} \cdot \pdv{\bm{a}}{\F} \right]
  \label{eq:dsigmai-dF}
\end{equation}
where $\hpdv{\F_{\mr{e}i}}{{\F_{\mr{p}i}}}$,  $\hpdv{\stress_i}{\Fe}$ and  $\hpdv{\stress_i}{\bm{a}}$ are given by \Cref{eq:dFe-dFp,eq:dsigma-dFe,eq:dsigmai-da}. The derivatives $\hpdv{\F_{\mr{e}i}}{{\F}}$ and $\hpdv{\bm{a}}{\F}$ can be found by differentiating \Cref{eq:deformation-decomposition,eq:a}. Furthermore, $\partial {\F_{\mr{p}i}} / \partial \F$ in \Cref{eq:dsigmai-dF} reads
\begin{equation}
  \pdv{\F_{\mr{p}i}}{\F} = \left.\pdv{\F_{\mr{p}i}}{\F}\right|_{a_\sigma} + \pdv{\F_{\mr{p}i}}{a_\sigma} \otimes \pdv{a_\sigma}{\F}
    \label{eq:dFpi-dF}
\end{equation} 
where $\hpdv{\textbf{F}_{\mr{p}i}}{a_\sigma}$ is already given by \Cref{eq:dFp-da}. Furthermore, $\left.\hpdv{\F_{\mr{p}i}}{\F}\right|_{a_\sigma}$ is derived through an additional consistency condition of the \emph{internal} scheme. 
For every fixed $a_\sigma$ and varying $\F$, the internal residual for mode $i$ vanishes between \emph{global} iterations. Therefore
\begin{equation}
  \delta \textbf{R}_{\F_{\mr{p}i}}  = \pdv{\textbf{R}_{\F_{\mr{p}i}}}{\textbf{F}_{\mr{p}i}} : \delta {\F}_{\mr{p}i} + \pdv{\textbf{R}_{\F_{\mr{p}i}}}{\F} : \delta \F = 0
\end{equation}
The derivative $\left.\hpdv{\F_{\mr{p}i}}{\F}\right|_{a_\sigma}$ can be found by rewriting this expression
\begin{equation}
  \delta \F_{\mr{p}i} = - \underbrace{\left[\pdv{\textbf{R}_{\F_{\mr{p}i}}}{\textbf{F}_{\mr{p}i}}\right]^{-1} : {\pdv{\textbf{R}_{\F_{\mr{p}i}}}{\F}}}_{\left.\pdv{\F_{\mr{p}i}}{\F}\right|_{a_\sigma}} : \delta \F= 0
  \label{eq:dFp-dF}
\end{equation}
where the first term on the RHS is again the Jacobian of the \emph{internal} residual (see \Cref{eq:internal-jacobian}) and the second term on the RHS is given in \ref{sec:derivatives-tangent}.

The third derivative $\hpdv{a}{\F}$ on the RHS of \Cref{eq:dFpi-dF}, which is the same for each mode, is obtained with a single \emph{consistency condition} of the \emph{external} scheme. At every \emph{global} iteration, the \emph{external} residual vanishes. Therefore
\begin{equation}
  \delta R_{a_\sigma} = \pdv{R_{a_\sigma}}{\F}: \delta \F +\pdv{R_{a_\sigma}}{a_\sigma} \delta a_\sigma = 0
\end{equation}
Rewriting this equation yields
\begin{equation}
  \delta a_\sigma  = \underbrace{- \left[ \pdv{{R_a}_\sigma}{a_\sigma} \right]^{-1}\pdv{R_{a_\sigma}}{\F}}_{\pdv{a}{\F}}: \delta \F 
      \label{eq:dashift-dF}
\end{equation}
where the derivative $\hpdv{a}{\F}$ is identified.
Note that $\partial R_{a_\sigma} / \partial a_\sigma$ is the Jacobian of the \emph{external} scheme (\Cref{eq:external-jac}). The second term is given in \ref{sec:derivatives-tangent}.

\subparagraph{Remark 6} \noindent In total, $2N+1$ \emph{consistency conditions} are used to derive the tangent modulus.

\section{Parameter identification}
\label{sec:analytical-parameter-identification}

\begin{figure}
   \centering
  \begin{tikzpicture}
  \node[] at (0,0)
    {
     {\def\svgwidth{0.8\columnwidth}{\scalebox{1.1}{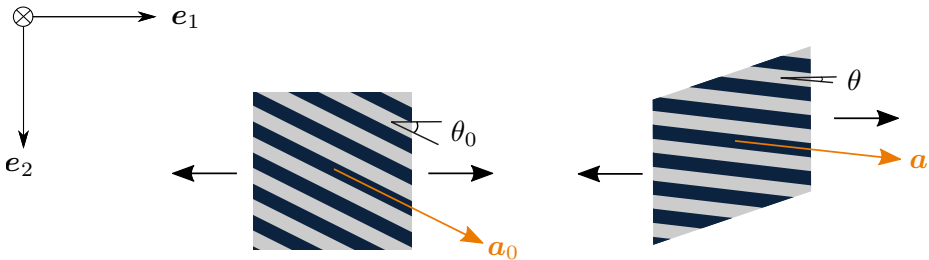}}}
    };
  \end{tikzpicture}
  \caption{Fiber reinforced polymer composite under off-axis tensile loading.}
  \label{fig:off-axis-global-deformations}
\end{figure}
 
To determine the (single-mode) yield parameters of the mesoscopic constitutive model, we consider a material point under uniaxial tension and compression with off-axis angle $\theta_0$ at constant strain rate $\dot{\varepsilon}$ (see \Cref{fig:off-axis-global-deformations}). 
In addition, we assume small deformations at the moment of yielding, such that: $\hat{\bm{a}} = \bm{a}_0$, $\stress = \bm{\Sigma} = \bm{\Sigma}^\mr{sym}$, and $\sigmaEq=\bar{\Sigma}$.
Furthermore, we choose an orthonormal basis $\{\bm{e}_i\}_{i=1,2,3}$ where unit vector $\bm{e}_1$ is aligned with the load direction. 
The flow rule (\Cref{eq:flow-rule}) gives the rate of plastic deformation in the load direction
\begin{equation}
  D^{\mr{p}}_{11} = \frac{\sigma_0}{\eta_0} \sinh\left(\frac{\sigmaEq}{\sigma_0}\right) \exp\left( - \mu_\mr{p}{\frac{I_3}{\sigma_0}}\right) \pdv{\sigmaEq}{\sigma_{11}} 
  \label{eq:1D-flow-rule}
\end{equation}
Plastic and elastic deformations develop simultaneously until the rate of plastic deformation is equal to the applied strain rate ($D^\mr{p}_{11}=\dot{\epsilon}$) upon which the stress reaches a plateau\footnote{Under large deformations, a geometric hardening or softening response may occur due to re-orientation of the fibers}, which marks the moment of yielding. 
When the material yields, $\sigmaEq \gg \sigma_0$ and the hyperbolic sine function can be approximated with an exponential function
\begin{equation}
  \dot{\epsilon} \approx \frac{\sigma_0}{2\eta_0} \exp\left(\frac{\sigmaEq-\mu_\mr{p} I_3}{\sigma_0}\right) \pdv{\sigmaEq}{\sigma_{11}} 
  \label{eq:analytical-yield}
\end{equation}
This equation provides an analytical relation between the applied strain rate $\dot{\varepsilon}$ and the equivalent stress $\sigmaEq$ at the moment of yielding.

\subsection{Transverse tension and compression}
\label{sec:transverse-tension-compression}

The parameters $\mu_\mr{p}$, $\sigma_0$ and $\eta_{0}$ can be determined from stress-strain curves of uniaxial tension and compression under off-axis angle $\theta_0=90^\circ$ at equal strain rates. For this angle, $I_2$ is zero and $\alpha_2$ is eliminated from the equations. The transversely isotropic stress invariants at the moment of yielding read
\begin{equation}
  I_1 = \frac{\sigma^2_{\mr{y},90}}{4}, \quad I_2 = 0,
  \quad I_3 = \begin{cases}
    \sigma_{\mr{y},90\mr{t}} & \text{in tension} \\
    -\sigma_{\mr{y},90\mr{c}} & \text{in compression}
  \end{cases}
\end{equation}
where $\sigma_{\mr{y},90}$ is the yield stress at $\theta=90^\circ$.
Substitution in \Cref{eq:total-equiv-stress,eq:plastic-normal2} and rewriting \Cref{eq:analytical-yield} provides the following expressions of the yield stresses in tension $\sigma_{\mr{y},90\mr{t}}$ and compression $\sigma_{\mr{y},90\mr{c}}$
\begin{align}
  \sigma_{\mr{y},90\mr{t}} &= \frac{\sigma_0}{\frac{1}{\sqrt{2}} + \mu_\mr{p}} \ln\left(2 \sqrt{2} \frac{\eta_0}{\sigma_0}   \dot{\epsilon}\right)\\
  \sigma_{\mr{y},90\mr{c}} &= \frac{\sigma_0}{\frac{1}{\sqrt{2}} - \mu_\mr{p}}
 \ln\left(2 \sqrt{2} \frac{\eta_0}{\sigma_0}   |\dot{\epsilon}|\right)
  \label{eq:compression-yield-strainrate}
\end{align}
When the yield stresses $\sigma_{\mr{y},90\mr{t}}$ and $\sigma_{\mr{y},90\mr{c}}$ are known, $\mu_\mr{p}$ is solved for, which gives the following closed-form relation
\begin{equation}
  \mu_\mr{p} = \frac{1}{\sqrt{2}}\left( \frac{\sigma_{\mr{y,90c}}-\sigma_{y,90t}}{\sigma_{y,90c}+\sigma_{y,90t}}\right)
   \label{eq:mu-id}
\end{equation}
With $\mu_\mr{p}$ known, $\sigma_0$ and $\eta_0$ are determined from an Eyring plot for uniaxial compression. This requires at least two compression curves at different strain rates. \Cref{eq:compression-yield-strainrate} is rearranged as
\begin{equation}
  \sigma_{\mr{y},90\mr{c}} = \underbrace{\frac{\sigma_0 \ln(10)}{\frac{1}{\sqrt{2}} - \mu_\mr{p}}}_{\mr{slope}\,m} \left[\log_{10}(|\dot{\epsilon}|) +
    \log_{10}\left(2 \sqrt{2} \frac{\eta_0}{\sigma_0}\right)\right]
\label{eq:analytical-eyring}
\end{equation}
where $m$ is the slope in a semi-log plot of yield stress $\sigma_{\mr{y},90\mr{c}}$ \textit{vs} strain rate $\dot{\varepsilon}$. From the slope, $\sigma_0$ and $\eta_0$ are given by
\begin{align}
  \sigma_0 &= m \left(\frac{ \frac{1}{\sqrt{2}} - \mu_\mr{p}}{\ln(10)}\right)
  \label{eq:sigma0-id} \\
  \eta_0 &= \frac{\sigma_0 \, 10^{\frac{\sigma_{\mr{y},90c}}{m}}}{2\sqrt{2} \, \dot{\varepsilon}} \label{eq:eta0-id}
\end{align}

\subsection{Off-axis loading in tension}
\label{sec:off-axis-tension}
Parameter $\alpha_2$ can be obtained from any other test where $I_2$ is non-zero, for example the $\theta_0=30^\circ$ case. By following the same steps as before, the analytical yield stress for this angle reads
\begin{equation}
  \sigma_{\mr{y},30\mr{t}} = \frac{4 \sigma_0}{\mu_\mr{p} + \sqrt{\frac{1}{2} + 6 \alpha_2}}
  \ln\left(\frac{8}{\sqrt{\frac{1}{2} + 6 \alpha_2}} \frac{\eta_0}{\sigma_0} \dot{\epsilon}\right)
    \label{eq:a2-id}
\end{equation}
which is a nonlinear equation in its argument $\alpha_2$ that can be solved numerically, given $\eta_0, \sigma_0, \mu_\mr{p}, \sigma_{\mr{y},30\mr{t}}$ and corresponding strain rate $\dot{\epsilon}$.

\section{Numerical homogenization of a micromodel}
\label{sec:calibration}
\begin{figure}
   \centering
   \hspace{1.0cm}
  \begin{tikzpicture}

   \node[] at (-7.1,0.9)
    {
     {\def\svgwidth{0.15\columnwidth}{\scalebox{1.0}{
\begingroup%
  \makeatletter%
  \providecommand\color[2][]{%
    \errmessage{(Inkscape) Color is used for the text in Inkscape, but the package 'color.sty' is not loaded}%
    \renewcommand\color[2][]{}%
  }%
  \providecommand\transparent[1]{%
    \errmessage{(Inkscape) Transparency is used (non-zero) for the text in Inkscape, but the package 'transparent.sty' is not loaded}%
    \renewcommand\transparent[1]{}%
  }%
  \providecommand\rotatebox[2]{#2}%
  \newcommand*\fsize{\dimexpr\f@size pt\relax}%
  \newcommand*\lineheight[1]{\fontsize{\fsize}{#1\fsize}\selectfont}%
  \ifx\svgwidth\undefined%
    \setlength{\unitlength}{85.67503621bp}%
    \ifx\svgscale\undefined%
      \relax%
    \else%
      \setlength{\unitlength}{\unitlength * \real{\svgscale}}%
    \fi%
  \else%
    \setlength{\unitlength}{\svgwidth}%
  \fi%
  \global\let\svgwidth\undefined%
  \global\let\svgscale\undefined%
  \makeatother%
  \begin{picture}(1,0.88912826)%
    \lineheight{1}%
    \setlength\tabcolsep{0pt}%
    \put(0,0){\includegraphics[width=\unitlength,page=1]{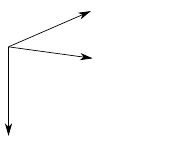}}%
    \put(-0.00581322,0.0165165){\color[rgb]{0,0,0}\makebox(0,0)[lt]{\lineheight{1.25}\smash{\begin{tabular}[t]{l}$\bm{e}_3$\end{tabular}}}}%
    \put(0.56865988,0.52952544){\color[rgb]{0,0,0}\makebox(0,0)[lt]{\lineheight{1.25}\smash{\begin{tabular}[t]{l}$\bm{e}_2$\end{tabular}}}}%
    \put(0.54403898,0.83592045){\color[rgb]{0,0,0}\makebox(0,0)[lt]{\lineheight{1.25}\smash{\begin{tabular}[t]{l}$\bm{e}_1$\end{tabular}}}}%
  \end{picture}%
\endgroup%
}}}
    };

  \node[] at (0.0,0.9)
    {
     {\def\svgwidth{0.45\columnwidth}{\scalebox{1.0}{
\begingroup%
  \makeatletter%
  \providecommand\color[2][]{%
    \errmessage{(Inkscape) Color is used for the text in Inkscape, but the package 'color.sty' is not loaded}%
    \renewcommand\color[2][]{}%
  }%
  \providecommand\transparent[1]{%
    \errmessage{(Inkscape) Transparency is used (non-zero) for the text in Inkscape, but the package 'transparent.sty' is not loaded}%
    \renewcommand\transparent[1]{}%
  }%
  \providecommand\rotatebox[2]{#2}%
  \newcommand*\fsize{\dimexpr\f@size pt\relax}%
  \newcommand*\lineheight[1]{\fontsize{\fsize}{#1\fsize}\selectfont}%
  \ifx\svgwidth\undefined%
    \setlength{\unitlength}{335.43552188bp}%
    \ifx\svgscale\undefined%
      \relax%
    \else%
      \setlength{\unitlength}{\unitlength * \real{\svgscale}}%
    \fi%
  \else%
    \setlength{\unitlength}{\svgwidth}%
  \fi%
  \global\let\svgwidth\undefined%
  \global\let\svgscale\undefined%
  \makeatother%
  \begin{picture}(1,0.5186683)%
    \lineheight{1}%
    \setlength\tabcolsep{0pt}%
    \put(0,0){\includegraphics[width=\unitlength,page=1]{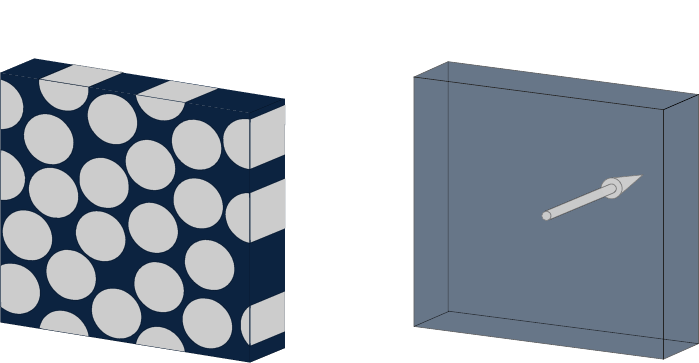}}%
    \put(0.920449,0.2754955){\color[rgb]{1,1,1}\makebox(0,0)[lt]{\lineheight{1.25}\smash{\begin{tabular}[t]{l}$\bm{a}_0$\end{tabular}}}}%
  \end{picture}%
\endgroup%
}}}
    };

    \node[anchor=west] (note) at (-6,1.8) {isotropic EGP};
    \draw[-latex, thick, TUDblue] (note) to[out=0, in=160] (-2.50,1.30);
    \node[anchor=west] (note2) at (-6,0.5) {hyperelastic};
    \draw[-latex, thick, TUDblue] (note2) to[out=0, in=150] (-3.30,-0.10);
    \node[] at (-2,-1.5) {\textbf{micro}};
    \node[] at (2,-1.5) {\textbf{meso}};
    \node[] (note3) at (5.5,0.7) {Invariant-based EGP};
    \draw[-latex, thick, TUDblue] (note3.west) to[out=180, in=0] (2.80,0.37);

  \end{tikzpicture}
  \caption{Micromodel with hyperelastic fibers and isotropic EGP model for the matrix \textit{vs} mesomodel with proposed invariant-based EGP model and fiber direction vector $\bm{a}_0$.}
  \label{fig:rve-macro}
\end{figure}
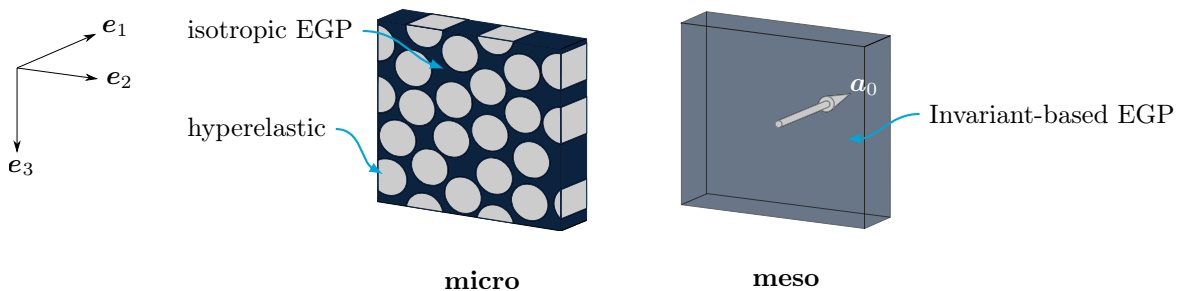

The parameters of the mesoscopic material model are determined by homogenizing a previously calibrated micromodel, with periodic boundary conditions, for carbon/PEEK \cite{kovacevicRateDependentFailure2022}. The micromodel comprises of hyperelastic transversely isotropic fibers and viscoplastic polymer matrix, where the latter is modeled with the original isotropic EGP model \cite{tervoortMultimode1996,vanbreemenExtending2011}. The micromodel and mesomodel are schematically depicted in \Cref{fig:rve-macro}. 

\subsection{Boundary conditions for off-axis loading}
Applying off-axis loads to the micromodel (as shown in \Cref{fig:off-axis-global-deformations}) is not straightforward. Since periodic boundary conditions are applied, it is not possible to vary the fiber angle inside the micromodel, which would violate the assumption of continuous fibers as imposed by the periodicity. Instead, off-axis loading is achieved by aligning the micromodel with the fibers, while a global deformation is applied in the local frame of the micromodel. Since the local frame changes under off-axis loading due to re-orientation of the fibers (see \Cref{fig:off-axis-global-deformations}), a special constraint equation is used that accounts for these re-orientations \cite{kovacevicStrainRateArclength2022, kovacevicMicromechanical2024}. 

In contrast, global deformations can straightforwardly be applied on a single element with the mesoscopic model. Off-axis loading is then achieved by varying the initial fiber direction vector $\bm{a}_0$, while applying the load in the $\bm{e}_1$-direction. 
Although the methods to apply boundary conditions on the micromodel and the mesomodel are different, the resulting (global) deformations are the same.

\subsection{Elasticity parameters}
The elasticity parameters of the mesoscopic material model are determined by subjecting the micromodel to three basic load cases: longitudinal tension, longitudinal shear and transverse shear. 
The transversely isotropic elasticity constants are given in \Cref{tab:elasticity-params}.

\begin{table}
\centering
\renewcommand{\arraystretch}{1.3}
\caption{Elasticity constants}
\label{tab:elasticity-params}
\begin{tabular}{ccccc}
\hline
$E_1$ [GPa] & $E_2$ [GPa] & $G_{12}$ [GPa] & $\nu_{21}$  \\
\hline
55.5 & 7.4 & 4.8 &  0.016  \\
\hline
\end{tabular}
\end{table}

\begin{table}
\centering
\renewcommand{\arraystretch}{1.3}
\caption{Plasticity parameters}
\label{tab:invariant-egp-params}
\begin{tabular}{cccccc}
\hline
$\mu_\mr{p} $ & $\sigma_0 $ [MPa]  & $\eta_{0}$ [MPa s] & $\alpha_2$ \\
\hline
0.053 & 1.71 & \SI{5.90e29}{} & 1.147  \\
\hline
\end{tabular}
\end{table}

\begin{table}
\centering
\caption{Relaxation spectrum}
\label{tab:relaxation-spectrum}
\begin{tabular}{
  c
  c
  S[scientific-notation=true, table-format=1.3e+2]
  c
  c
  S[scientific-notation=true, table-format=1.3e+2]}
\toprule
mode $i$ & {$m_{i}$ [-]} & {$\eta_{0i}$ [\si{\mega\pascal\second}]} &
mode $i$ & {$m_{i}$ [-]} & {$\eta_{0i}$ [\si{\mega\pascal\second}]} \\
\midrule
1  & 0.020 & 1.002e+06  & 13 & 0.014 & 2.453e+24 \\
2  & 0.033 & 1.486e+09  & 14 & 0.023 & 1.131e+25 \\
3  & 0.040 & 1.025e+12  & 15 & 0.014 & 1.654e+25 \\
4  & 0.053 & 1.963e+14  & 16 & 0.016 & 3.367e+25 \\
5  & 0.051 & 2.726e+16  & 17 & 0.018 & 7.969e+25 \\
6  & 0.054 & 1.089e+18  & 18 & 0.021 & 1.920e+26 \\
7  & 0.056 & 6.664e+19  & 19 & 0.006 & 9.983e+25 \\
8  & 0.034 & 3.867e+20  & 20 & 0.029 & 7.309e+26 \\
9  & 0.037 & 6.447e+21  & 21 & 0.052 & 4.257e+27 \\
10 & 0.031 & 4.479e+22  & 22 & 0.011 & 1.396e+27 \\
11 & 0.034 & 2.799e+23  & 23 & 0.029 & 6.464e+27 \\
12 & 0.032 & 2.048e+24  & 24 & 0.292 & 5.920e+29 \\
\bottomrule
\end{tabular}
\end{table}

\subsection{Plasticity parameters}
\begin{figure}
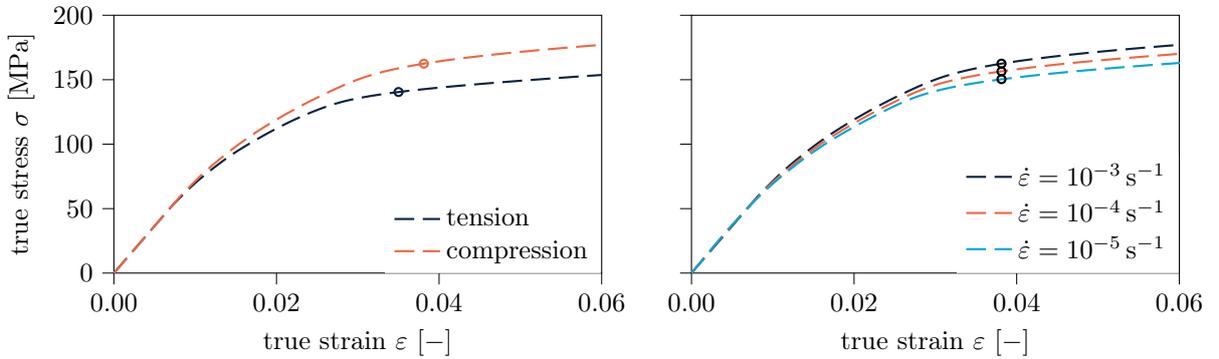

  \hspace{-0.8cm}
  \begin{tikzpicture}

  \node[] at (3.8,0)
    {
      \input{figures/90-compression.tex}
    };

  \node[] at (-3.8,0)
    {
      \input{figures/90-tension-compression.tex}
    };

  \end{tikzpicture}
  \caption{Input curves generated with micromodel at $\theta_0=90^\circ$: (\textit{left}) transverse tension and compression under strain rate $\dot{\varepsilon}=10^{-3}\,\si{\per\second}$ and (\textit{right}) transverse compression under three different strain rates. The yield stresses are indicated with a dot.}
  \label{fig:90-calibration-curve}
\end{figure}

The mesoscopic yield parameters are obtained with the analytical expressions derived in \Cref{sec:analytical-parameter-identification}.
To obtain the pressure-dependency parameter $\mu_\mr{p}$, the micromodel is subjected to uniaxial transverse compression and tension under true strain rate $\dot{\varepsilon}=10^{-3}\,\si{\per\second}$. 
For finding $\eta_0$ and $\sigma_0$, the micromodel is subjected to three strain rates under transverse compression. 
The stress-strain curves are shown in \Cref{fig:90-calibration-curve}.
Note that these curves do not reach a plateau due to hardening, which obscures a clear yield point. 
In this work, the point at which the stress starts to increase almost linearly is chosen as the 'yield' stress. The resulting mesoscopic parameters are 
tabulated in \Cref{tab:invariant-egp-params}. The fit of the Eyring curve (\Cref{eq:analytical-eyring}) with the transverse compression yield data is shown in \Cref{fig:Eyring-fit}.

\begin{figure}
   \centering
  \begin{tikzpicture}
  \node[] at (0,0)
    {
\begin{tikzpicture}

\definecolor{crimson2246049}{RGB}{224,60,49}
\definecolor{darkgray176}{RGB}{176,176,176}
\definecolor{darkturquoise0166214}{RGB}{0,166,214}

\begin{axis}[
height=5cm,
legend cell align={left},
legend style={
  fill opacity=0.8,
  draw opacity=1,
  text opacity=1,
  at={(0.97,0.03)},
  anchor=south east,
  draw=none,
  fill=none
},
minor xtick={},
minor ytick={},
scaled ticks=false,
tick align=outside,
tick pos=left,
width=8cm,
x grid style={darkgray176},
xlabel={true strain \(\displaystyle \varepsilon~[-]\)},
xmin=0, xmax=0.04,
xtick style={color=black},
xtick={0,0.01,0.02,0.03,0.04},
xticklabels={
  \(\displaystyle {0.00}\),
  \(\displaystyle {0.01}\),
  \(\displaystyle {0.02}\),
  \(\displaystyle {0.03}\),
  \(\displaystyle {0.04}\)
},
y grid style={darkgray176},
ylabel={true stress  \(\displaystyle \sigma~[\mathrm{MPa}]\)},
ymin=0, ymax=220,
ytick style={color=black},
ytick={0,50,100,150,200},
yticklabels={
  \(\displaystyle {0}\),
  \(\displaystyle {50}\),
  \(\displaystyle {100}\),
  \(\displaystyle {150}\),
  \(\displaystyle {200}\)
}
]
\addplot [line width=2.4pt, darkturquoise0166214, dash pattern=on 6pt off 9.9pt]
table {%
0 0
0.00337841423 56.5957706
0.00458552101 74.0742764
0.00590150279 89.9677237
0.00736436079 104.431184
0.00966548131 121.502477
0.0118273862 134.206909
0.0139296272 143.053295
0.0160318683 150.100196
0.0181341093 155.123555
0.0202363503 158.420879
0.0223385914 160.633429
0.0244408324 162.367377
0.0265430735 163.907702
0.0286453145 165.367578
0.0307475555 166.80488
0.0328497966 168.254917
0.0349520376 169.73884
};
\addlegendentry{micro}
\addplot [very thick, crimson2246049]
table {%
0 0
0.00100000000000004 17.4136624811981
0.00300000000000006 50.5084865464345
0.00582842712474616 88.1141900707174
0.00762842712474626 105.939024049652
0.0101740115370177 123.983338336233
0.0119740115370178 134.390942634357
0.0135013621843807 141.343630668411
0.0147973621843808 146.04928689791
0.0166301829612162 150.809264754244
0.0181853829612162 153.961195593708
0.0203847678934189 157.11873970226
0.0221847678934189 159.228122269024
0.0247303523056905 161.613026720172
0.0277303523056905 163.916108713338
0.0307303523056905 165.891095497146
0.0337303523056905 167.711337720276
0.0367303523056905 169.499406161837
0.0397303523056904 171.320546504104
0.0427303523056905 173.206897083245
0.0457303523056905 175.175969475356
0.0487303523056904 177.239467251204
0.0517303523056904 179.407085866288
0.0547303523056906 181.6881619204
0.0577303523056905 184.092446759298
0.0607303523056905 186.630529159576
0.0637303523056905 189.314144924933
0.0667303523056905 192.15636569345
0.0697303523056904 195.171908113809
0.0727303523056904 198.377428402253
0.0757303523056905 201.791837742054
0.0787303523056905 205.436715791879
0.0817303523056905 209.336808626901
0.0847303523056906 213.520639799566
0.0877303523056904 218.021267611665
};
\addlegendentry{meso}
\end{axis}

\end{tikzpicture}
    };
  \end{tikzpicture}
  \caption{Multimode calibration with uniaxial tension under $\theta_0=30^\circ$ and $\dot{\varepsilon}=10^{-3}\,\si{\per\second}$: output curve with mesomodel (\textit{meso}) \textit{vs} input curve with micromodel (\textit{micro}).}
  \label{fig:multimode-calibration-curve}
\end{figure}
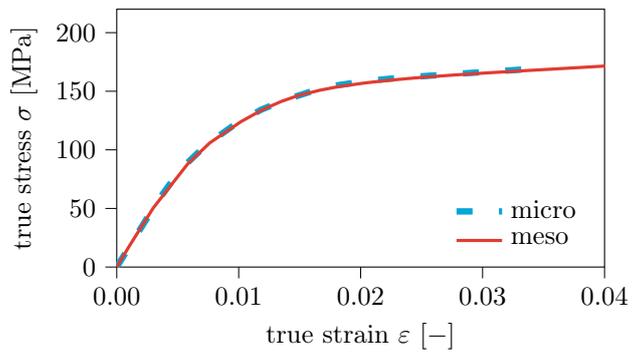

The micromodel is subjected to uniaxial tension under off-axis angle $\theta_0=30^\circ$ and true strain rate $\dot{\varepsilon}= 10^{-3}\,\si{\per\second}$ . The parameter $\alpha_2$ is first determined by solving \Cref{eq:a2-id}. With all single-mode parameters known, a multimode relaxation spectrum, with $24$ modes,\footnote{It is recommended in Ref. \cite{amiri-radAnisotropic2019,vanbreemenExtending2011} to include one mode per decade in the relaxation spectrum, ensuring an accurate pre-yield and creep response. A smaller number of modes may introduce spurious oscillations in the stress-strain curve \cite{vanbreemenExtending2011}.}
is determined by following the procedure as outlined in \Cref{sec:multimode-model}.
Applying the method as described in \cite{vanbreemenExtending2011, amiri-radAnisotropic2019} to the present mesoscopic model, resulted in a slight mismatch between the input and output results. 
Therefore, the input curve is iteratively adjusted such that the output curve matched with the original input curve. 
The relaxation spectrum is tabulated in \Cref{tab:relaxation-spectrum}. With the ratios $\{m_i\}$, the elasticity parameters are obtained for each mode with \Cref{eq:stiffness-distribution} and \Cref{tab:elasticity-params}.
The resulting stress-strain curve is shown in \Cref{fig:multimode-calibration-curve}.

\begin{figure}
   \centering
  \begin{tikzpicture}
  \node[] at (0,0)
    {
\begin{tikzpicture}

\definecolor{darkgray176}{RGB}{176,176,176}
\definecolor{darkturquoise0166214}{RGB}{0,166,214}

\begin{axis}[
height=5cm,
legend cell align={left},
legend style={
  fill opacity=0.8,
  draw opacity=1,
  text opacity=1,
  at={(0.97,0.03)},
  anchor=south east,
  draw=none
},
log basis x={10},
minor xtick={2e-07,3e-07,4e-07,5e-07,6e-07,7e-07,8e-07,9e-07,2e-06,3e-06,4e-06,5e-06,6e-06,7e-06,8e-06,9e-06,2e-05,3e-05,4e-05,5e-05,6e-05,7e-05,8e-05,9e-05,0.0002,0.0003,0.0004,0.0005,0.0006,0.0007,0.0008,0.0009,0.002,0.003,0.004,0.005,0.006,0.007,0.008,0.009,0.02,0.03,0.04,0.05,0.06,0.07,0.08,0.09,0.2,0.3,0.4,0.5,0.6,0.7,0.8,0.9},
minor ytick={},
tick align=outside,
tick pos=left,
width=8cm,
x grid style={darkgray176},
xlabel={true strain rate \(\displaystyle \dot{\varepsilon}~\) [s\(\displaystyle ^{-1}\)]},
xmajorgrids,
xmin=7.94328234724282e-06, xmax=0.00125892541179417,
xminorgrids,
xmode=log,
xtick style={color=black},
xtick={1e-07,1e-06,1e-05,0.0001,0.001,0.01,0.1},
xticklabels={
  \(\displaystyle {10^{-7}}\),
  \(\displaystyle {10^{-6}}\),
  \(\displaystyle {10^{-5}}\),
  \(\displaystyle {10^{-4}}\),
  \(\displaystyle {10^{-3}}\),
  \(\displaystyle {10^{-2}}\),
  \(\displaystyle {10^{-1}}\)
},
y grid style={darkgray176},
ylabel={yield stress \(\displaystyle \sigma_{\mathrm{y}}~[\mathrm{MPa}]\)},
ymajorgrids,
ymin=145, ymax=170,
yminorgrids,
ytick style={color=black},
ytick={145,150,155,160,165,170},
yticklabels={
  \(\displaystyle {145}\),
  \(\displaystyle {150}\),
  \(\displaystyle {155}\),
  \(\displaystyle {160}\),
  \(\displaystyle {165}\),
  \(\displaystyle {170}\)
}
]
\addplot [thick, darkturquoise0166214]
table {%
1e-05 150.4246438366
0.00012 156.920816164043
0.00023 158.621616048862
0.00034 159.643439101426
0.00045 160.376219091177
0.00056 160.94792778524
0.00067 161.416770177372
0.00078 161.814178860415
0.00089 162.159070711549
0.001 162.463719480724
};
\addlegendentry{fit}
\addplot [semithick, black, mark=o, mark size=2.5, mark options={solid,fill opacity=0}, only marks]
table {%
1e-05 150.42464384
0.0001 156.44418166
0.001 162.46371948
};
\addlegendentry{micro}
\end{axis}

\end{tikzpicture}
    };

  \end{tikzpicture}
  \caption{Eyring fit (\Cref{eq:analytical-eyring}) of yield stress versus strain rate for $\theta_0=90^\circ$ in compression.}
  \label{fig:Eyring-fit}
\end{figure}
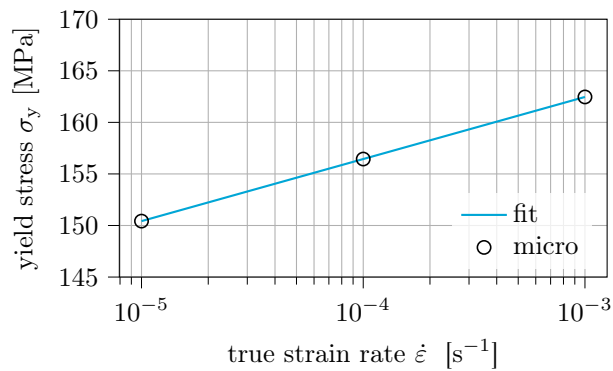

\subparagraph{Remark 7}
Other invariant-based (Perzyna-type) viscoplasticity models \cite{koerberExperimental2018, gerbaudInvariant2019, rodrigueslopesInvariantbased2022}, more suitable for unidirectional \emph{thermosetting} polymer composites, require six hardening functions as inputs (obtained from bi-axial tension/compression, longitudinal shear, transverse shear and uniaxial tension/compression tests) to describe the nonlinear rate-dependent plastic response. However, obtaining transverse shear and biaxial test data through experiments is not straightforward. Therefore, these hardening functions are usually deduced from other tests, engineering assumptions or micromechanical models \cite{voglerModeling2013}. With the present invariant-based non-Newtonian flow model for \emph{thermoplastic} polymer composites, the yield stress is determined by the mode with the highest initial viscosity (see \Cref{fig:single-multi}), and thus, only four parameters are required. These parameters can be determined from a small number of off-axis constant strain-rate tests as shown in this section with a micromodel, or from off-axis coupon tests with oblique ends under (almost) uniform stress states \cite{sunOblique1993}. Subsequently, the pre-yield nonlinearity is described by a relaxation spectrum, which can be determined from a \emph{single} stress-strain curve under off-axis loading. Therefore, a significant reduction in the amount of necessary inputs is achieved with the present invariant-based constitutive model.

\section{Results}
\label{sec:results}
The performance of the mesoscopic constitutive model in simulating rate-dependent plasticity and creep is studied in this section. First, its capability in representing a material point of a composite under off-axis loading is assessed with a single element, under the assumption of a uniform deformation (see \Cref{fig:off-axis-global-deformations}). 
Subsequently, the model is applied to the simulation of ply-level off-axis specimens and compared against experiments \cite{sundararajanMatrix2024}.

\subsection{Constant strain rate}
\label{sec:const-strain-rate}
The microscale and mesoscale model are subjected to constant true strain rates $\epsdot\left(\si{\per\second}\right)\in\{10^{-5}, 10^{-4}, 10^{-3}\}$ under off-axis angles $\theta\left(^\circ\right)\in\{90, 45, 30, 15, 0\}$ in tension and compression. 

\begin{figure}
   \centering
  \begin{tikzpicture}
  \node[] at (0,0)
    {
      \input{figures/transverse-isotropy.tex}
    };
  \node at (1.1,-0.3){{\def\svgwidth{0.10\columnwidth}{\scalebox{1.0}{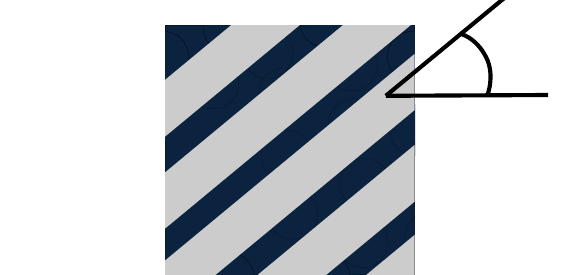}}}};
  \node at (1.1,-0.3){{\def\svgwidth{0.10\columnwidth}{\scalebox{1.0}{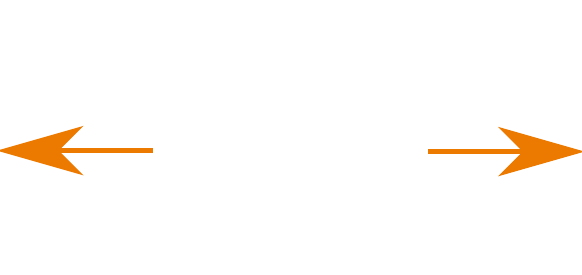}}}};
  \node at (2.0,-0.0) {$\theta_0$};

  \node at (-2.0,1.3){{\def\svgwidth{0.10\columnwidth}{\scalebox{1.0}{\input{inkscape/rve-angle.pdf_tex}}}}};
  \node at (-2.0,1.3){{\def\svgwidth{0.10\columnwidth}{\scalebox{1.0}{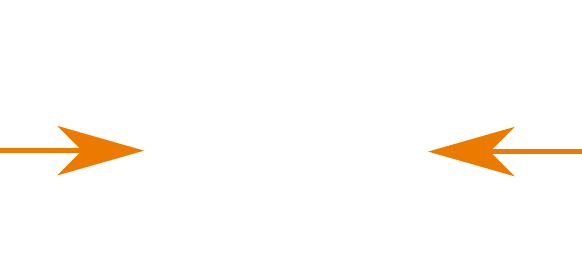}}}};
  \node at (-1.1,1.6) {$\theta_0$};

  \end{tikzpicture}
  \caption{Stress-strain curves under various initial off-axis angles $\theta_0$ and constant strain rate $\dot{\varepsilon}=10^{-3}\,\si{\per\second}$ in tension and compression: micromodel (\emph{dashed} line) \textit{vs} mesomodel (\emph{solid} line).}
  \label{fig:examples-direction-dependence}
\end{figure}

\subparagraph{Direction-dependence}
\Cref{fig:examples-direction-dependence} shows the stress-strain curves with $\epsdot=10^{-3}\,\si{\per\second}$ and various off-axis angles $\theta_0$. It is observed that the strongly anisotropic response of the micromodel is well represented with the mesoscale model: under $\theta_0=0^\circ$, the response is elastic, whereas under off-axis loading, it is viscoplastic. It is worth noting that the rather simple approach, as described in \Cref{sec:multimode-model}, of finding a relaxation spectrum with a single stress-strain curve, gives a good pre-yield response for all off-axis angles and strain rates.

\begin{figure}
   \centering
  \begin{tikzpicture}
  \node[] at (0,0)
    {
      \input{figures/fibangle-strain.tex}
    };
  \node at (2.5,1.7){{\def\svgwidth{0.10\columnwidth}{\scalebox{1.0}{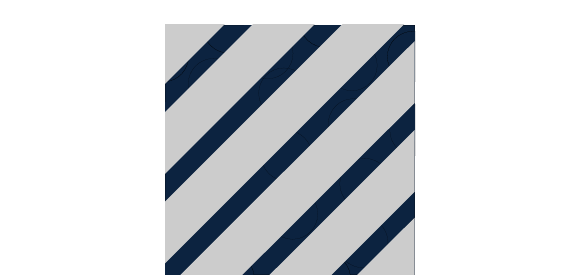}}}};
  \node at (2.5,1.7){{\def\svgwidth{0.10\columnwidth}{\scalebox{1.0}{\input{inkscape/off-axis-rve-tension-arrows.pdf_tex}}}}};
  \node at (2.5, 1.0) {$\theta_0=45^\circ$};

  \node at (2.5,0.2){{\def\svgwidth{0.10\columnwidth}{\scalebox{1.0}{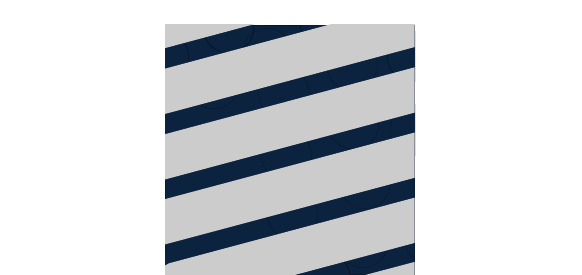}}}};
  \node at (2.5,0.2){{\def\svgwidth{0.10\columnwidth}{\scalebox{1.0}{\input{inkscape/off-axis-rve-tension-arrows.pdf_tex}}}}};
  \node at (2.5,-0.5) {$\theta_0=15^\circ$};
  
  \draw[-] (0.5, 0.9 ) -- (1.5, 1.2);
  \draw[-] (0.5,-0.45) -- (1.5,-0.2);

  \end{tikzpicture}
  \caption{Evolution of the off-axis angle $\theta$ with the mesoscale model for two \emph{initial} off-axis angles $\theta_0\left(^\circ\right)\in\{15,30\}$.}
  \label{fig:fibangle-strain}
\end{figure}

With both the micromodel and the mesomodel under off-axis angle $\theta_0=15^\circ$ in tension, an increasing stiffness (hardening) is observed in the post-yield regime, whereas under compression, a softening response is obtained.  
When off-axis tensile loads are applied to the composite material, the fibers progressively align with the load direction (see \Cref{fig:off-axis-global-deformations}). This re-orientation of the fibers is captured by the mesoscale model and is numerically depicted in \Cref{fig:fibangle-strain}.
In contrast, under compression, the opposite effect takes place where the off-axis angle increases, leading to a softening response. 
The agreement between the two models indicates that the re-orientation of the fibers is captured just as well in the mesoscopic constitutive model as in the micromodel where the fibers are explicitly modeled.

\begin{figure}
   \hspace{-0.7cm}
  \begin{tikzpicture}
  \node[] at (-3.8,0)
    {
      \input{figures/rate-dependence-15d-45d.tex}
    };
  \node[] at (3.8,0)
    {
      \input{figures/rate-dependence-90d-30d.tex}
    };

  \node (45deg) at (-1.0,0.9){{\def\svgwidth{0.10\columnwidth}{\scalebox{1.0}{\input{inkscape/off-axis-rve-tension-p45.pdf_tex}}}}};
  \node at (-1.0,0.9){{\def\svgwidth{0.10\columnwidth}{\scalebox{1.0}{\input{inkscape/off-axis-rve-tension-arrows.pdf_tex}}}}};
  \node at (-1.0,0.2) {$\theta_0=45^\circ$};

  \node (45deg) at (-5.7,1.7){{\def\svgwidth{0.10\columnwidth}{\scalebox{1.0}{\input{inkscape/off-axis-rve-tension-p15.pdf_tex}}}}};
  \node at (-5.7,1.7){{\def\svgwidth{0.10\columnwidth}{\scalebox{1.0}{\input{inkscape/off-axis-rve-tension-arrows.pdf_tex}}}}};
  \node at (-5.7,1.0) {$\theta_0=15^\circ$};

  \node (30deg) at (1.9,1.7){{\def\svgwidth{0.10\columnwidth}{\scalebox{1.0}{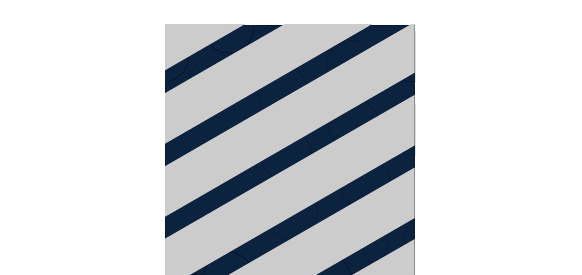}}}};
  \node at (1.9,1.7){{\def\svgwidth{0.10\columnwidth}{\scalebox{1.0}{\input{inkscape/off-axis-rve-tension-arrows.pdf_tex}}}}};
  \node at (1.9,1.0) {$\theta_0=30^\circ$};

  \node (90deg) at (3.4,-0.3){{\def\svgwidth{0.10\columnwidth}{\scalebox{1.0}{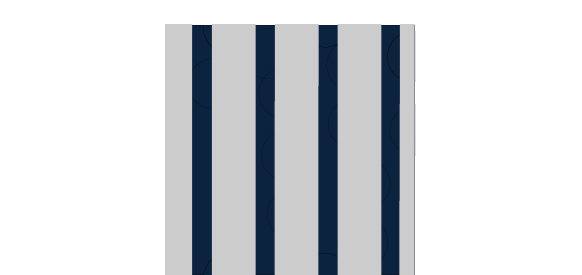}}}};
  \node at (3.4,-0.3){{\def\svgwidth{0.10\columnwidth}{\scalebox{1.0}{\input{inkscape/off-axis-rve-tension-arrows.pdf_tex}}}}};
  \node at (3.4,-1.0) {$\theta_0=90^\circ$};
 
  \end{tikzpicture}
  \caption{Rate-dependence under uniaxial tension with various initial off-axis angles $\theta_0$ and strain rates $\dot{\varepsilon}$: micromodel (\emph{dashed} line) vs mesomodel (\emph{solid} line). The yield stresses are indicated with a dot.}
  \label{fig:examples-rate-dependence}
\end{figure}

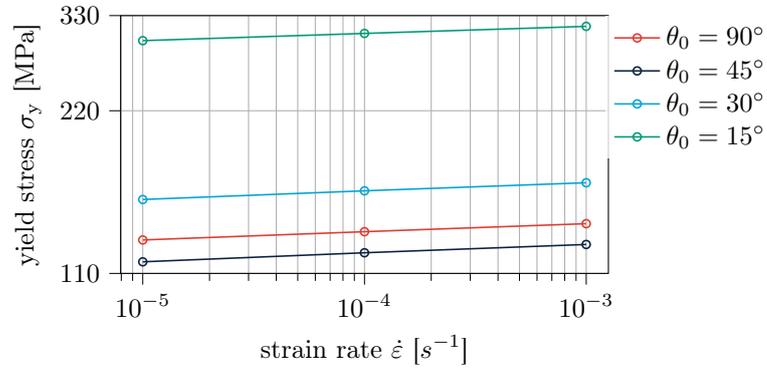
\begin{figure}
  \hspace{3cm}
  \begin{tikzpicture}
  \node[] at (0,0)
    {
\begin{tikzpicture}

\definecolor{crimson2246049}{RGB}{224,60,49}
\definecolor{darkcyan0155119}{RGB}{0,155,119}
\definecolor{darkgray176}{RGB}{176,176,176}
\definecolor{darkturquoise0166214}{RGB}{0,166,214}
\definecolor{midnightblue123564}{RGB}{12,35,64}

\begin{axis}[
height=5cm,
legend cell align={left},
legend style={fill opacity=0.8, draw opacity=1, text opacity=1, at={(1.35,1)}, draw=none},
log basis x={10},
minor xtick={2e-07,3e-07,4e-07,5e-07,6e-07,7e-07,8e-07,9e-07,2e-06,3e-06,4e-06,5e-06,6e-06,7e-06,8e-06,9e-06,2e-05,3e-05,4e-05,5e-05,6e-05,7e-05,8e-05,9e-05,0.0002,0.0003,0.0004,0.0005,0.0006,0.0007,0.0008,0.0009,0.002,0.003,0.004,0.005,0.006,0.007,0.008,0.009,0.02,0.03,0.04,0.05,0.06,0.07,0.08,0.09,0.2,0.3,0.4,0.5,0.6,0.7,0.8,0.9},
minor ytick={},
tick align=outside,
tick pos=left,
width=8cm,
x grid style={darkgray176},
xlabel={strain rate \(\displaystyle \dot{\varepsilon}\) [\(\displaystyle s^{-1}\)]},
xmajorgrids,
xmin=7.94328234724282e-06, xmax=0.00125892541179417,
xminorgrids,
xmode=log,
xmode=log,
xtick style={color=black},
xtick={1e-07,1e-06,1e-05,0.0001,0.001,0.01,0.1},
xticklabels={
  \(\displaystyle {10^{-7}}\),
  \(\displaystyle {10^{-6}}\),
  \(\displaystyle {10^{-5}}\),
  \(\displaystyle {10^{-4}}\),
  \(\displaystyle {10^{-3}}\),
  \(\displaystyle {10^{-2}}\),
  \(\displaystyle {10^{-1}}\)
},
y grid style={darkgray176},
ylabel={yield stress \(\displaystyle \sigma_{\mathrm{y}}\) [MPa]},
ymajorgrids,
ymin=110, ymax=330,
yminorgrids,
ymode=log,
ytick style={color=black},
ytick={110,220,330},
yticklabels={\(\displaystyle {110}\),\(\displaystyle {220}\),\(\displaystyle {330}\)}
]
\addplot [semithick, crimson2246049, mark=o, mark size=1.5, mark options={solid,fill opacity=0}]
table {%
1e-05 126.85
0.0001 131.4
0.001 135.97
};
\addlegendentry{$\theta_0=90^\circ$}
\addplot [semithick, midnightblue123564, mark=o, mark size=1.5, mark options={solid,fill opacity=0}]
table {%
1e-05 115.57
0.0001 120.12
0.001 124.44
};
\addlegendentry{$\theta_0=45^\circ$}
\addplot [semithick, darkturquoise0166214, mark=o, mark size=1.5, mark options={solid,fill opacity=0}]
table {%
1e-05 150.72
0.0001 156.4
0.001 161.88
};
\addlegendentry{$\theta_0=30^\circ$}
\addplot [semithick, darkcyan0155119, mark=o, mark size=1.5, mark options={solid,fill opacity=0}]
table {%
1e-05 296.58
0.0001 305.83
0.001 315.18
};
\addlegendentry{$\theta_0=15^\circ$}
\end{axis}

\end{tikzpicture}
    };

  \end{tikzpicture}
  \caption{Yield stress as function of strain rate for various off-axis angles $\theta_0$ with mesoscale model.}
  \label{fig:Eyring-plot-1elem}
\end{figure}

\subparagraph{Rate-dependence}
The stress-strain curves with off-axis angles $\theta_0\left(^\circ\right)\in\{15,30,45,90\}$ and constant strain rates $\epsdot\left(\,\si{\per\second}\right)\in\{10^{-5}, 10^{-4}, 10^{-3}\}$ in tension are shown in \Cref{fig:examples-rate-dependence}. It can be observed that the rate-dependence, which describes an increasing yield stress with increasing strain rate, is accurately reflected by the mesoscale model. The yield stresses from the mesoscale model are indicated in the figures and plotted against strain rates $\dot{\varepsilon}$ on a double logarithmic scale for each off-axis angle $\theta_0$ in \Cref{fig:Eyring-plot-1elem}. 
In line with experimental observations for unidirectional polymer composites \cite{erartsinTime2022}, the curves are parallel, 
indicating a factorizable dependence of yield stress on strain rate $\dot{\varepsilon}$ and off-axis angle $\theta_0$.

\begin{figure}
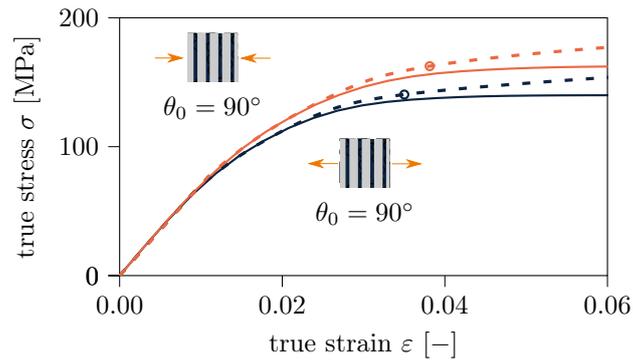

  \hspace{3cm}
  \begin{tikzpicture}
  \node[] at (0,0)
    {
      \input{figures/pressure-dependence.tex}
    };
  
  \node at (0.5,0.3){{\def\svgwidth{0.10\columnwidth}{\scalebox{1.0}{\input{inkscape/off-axis-rve-tension-p90.pdf_tex}}}}};
  \node at (0.5,0.3){{\def\svgwidth{0.10\columnwidth}{\scalebox{1.0}{\input{inkscape/off-axis-rve-tension-arrows.pdf_tex}}}}};
  \node at (0.5,-0.4) {$\theta_0=90^\circ$};

  \node at (-1.5,1.7){{\def\svgwidth{0.10\columnwidth}{\scalebox{1.0}{\input{inkscape/off-axis-rve-tension-p90.pdf_tex}}}}};
  \node at (-1.5,1.7){{\def\svgwidth{0.10\columnwidth}{\scalebox{1.0}{\input{inkscape/off-axis-rve-compression-arrows.pdf_tex}}}}};
  \node at (-1.5,1.0) {$\theta_0=90^\circ$};

  \end{tikzpicture}
  \caption{Transverse tension and compression under off-axis angle $\theta_0=90^\circ$: micromodel (\emph{dashed} line) with yield stresses (indicated with a dot) vs mesomodel (\emph{solid} line).}
  \label{fig:examples-pressure-dependence}
\end{figure}

\subparagraph{Pressure-dependence}
The stress-strain curves of the micromodel and the mesomodel under transverse tension and compression with $\dot{\varepsilon}=10^{-3}\,\si{\per\second}$ are shown in \Cref{fig:examples-pressure-dependence}. It can be observed that the response is accurate until the yield point. However, after yielding, a hardening response is observed with the micromodel. In the isotropic EGP for the matrix material of the micromodel, an (elastic) hardening contribution is present \cite{kovacevicMicromechanical2024}, which is currently not included in the mesoscale model. Under transverse tensile loading, Carbon/PEEK fractures before a fully developed plastic response is reached due to large hydrostatic stresses in the polymer matrix. The post-yield hardening response is therefore less relevant under tensile loading. However, for a more accurate post-yield response under transverse compression, an (anisotropic) hardening contribution can be included to account for this effect.

\begin{figure}
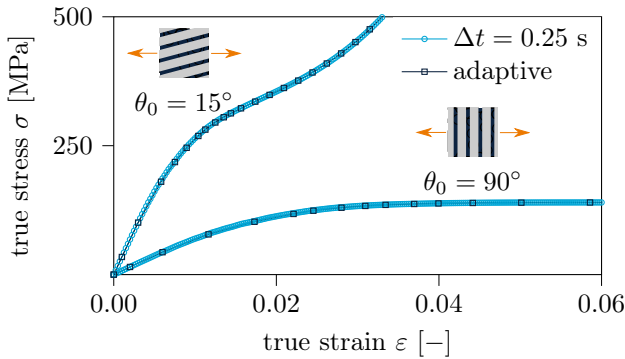

   \centering
  \begin{tikzpicture}
  \node[] at (0,0)
    {
      \input{figures/timestep-dependence.tex}
    };
  \node (15deg) at (-1.8,1.75){{\def\svgwidth{0.10\columnwidth}{\scalebox{1.0}{\input{inkscape/off-axis-rve-tension-p15.pdf_tex}}}}};
  \node at (-1.8,1.75){{\def\svgwidth{0.10\columnwidth}{\scalebox{1.0}{\input{inkscape/off-axis-rve-tension-arrows.pdf_tex}}}}};
  \node at (-1.8,1.05) {$\theta_0=15^\circ$};

  \node (90deg) at (2.0,0.7){{\def\svgwidth{0.10\columnwidth}{\scalebox{1.0}{\input{inkscape/off-axis-rve-tension-p90.pdf_tex}}}}};
  \node at (2.0,0.7){{\def\svgwidth{0.10\columnwidth}{\scalebox{1.0}{\input{inkscape/off-axis-rve-tension-arrows.pdf_tex}}}}};
  \node at (2.0,0.0) {$\theta_0=90^\circ$};
  \end{tikzpicture}
  \caption{Time-step dependence: stress-strain curves with small constant time-steps and adaptive stepping. The markers denote the time-steps. }
  \label{fig:timestep-dependence}
\end{figure}

\subparagraph{Time-step dependence}
The time-step dependence of the time integration scheme, with the Pad\'{e} approximation (\Cref{eq:pade-approx}), is assessed by comparing the response obtained with adaptive stepping based on global iterations \cite{hofmanNumerical2024_}, to the response with (constant) small time increments. 
For this purpose, simulations with off-axis constant strain rates $\dot{\varepsilon}=10^{-3}\,\si{\per\second}$ under $\theta_0=15^\circ$ and $90^\circ$ are used for the comparison.
The simulation with small time steps is performed with $\Delta t = 0.25\,\si{\second}$, resulting in strain increments $\Delta \varepsilon=2.5\times10^{-4}$. \Cref{fig:timestep-dependence} shows the stress-strain curves, from which it is concluded that time-step dependence of the time integration scheme is negligible. In combination with the fully consistent tangent stiffness, adaptive stepping based on global iterations is possible for efficient simulations with high accuracy.

\subsection{Creep}
An important feature of the EGP model is the capability to simulate not only rate-dependent plasticity but also creep in polymers. This also holds for the present mesoscopic version for polymer composites. To assess the performance under creep, the micro- and meso-scale models are subjected to a constant tensile \emph{engineering} stress rate until a specified stress level is reached in $\SI{10}{s}$. After this phase, the engineering stress is kept constant. 

\begin{figure}
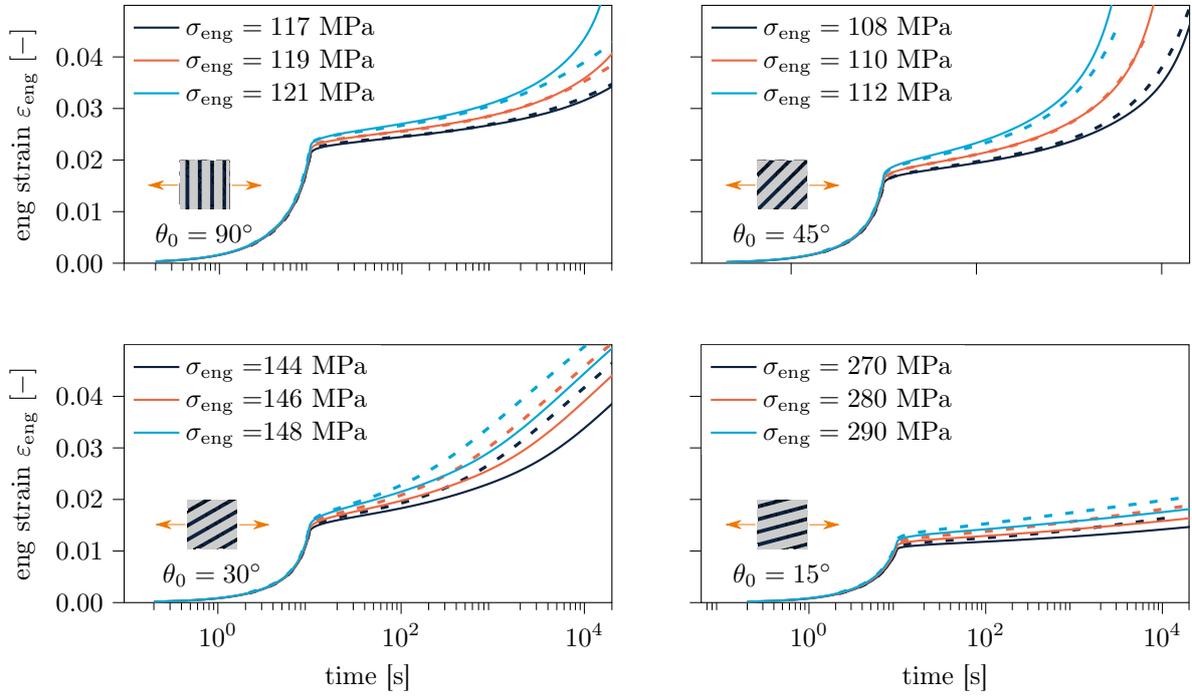

  \hspace{-0.5cm}
  \begin{tikzpicture}
  \node[] at (3.8,0)
    {
      \input{figures/creep-45.tex}
    };
  \node at (2.4,-0.0){{\def\svgwidth{0.10\columnwidth}{\scalebox{1.0}{\input{inkscape/off-axis-rve-tension-p45.pdf_tex}}}}};
  \node at (2.4,-0.0){{\def\svgwidth{0.10\columnwidth}{\scalebox{1.0}{\input{inkscape/off-axis-rve-tension-arrows.pdf_tex}}}}};
  \node at (2.4,-0.70) {$\theta_0=45^\circ$};

  \node[] at (-3.8,0)
    {
      \input{figures/creep-90.tex}
    };
  \node at (-5.2,-0.0){{\def\svgwidth{0.10\columnwidth}{\scalebox{1.0}{\input{inkscape/off-axis-rve-tension-p90.pdf_tex}}}}};
  \node at (-5.2,-0.0){{\def\svgwidth{0.10\columnwidth}{\scalebox{1.0}{\input{inkscape/off-axis-rve-tension-arrows.pdf_tex}}}}};
  \node at (-5.2,-0.70) {$\theta_0=90^\circ$};

  \node[] at (3.8,-4.5)
    {
      \input{figures/creep-15.tex}
    };

  \node at (2.4,-4.5){{\def\svgwidth{0.10\columnwidth}{\scalebox{1.0}{\input{inkscape/off-axis-rve-tension-p15.pdf_tex}}}}};
  \node at (2.4,-4.5){{\def\svgwidth{0.10\columnwidth}{\scalebox{1.0}{\input{inkscape/off-axis-rve-tension-arrows.pdf_tex}}}}};
  \node at (2.4,-5.2) {$\theta_0=15^\circ$};

  \node[] at (-3.8,-4.5)
    {
      \input{figures/creep-30.tex}
    };

  \node at (-5.1,-4.5){{\def\svgwidth{0.10\columnwidth}{\scalebox{1.0}{\input{inkscape/off-axis-rve-tension-p30.pdf_tex}}}}};
  \node at (-5.1,-4.5){{\def\svgwidth{0.10\columnwidth}{\scalebox{1.0}{\input{inkscape/off-axis-rve-tension-arrows.pdf_tex}}}}};
  \node at (-5.1,-5.2) {$\theta_0=30^\circ$};

  \end{tikzpicture}
 
  \caption{Creep response under various off-axis angles $\theta_0$ and engineering stress levels $\sigma_{\mr{eng}}$: micromodel (\emph{dashed} line) vs mesomodel (\emph{solid} line).}
  \label{fig:examples-creep}
\end{figure}

The engineering strain as a function of time is shown in \Cref{fig:examples-creep} for four off-axis angles $\theta\left(^\circ\right)\in\{90, 45, 30, 15\}$. For each angle, three different engineering stress levels are applied, as indicated in the figures. It can be observed that for all off-axis angles, the strains of the mesomodel during the ramp-up to the maximum engineering stress are in close agreement with those of the micromodel. This is expected since the mesoscale model parameters were determined with (short-term) constant-strain rate data (\Cref{sec:calibration}). After reaching the maximum applied stress level, the creep response with $\theta_0=45^\circ$ is very similar to that of the micromodel. However, for the other angles, the match is adequate but not as good as with $\theta_0=45^\circ$. It is somewhat surprising that, although the $\theta_0=30^\circ$ off-axis angle has been used for determining the multi-mode relaxation spectrum (see \Cref{fig:multimode-calibration-curve}), the match in creep is worse than with the other off-axis angles. A parameter identification procedure which includes creep data, \eg through a compliance-time master curve from a series of creep tests at different stress levels \cite{tervoortMultimode1996}, may improve the creep response.

\subsection{Unidirectional ply under off-axis tensile loading}
\begin{figure}
   \centering
  \begin{tikzpicture}
  \node[] at (0,0)
    {
     {\def\svgwidth{0.6\columnwidth}{\scalebox{1.1}{
\begingroup%
  \makeatletter%
  \providecommand\color[2][]{%
    \errmessage{(Inkscape) Color is used for the text in Inkscape, but the package 'color.sty' is not loaded}%
    \renewcommand\color[2][]{}%
  }%
  \providecommand\transparent[1]{%
    \errmessage{(Inkscape) Transparency is used (non-zero) for the text in Inkscape, but the package 'transparent.sty' is not loaded}%
    \renewcommand\transparent[1]{}%
  }%
  \providecommand\rotatebox[2]{#2}%
  \newcommand*\fsize{\dimexpr\f@size pt\relax}%
  \newcommand*\lineheight[1]{\fontsize{\fsize}{#1\fsize}\selectfont}%
  \ifx\svgwidth\undefined%
    \setlength{\unitlength}{299.64457354bp}%
    \ifx\svgscale\undefined%
      \relax%
    \else%
      \setlength{\unitlength}{\unitlength * \real{\svgscale}}%
    \fi%
  \else%
    \setlength{\unitlength}{\svgwidth}%
  \fi%
  \global\let\svgwidth\undefined%
  \global\let\svgscale\undefined%
  \makeatother%
  \begin{picture}(1,0.37568937)%
    \lineheight{1}%
    \setlength\tabcolsep{0pt}%
    \put(0,0){\includegraphics[width=\unitlength,page=1]{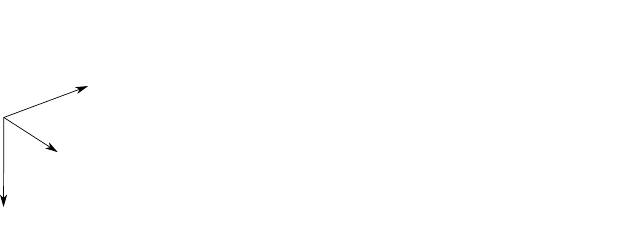}}%
    \put(0.00210729,0.02291091){\color[rgb]{0,0,0}\rotatebox{-3.0367802}{\makebox(0,0)[lt]{\lineheight{1.25}\smash{\begin{tabular}[t]{l}$\bm{e}_3$\end{tabular}}}}}%
    \put(0.10213381,0.11919156){\color[rgb]{0,0,0}\rotatebox{-3.0367802}{\makebox(0,0)[lt]{\lineheight{1.25}\smash{\begin{tabular}[t]{l}$\bm{e}_2$\end{tabular}}}}}%
    \put(0.14928027,0.23466735){\color[rgb]{0,0,0}\rotatebox{-1.36511}{\makebox(0,0)[lt]{\lineheight{1.25}\smash{\begin{tabular}[t]{l}$\bm{e}_1$\end{tabular}}}}}%
    \put(0,0){\includegraphics[width=\unitlength,page=2]{ply-mesh-v3.pdf}}%
    \put(0.53257822,0.17490349){\color[rgb]{0,0,0}\makebox(0,0)[lt]{\lineheight{1.25}\smash{\begin{tabular}[t]{l}$\theta_0$\end{tabular}}}}%
  \end{picture}%
\endgroup%
}}}
    };
  \end{tikzpicture}
  \caption{Ply simulation: mesh and initial off-axis angle $\theta_0$.}
  \label{fig:ply-mesh}
\end{figure}

So far, material point analyses have been carried out with the mesoscale model. 
In this section, the mesoscale model is used for the simulation of a unidirectional ply with dimensions $120\times15\times1.8\,\mr{mm}$. The three-dimensional mesh, consisting of 240 trilinear finite elements, is shown in \Cref{fig:ply-mesh}. 
On each end of the specimen, the displacements in the $\bm{e}_2$-and $\bm{e}_3$-direction are fixed, mimicking the constraining effect of the grips in the experimental test \cite{sundararajanMatrix2024}. In the $\bm{e}_1$-direction, a constant \emph{engineering} strain rate of 
$\dot{\varepsilon}_{\mr{eng}}=10^{-4} \,\si{\per\second}$ is applied. Simulations are performed with four initial off-axis angles $\theta_0\left(^\circ\right)\in\{15, 30, 45, 90\}$. 

\begin{figure}
   \centering
  \begin{tikzpicture}
  \node[] at (0,0)
    {
      \input{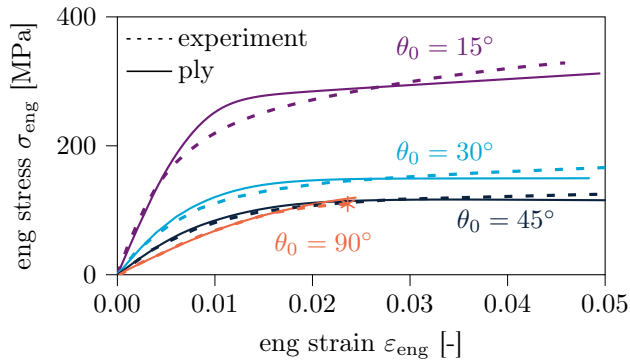}
    };
  \node at (1.6,1.8) {$\textcolor{Paars}{{\theta_0=15^\circ}}$};
  \node at (1.6,0.4) {$\textcolor{TUDblue}{{\theta_0=30^\circ}}$};
  \node at (2.4,-0.5) {$\textcolor{Donkerblauw}{\theta_0=45^\circ}$};
  \node at (0.0,-0.8) {$\textcolor{Oranje}{\theta_0=90^\circ}$};
  \end{tikzpicture}
\caption{Ply simulations \textit{vs} experiments with $\theta_0\left(^\circ\right)=\{15,30,45,90\}$. The experiment with $\theta_0=90^\circ$ fractured (indicated with {\large$\textcolor{Oranje}{*}$})
before plasticity fully developed. 
}
  \label{fig:ply-vs-exp}
\end{figure}

The stress-strain curves are shown in \Cref{fig:ply-vs-exp}.
The ply simulations give an excellent match with the experiments for $\theta_0=30^\circ$, $45^\circ$ and $90^\circ$. For $\theta_0=15^\circ$, the numerical response is similar to the experiment, although the pre-yield stiffness is over-predicted and the post-yield hardening response is under-predicted. 

\begin{figure}
  \hspace{-2.2cm}
  \begin{tikzpicture}
    \node[] at (2.2,0.5)
    {
\begin{tikzpicture}

\definecolor{darkcyan0118194}{RGB}{0,118,194}
\definecolor{darkgray176}{RGB}{176,176,176}

\begin{axis}[
height=5cm,
legend cell align={left},
legend style={
  fill opacity=0.8,
  draw opacity=1,
  text opacity=1,
  at={(0,1)},
  anchor=north west,
  draw=none,
  fill=none
},
minor xtick={},
minor ytick={},
scaled ticks=false,
tick align=outside,
tick pos=left,
width=8cm,
x grid style={darkgray176},
xlabel={eng strain \(\displaystyle \varepsilon_{\mathrm{eng}}\) [-]},
xmin=0, xmax=0.03,
xtick style={color=black},
xtick={0,0.01,0.02,0.03},
xticklabels={
  \(\displaystyle {0.00}\),
  \(\displaystyle {0.01}\),
  \(\displaystyle {0.02}\),
  \(\displaystyle {0.03}\)
},
y grid style={darkgray176},
ylabel={eng stress \(\displaystyle \sigma_{\mathrm{eng}}\) [MPa]},
ymin=0, ymax=500,
ytick style={color=black},
ytick={0,250,500},
yticklabels={\(\displaystyle {0}\),\(\displaystyle {250}\),\(\displaystyle {500}\)}
]
\addplot [thick, darkcyan0118194, forget plot]
table {%
0.0001 3.51507385462922
0.000699999999999999 24.6177490748455
0.00206568542494924 71.8266725965037
0.00368391918985787 124.220697948851
0.0047021529547665 153.530612545926
0.00572038671967512 179.378894558763
0.00673862048458375 201.890314915783
0.00775685424949238 221.258224676842
0.00877508801440102 237.23957041817
0.00979332177930964 249.920653447901
0.0108115555442183 259.4929271726
0.0118297893091269 266.429868259805
0.0128480230740355 271.348416677179
0.0138662568389442 274.86931596303
0.0148844906038528 277.43502836127
0.0159027243687614 279.35452421517
0.01692095813367 280.882219359286
0.0179391918985787 282.190243870149
0.0189574256634873 283.373271416145
0.0199756594283959 284.4817946771
0.0209938931933045 285.543351444249
0.0220121269582132 286.573459753088
0.0230303607231218 287.581525277563
0.0240485944880304 288.573703322047
0.025066828252939 289.554102719394
0.0260850620178476 290.525547483482
0.0271032957827563 291.490112772424
0.0281215295476649 292.449373573124
0.0291397633125735 293.404509765262
0.0301579970774822 294.35640509147
0.0311762308423908 295.305752770255
0.0321944646072994 296.253114484136
0.0332126983722081 297.198956484326
0.0342309321371167 298.143672964828
0.0352491659020253 299.087598826685
0.036267399666934 300.03100135595
0.0372856334318426 300.974080206717
0.0383038671967513 301.916988694353
0.0393221009616599 302.859862185559
0.0403403347265685 303.802831131266
0.0413585684914772 304.74601848939
0.0423768022563858 305.689532378725
0.0433950360212945 306.633463033751
0.0444132697862031 307.577883617591
0.0454315035511117 308.522854468695
0.0464497373160204 309.468428183952
0.047467971080929 310.414653391485
0.0484862048458376 311.361574961858
0.0495044386107463 312.30923255056
};
\addplot [semithick, black, mark=o, mark size=1.5, mark options={solid,fill opacity=0}, only marks, forget plot]
table {%
0.00928420489685533 243.989493470704
0.0164118412512157 280.154394818873
0.0281215295476649 292.449373573124
};
\addplot [thick, darkcyan0118194, dash pattern=on 7.4pt off 3.2pt, forget plot]
table {%
0 0
0.0002 6.81761244732804
0.0004 13.6444563547957
0.0006 20.4799955045419
0.0008 27.3217920706435
0.001 34.1595143483833
0.0012 40.9585834054561
0.0014 47.669863882292
0.0016 54.3591475216287
0.0018 61.1276067507169
0.002 67.902521952612
0.0022 74.5447608993403
0.0024 80.9405118226341
0.0026 87.3168410168048
0.0028 93.9200088371401
0.003 100.526313951907
0.0032 106.834905806027
0.0034 112.779306450284
0.0036 118.811860287714
0.0038 125.0437989779
0.004 130.916293568872
0.0042 136.29254206487
0.0044 141.791798086538
0.0046 147.732967065699
0.0048 153.469644471226
0.005 158.588532664306
0.0052 163.478927933225
0.0054 168.599814061845
0.0056 173.566669740299
0.0058 178.109489860962
0.006 182.737864992794
0.0062 187.737609318952
0.0064 192.480455779897
0.0066 196.569881700913
0.0068 200.360281730009
0.007 204.300671008068
0.0072 208.4238694858
0.0074 212.572832882199
0.0076 216.753435779869
0.0078 220.788398231724
0.008 224.516800859812
0.0082 228.107508016215
0.0084 231.68389441655
0.0086 235.153091570222
0.0088 238.481955329034
0.009 241.739431270076
0.0092 244.946984558235
0.0094 248.109506305544
0.0096 251.229221274845
0.0098 254.237063735063
0.01 257.075300218404
0.0102 259.792700771736
0.0104 262.459911946923
0.0106 265.082602672189
0.0108 267.620844454257
0.011 270.051483798998
0.0112 272.381999381213
0.0114 274.628031235776
0.0116 276.797917912723
0.0118 278.89277938878
0.012 280.913381820723
0.0122 282.862804872956
0.0124 284.744824721225
0.0126 286.563153092177
0.0128 288.321692354694
0.013 290.024741660446
0.0132 291.676830973842
0.0134 293.282653436724
0.0136 294.846752901851
0.0138 296.373558036363
0.014 297.867249621957
0.0142 299.332258178881
0.0144 300.772741615326
0.0146 302.192502546817
0.0148 303.594954941974
0.015 304.983124906365
0.0152 306.359666194274
0.0154 307.726865074602
0.0156 309.086626006054
0.0158 310.440442840885
0.016 311.789373846636
0.0162 313.134040748623
0.0164 314.474665607336
0.0166 315.811148563138
0.0168 317.143178781666
0.017 318.470363251642
0.0172 319.792354782236
0.0174 321.108966857085
0.0176 322.420329260786
0.0178 323.726621655086
0.018 325.028460341717
0.0182 326.326711173391
0.0184 327.622443843336
0.0186 328.916877719962
0.0188 330.211324953588
0.019 331.507137907743
0.0192 332.805665161996
0.0194 334.108217844213
0.0196 335.416046228433
0.0198 336.730325414223
0.02 338.052148374068
0.0202 339.382524549712
0.0204 340.722382331398
0.0206 342.07257403147
0.0208 343.433882276315
0.021 344.807027032072
0.0212 346.192672726155
0.0214 347.591435119474
0.0216 349.003887727147
0.0218 350.430567686021
0.022 351.871981034496
0.0222 353.328607412412
0.0224 354.80090421378
0.0226 356.289310237789
0.0228 357.794249768501
0.023 359.316139293144
0.0232 360.855369909611
0.0234 362.412335350995
0.0236 363.987422983269
0.0238 365.581015314128
0.024 367.193491591711
0.0242 368.825229164276
0.0244 370.476604646516
0.0246 372.147994931533
0.0248 373.839778076573
0.025 375.552334083237
0.0252 377.286045588106
0.0254 379.041298476403
0.0256 380.818482429303
0.0258 382.617991413288
0.026 384.440224119228
0.0262 386.285584357081
0.0264 388.154481411857
0.0266 390.04733036503
0.0268 391.964552388958
0.027 393.906574998376
0.0272 395.873832306939
0.0274 397.866765234776
0.0276 399.885821708135
0.0278 401.931456838833
0.028 404.004133086848
0.0282 406.104320407203
0.0284 408.232496382076
0.0286 410.389146338811
0.0288 412.574763454475
0.029 414.789848847173
0.0292 417.03491165458
0.0294 419.310469099509
0.0296 421.617046542836
0.0298 423.955177523292
0.03 426.3254037843
0.0302 428.728275287221
0.0304 431.164350210798
0.0306 433.634194936369
0.0308 436.138384018204
0.031 438.677500138494
0.0312 441.252134046307
0.0314 443.862884479832
0.0316 446.510358071252
0.0318 449.195169233294
0.032 451.917940026965
0.0322 454.679300009234
0.0324 457.479886060268
0.0326 460.320342188895
0.0328 463.201319315797
0.033 466.123475033252
0.0332 469.087473340712
0.0334 472.093984355224
0.0336 475.143683995865
0.0338 478.237253641337
0.034 481.375379759843
0.0342 484.558753510493
0.0344 487.788070315536
0.0346 491.064029402517
0.0348 494.387333316056
0.035 497.758687398289
0.0352 501.178799237917
0.0354 504.648378087014
0.0356 508.168134245794
0.0358 511.738778414675
0.036 515.361021014081
0.0362 519.035571471646
0.0364 522.763137477389
0.0366 526.544424207026
0.0368 530.380133514193
0.037 534.270963092132
0.0372 538.217605605996
0.0374 542.220747796661
0.0376 546.281069557767
0.0378 550.399242986992
0.038 554.575931414064
0.0382 558.811788406747
0.0384 563.107456758
0.0386 567.46356745483
0.0388 571.880738623482
0.039 576.359574516901
0.0392 580.900664389975
0.0394 585.504581466159
0.0396 590.171881860792
0.0398 594.903103512915
0.04 599.698765123721
0.0402 604.559365105249
0.0404 609.4853805438
0.0406 614.477266181905
0.0408 619.535453423759
0.041 624.660349367964
0.0412 629.852335872636
0.0414 635.111768656665
0.0416 640.438976442218
0.0418 645.834259715925
0.042 651.297902086317
0.0422 656.830133313162
0.0424 662.431166734524
0.0426 668.101185698532
0.0428 673.840342261414
0.043 679.648756785521
0.0432 685.526517617305
0.0434 691.473680815118
0.0436 697.490269936119
0.0438 703.576275889278
0.044 709.731656860542
0.0442 715.956338312556
0.0444 722.250213061805
0.0446 728.613141433148
0.0448 735.044951492735
0.045 741.545439358052
0.0452 748.114369584467
0.0454 754.751475626299
0.0456 761.456460370611
0.0458 768.228996740965
0.046 775.06872836847
0.0462 781.975270326557
0.0464 788.948209926406
0.0466 795.987107568394
0.0468 803.091497646418
0.047 810.260889499721
0.0472 817.494768408697
0.0474 824.792596629171
0.0476 832.153814461155
0.0478 839.577841346937
0.048 847.064076991976
0.0482 854.611902510356
0.0484 862.220681579134
0.0486 869.889761606445
0.0488 877.618474904679
0.049 885.406139864751
0.0492 893.252062127639
0.0494 901.1555357485
0.0496 909.115844350325
0.0498 917.132262262543
0.05 925.204055640303
};
\addplot [semithick, black, dash pattern=on 5.55pt off 2.4pt]
table {%
-1 -1
};
\addlegendentry{single element}
\addplot [semithick, black]
table {%
-1 -1
};
\addlegendentry{ply}
\end{axis}

\end{tikzpicture}
    };

    \node at (-2,-3.4) {\includegraphics[clip, width=0.65\textwidth, trim = 1cm 20cm 0 35cm]{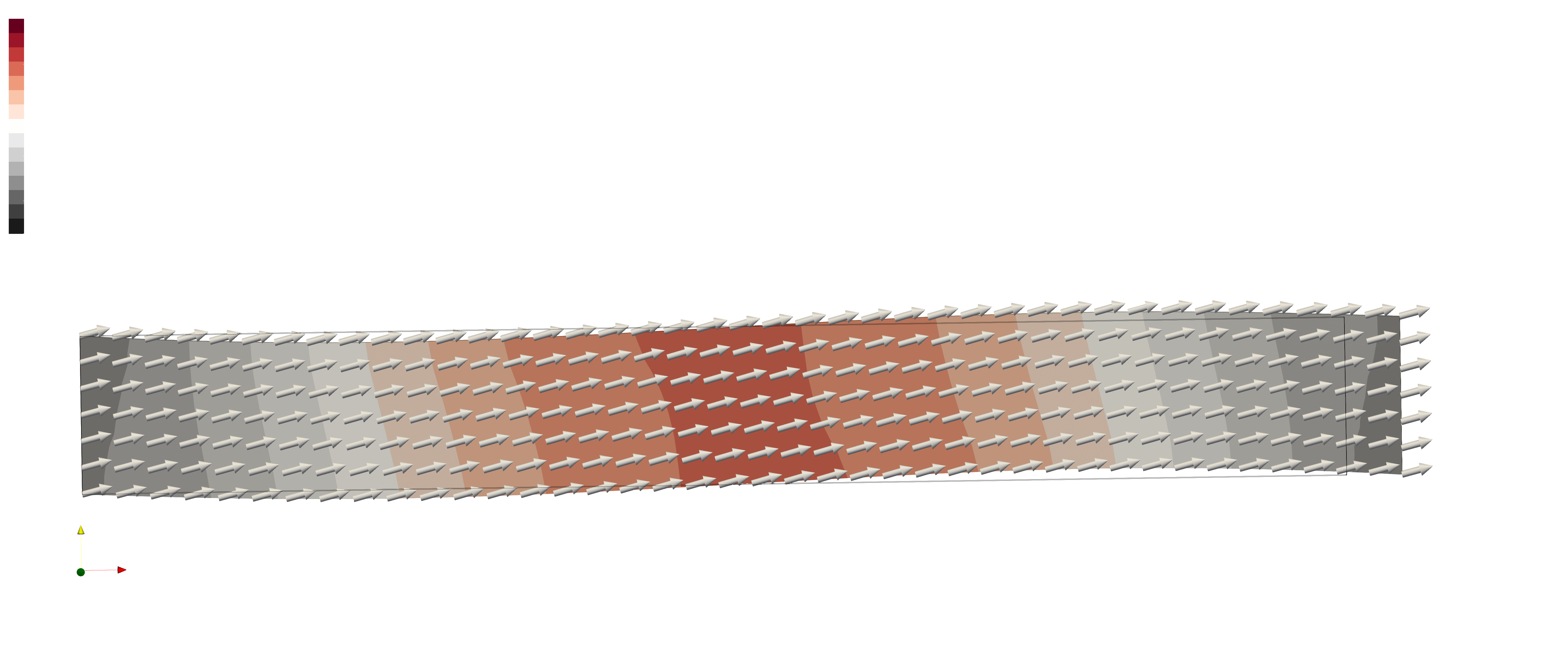}};
    \node at (-2,-5.4) {\includegraphics[clip, width=0.65\textwidth, trim = 1cm 20cm 0 35cm]{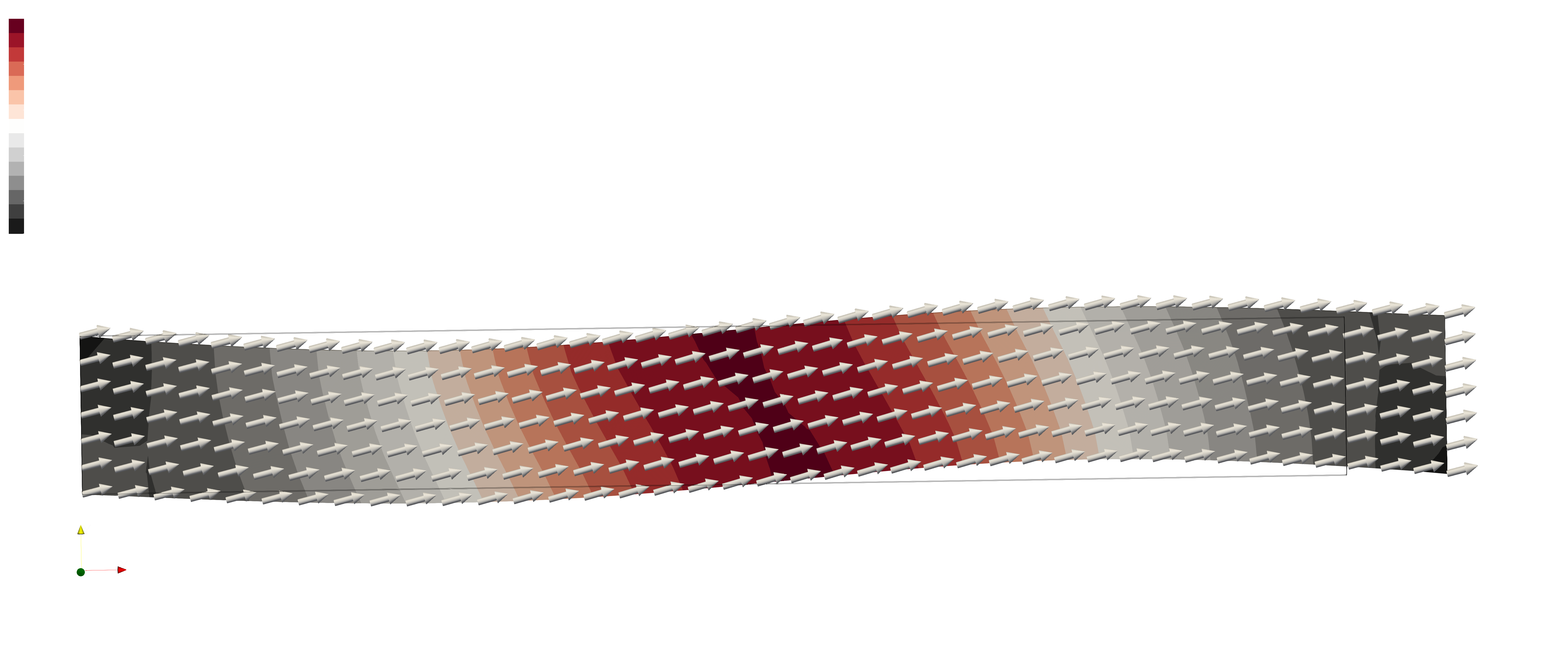}};
    \node at (-2,-7.4) {\includegraphics[clip, width=0.65\textwidth, trim = 1cm 20cm 0 35cm]{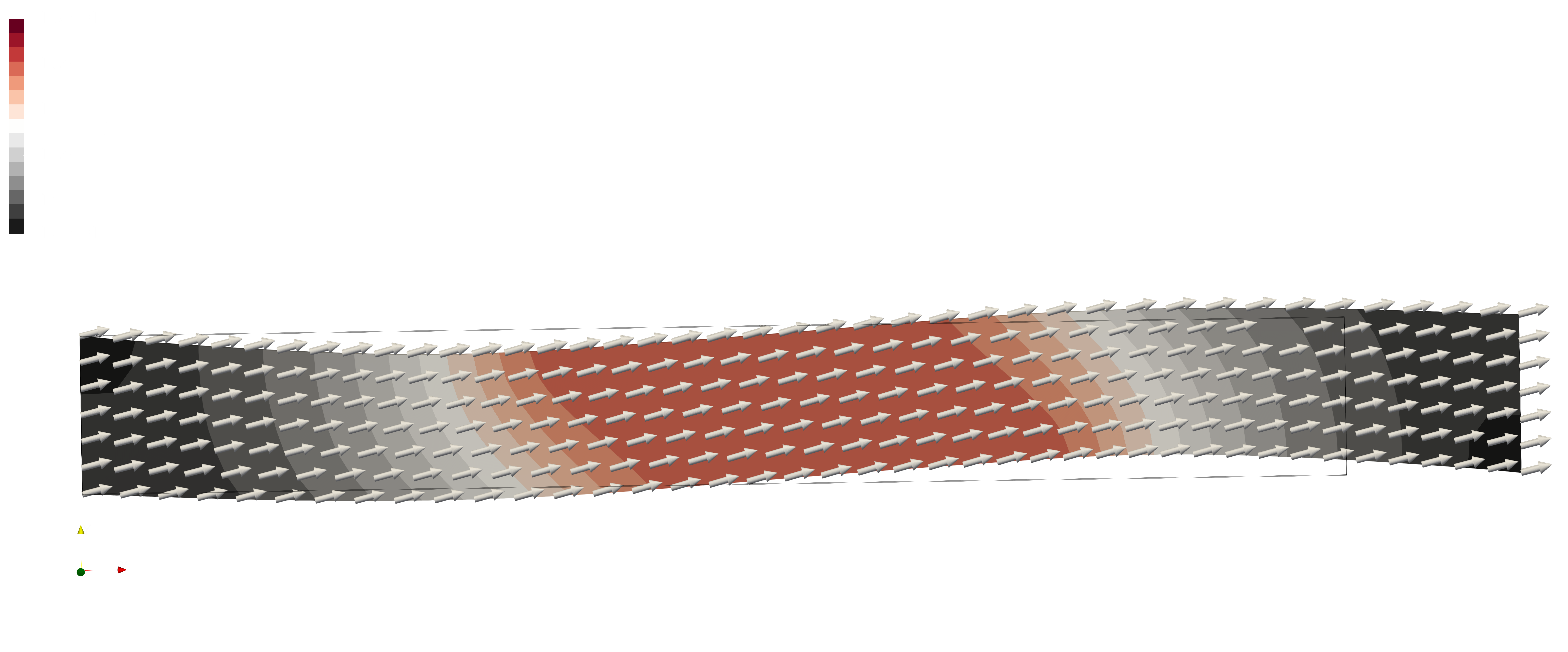}};

    \node at (7.4,-4.2) {\includegraphics[clip, width=0.61\textwidth, trim = 1cm 20cm 0 35cm]{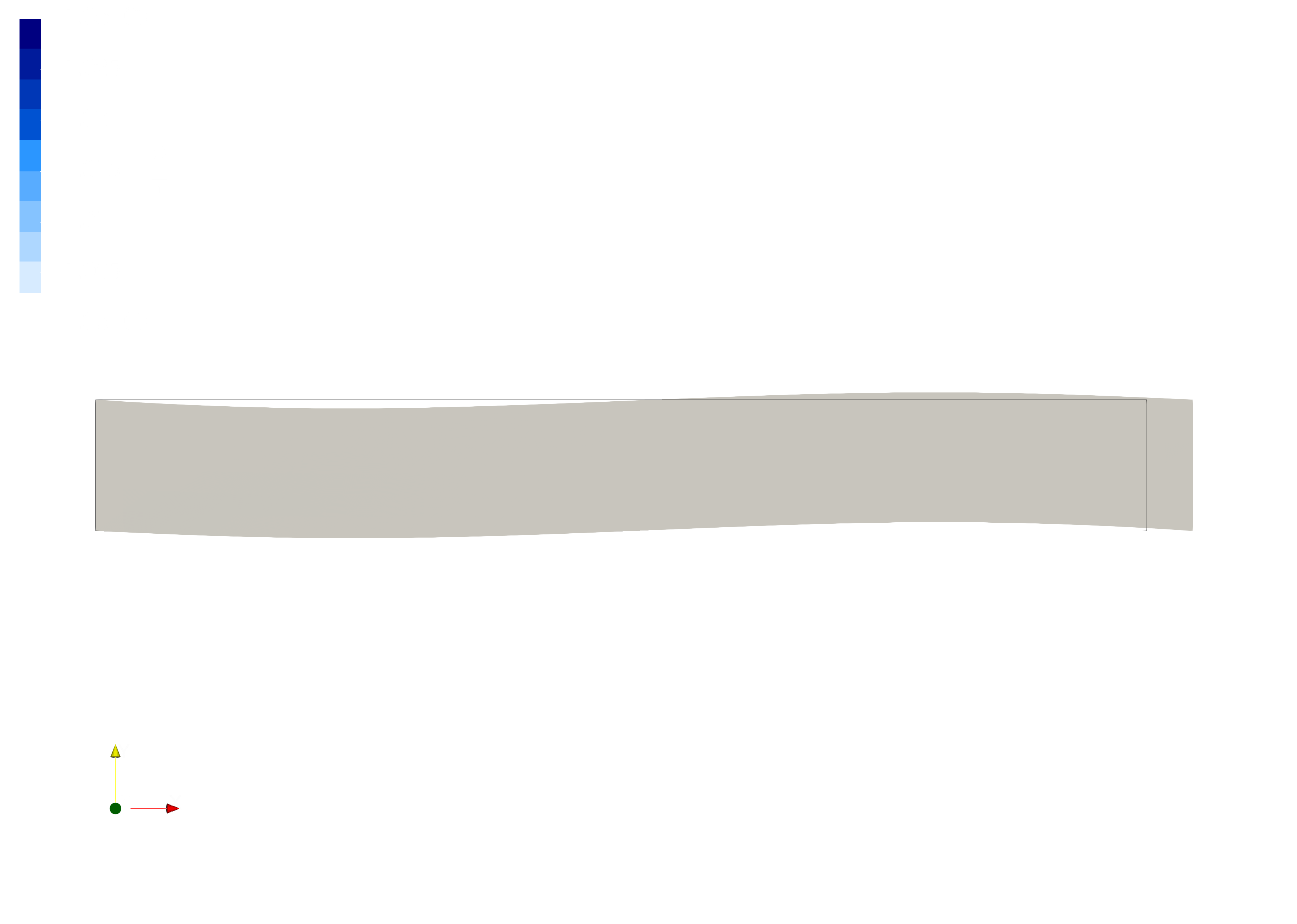}};
    \node at (7.4,-6.1) {\includegraphics[clip, width=0.61\textwidth, trim = 1cm 20cm 0 35cm]{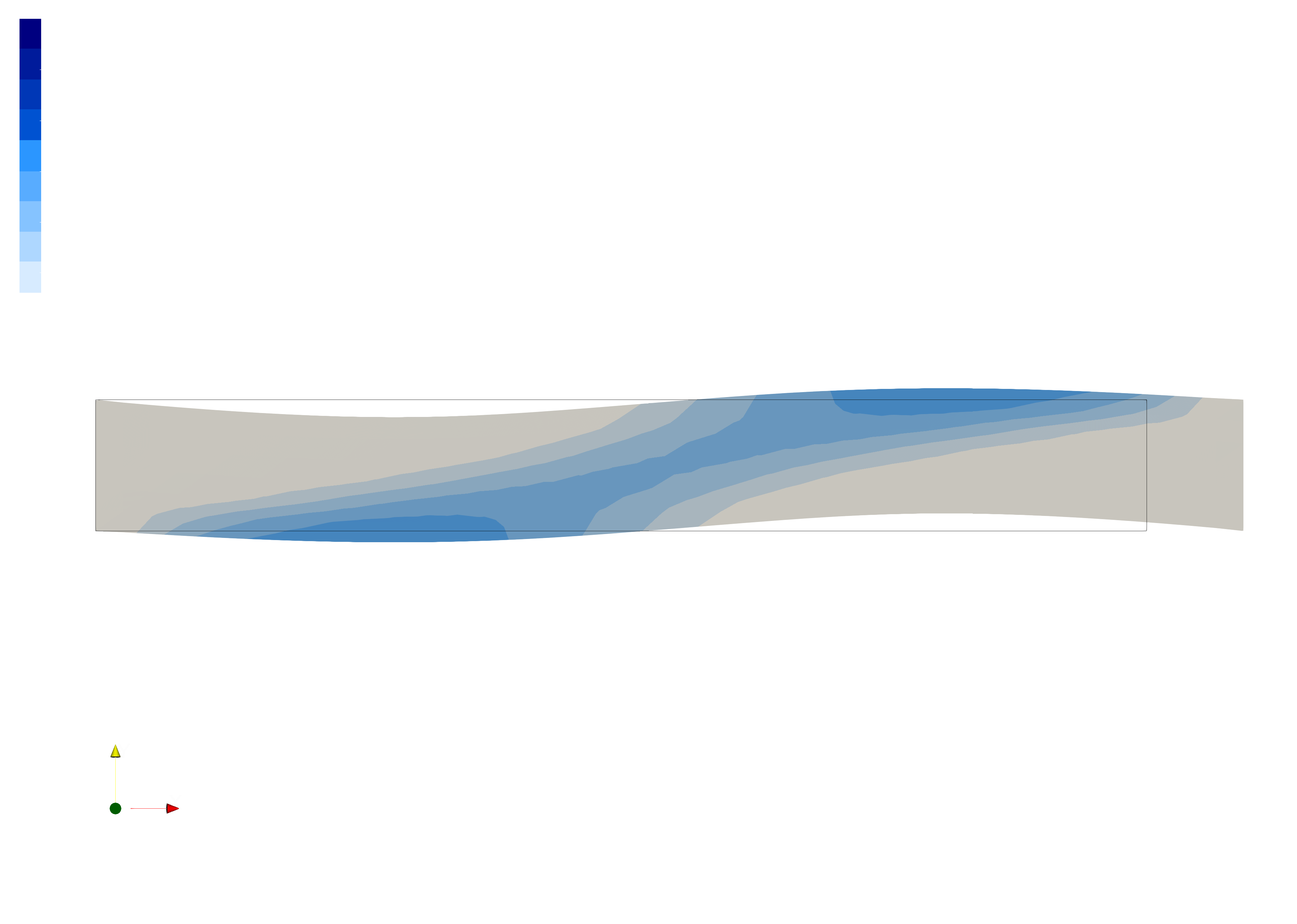}};
    \node at (7.4,-8.0) {\includegraphics[clip, width=0.61\textwidth, trim = 1cm 20cm 0 35cm]{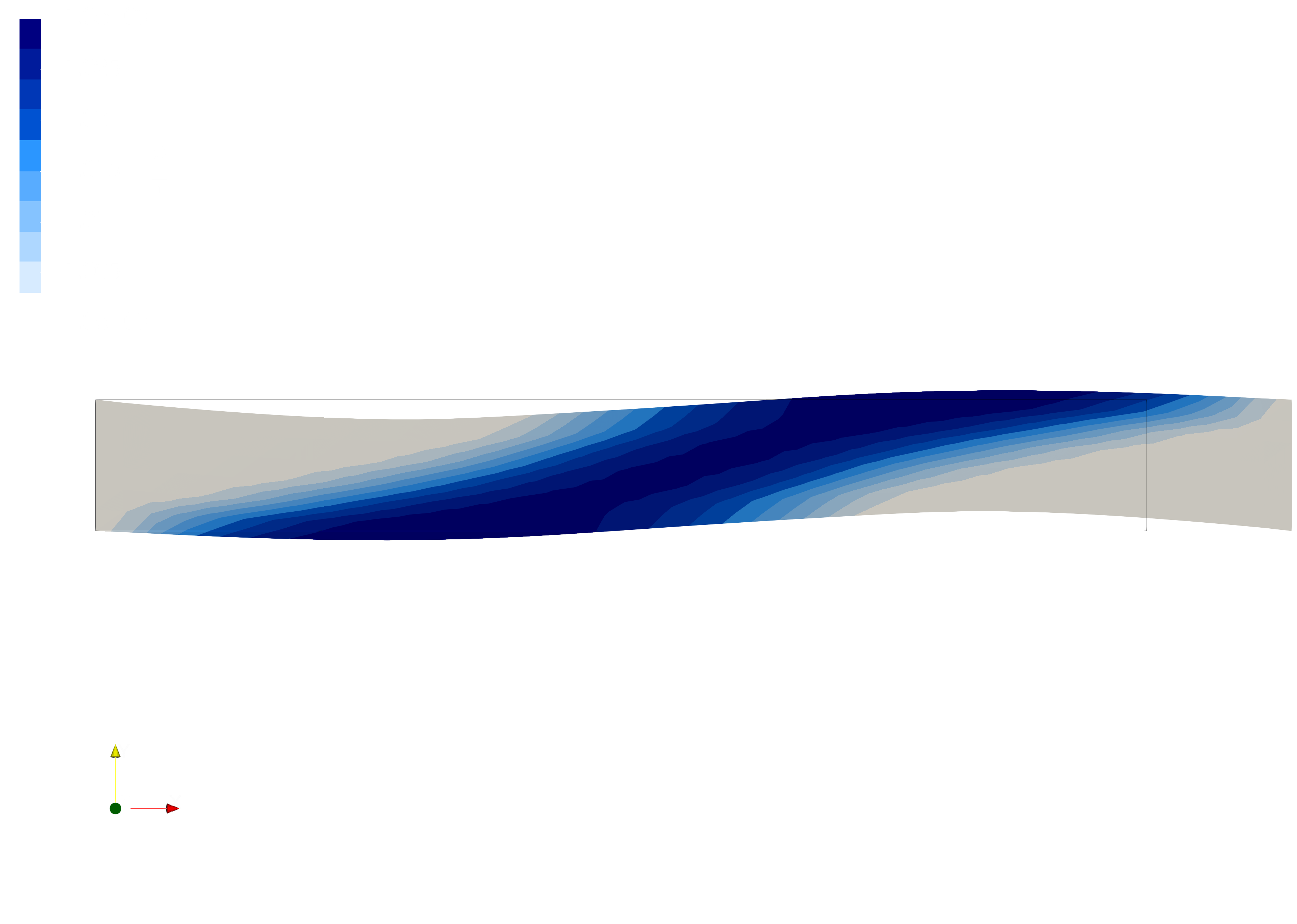}};

    \node at (-2.35,-9.1) {\def\svgwidth{0.4\columnwidth}{\scalebox{1.0}{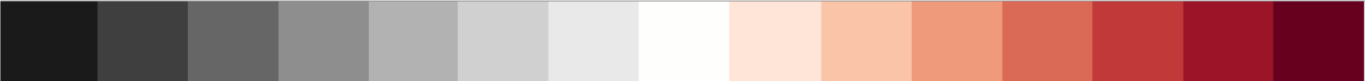}}};
    \node[] at ( 1.0,-9.1){$16$};
    \node[] at (-5.8,-9.1){$14$};
    \node[] at (-2.4,-9.6){$\theta$ ($^\circ$)};

    \node at (7.75,-9.1) {\def\svgwidth{0.4\columnwidth}{\scalebox{1.0}{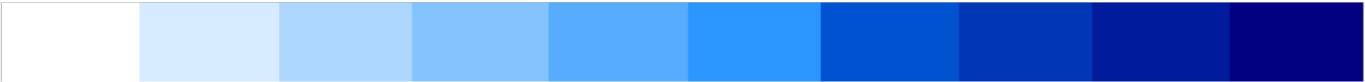}}};
    \node[] at (11.5,-9.1){$1.03$};
    \node[] at (4.1,-9.1){$1.00$};
    \node[] at (7.4,-9.6){$F^{\mr{p}}_{11}$};
    \node[] at (1.45,0.5){(a)};
    \node[] at (3.0,0.8){(b)};
    \node[] at (5.45,0.9){(c)};

    \node[] at (-7.0,-3.6) {(a)};
    \node[] at (-7.0,-5.6) {(b)};
    \node[] at (-7.0,-7.6) {(c)};

      \end{tikzpicture}

  \caption{Ply simulation with $\theta_0=15^\circ$: evolution of fiber angle $\theta$ (\emph{left}) and plastic deformation $F^{\mr{p}}_{11}$ in load direction for the mode with highest initial viscosity (\emph{right}) at indicated time instances on the stress-strain plot (\emph{top}). For comparison, a single element test with $\dot{\varepsilon}_{\mr{eng}}=10^{-4} \,\si{\per\second}$ is added to the stress-strain curve. Deformed mesh is magnified ($\times 5$).}
  \label{fig:theta-contours}
\end{figure}

It has been observed in \Cref{sec:const-strain-rate} that the mesomodel and micromodel under tension with $\theta_0=15^\circ$ showed a pronounced upswing of the stress after yielding (see \Cref{fig:examples-direction-dependence}), due to an increasing alignment of the fibers with the load direction. The same type of re-orientation is prevented by the grips in the coupon test.
This can be illustrated by plotting the evolution of off-axis angle ($\theta$) and plastic deformation component in $\bm{e}_1$-direction ($F^{\mr{p}}_{11}$), for the mode with the highest initial viscosity (mode 24 in \Cref{tab:relaxation-spectrum}), at three different time-steps (see \Cref{fig:theta-contours}). 
As the fibers tend to align with the load direction near the ends ($\theta < \theta_0$), the off-axis angles increase in the middle of the specimen ($\theta > \theta_0$), which is opposite to the direction of re-orientation as was previously seen with the single element test under tension (see \Cref{fig:fibangle-strain}). This increase of matrix-dominated loading, combined with the presence of stress concentrations in the ply specimen, results in an earlier development of plasticity with respect to the single element (see \Cref{fig:theta-contours}, \textit{top}). 

The deformations in the coupon test are inhomogeneous and cannot be used directly as material input. To obtain a more homogeneous deformation state, off-axis specimens with oblique tabs may be used  \cite{sunOblique1993}. The analytical parameter identification procedure outlined in \Cref{sec:analytical-parameter-identification} may then be directly applied to experimental data of off-axis constant strain rates, without requiring a pre-calibrated micromodel to generate inputs for the mesoscopic constitutive model.

\section{Conclusion}
A mesoscopic constitutive model for simulating rate-dependent plasticity and creep in unidirectional thermoplastic composites has been presented.
The model is an extension of a viscoplastic material model for isotropic polymers with an Eyring-type non-Newtonian flow rule. 
Strong anisotropy is incorporated through the use of three transversely isotropic stress invariants in the flow rule. As a result, plastic flow in fiber direction is removed and pressure-dependency of the polymer matrix is taken into account by extending the Eyring relation with anisotropic pressure dependence.
An important feature of the present invariant-based anisotropic viscoplasticity model is that it can describe both rate-dependent plasticity and creep in thermoplastic polymer composites with non-Newtonian flow.

The constitutive equations are implicitly integrated, which allows for the use of relatively large time steps. Furthermore, a consistent tangent stiffness modulus has been derived by linearizing the stress update algorithm.
The model requires four viscoplasticity-related input parameters to describe direction-, rate- and pressure-dependent plasticity and creep, obtained from a few stress-strain curves under off-axis loading. For an accurate pre-yield and creep response, multiple modes can be used with a relaxation spectrum determined from a \emph{single} stress-strain curve.  
In this manuscript, a micromodel for unidirectional carbon/PEEK is used to determine the mesocale model parameters. However, off-axis coupon tests with oblique ends may also be used. 

The mesoscopic constitutive model has been compared to a previously developed micromodel for unidirectional carbon/PEEK.
It has been shown that the mesoscale model gives a response similar to the micromodel under various strain rates and off-axis angles. 
However, under transverse compression, a hardening contribution can be included for an improved post-yield response. 
The model gives satisfactory results under creep, although not as good as under constant strain rates. This may indicate that the parameter identification procedure, solely based on (short-term) constant strain rate data, requires further improvements. 

Finally, the mesoscale model has been applied to the simulation of unidirectional composite coupon tests under off-axis strain rates and shows a good agreement with experiments. The development of the mesoscopic constitutive model, with a few model parameters, while retaining a high degree of the accuracy of a detailed micromodel, is an important step towards virtual testing of thermoplastic composite laminates. Further extensions can be made to cover multiple relaxation processes and to include temperature dependence.

\subparagraph{Summary of contributions}
\noindent The EGP model has been extended for unidirectional thermoplastic composites. Compared to other anisotropic versions of the EGP model for \emph{short} and \emph{long} fiber composites \cite{amiri-radAnisotropic2019,amiri-radImproved2021}, new features of the present model are:
\begin{itemize}
  \item Strong anisotropy is described by transversely isotropic stress invariants
  \item Plastic flow in fiber direction is removed
  \item The Eyring-type viscosity function is extended with anisotropic pressure dependence
  \item The constitutive equations are implicitly integrated and consistently linearized
  \item The model is formulated in global frame and does not require rotations to local frame
\end{itemize}
\noindent Compared to previous invariant-based Perzyna-type viscoplasticity models for unidirectional composites \cite{koerberExperimental2018,gerbaudInvariant2019, rodrigueslopesInvariantbased2022}:
\begin{itemize}
  \item An Eyring-type non-Newtonian flow rule, suitable for \emph{thermoplastic} composites, is used to describe both rate-dependent plasticity and creep
  \item Only four parameters and a relaxation spectrum are required, which can be obtained from a small number of off-axis tests (either with a micromodel or with off-axis coupon tests with oblique ends)
  \item The present anisotropic model allows for future extensions regarding the effects of aging \cite{klompenModeling2005}, temperature dependence \cite{tervoortConstitutive1997} and to cover multiple relaxation processes \cite{klompenNonlinear1999} through the Eyring relation
\end{itemize}

\newpage
Highlights
\begin{itemize}
  \item A viscoplasticity model for unidirectional thermoplastic composites has been developed
  \item A consistent linearization is performed to obtain the tangent stiffness modulus 
  \item A small number of off-axis tests is necessary to determine the material parameters
  \item A detailed micromodel for carbon/PEEK is used to compare the response
  \item An off-axis ply is simulated and compared against experiments
\end{itemize}

\appendix
\section{Jacobian internal Newton-Raphson scheme}
\label{sec:internal-jac}

\noindent The Jacobian for solving the plastic deformation gradient $\textbf{F}_{\mr{p}i}$ with the internal scheme of each mode $i$ is determined in this appendix. To improve readability, subscript $i$ is dropped and index notation is used.
The residual for \emph{each} mode reads
\begin{align}
  R^{\F_{\mr{p}i}}_{ij} &= F^\mr{p}_{ij} - f_{ij}
  \label{eq:int-residual}
\end{align}
where 
\begin{equation}
  f_{ij} = \Pi_{ik} F^{\mr{p},0}_{kj}
\end{equation}
with
\begin{align}
  \Pi_{ik} = \left(\underbrace{\delta_{il}-\frac{\dt}{2} \hat{D}^\mr{p}_{il}}_{Z_{il}}\right)^{-1} \left(\underbrace{\delta_{lk}+\frac{\dt}{2} \hat{D}^\mr{p}_{lk}}_{Y_{lk}}\right)
  \label{eq:xi-index}
\end{align}
Taking the derivative of \Cref{eq:int-residual} with respect to the plastic deformation gradient gives the Jacobian 
\begin{align}
  \pdv{R^{\F_{\mr{p}i}}_{ij}}{F^{\mr{p}}_{mn}} = \delta_{im} \delta_{jn} - \pdv{f_{ij}}{\Pi_{kl}}\pdv{\Pi_{kl}}{\hat{D}^\mr{p}_{uv}}
  \left[
    \pdv{\hat{D}^\mr{p}_{uv}}{\Sigma^\mr{sym}_{qp}}
    \pdv{\Sigma^{\mr{sym}}_{qp}}{F^\mr{e}_{rs}} 
    \pdv{F^{\mr{e}}_{rs}}{F^\mr{p}_{mn}}
  + \pdv{\hat{D}^\mr{p}_{uv}}{\hat{a}_w} \pdv{\hat{a}_w}{F^\mr{p}_{mn}}
\right]
  \label{eq:dR_dFp}
\end{align}
where
\begin{align}
  \pdv{f_{ij}}{\Pi_{mn}} &=  \delta_{im} F^{\mr{p},0}_{nj} 
  \label{eq:df-dXi} \\
  \pdv{\Pi_{ij}}{\hat{D}^\mr{p}_{mn}} &= \pdv{Z_{ik}^{-1}}{\hat{D}^\mr{p}_{mn}} Y_{kj} + 
  Z_{ik} \pdv{Y_{kj}}{\hat{D}^\mr{p}_{mn}} \label{eq:dFincr-dD}
\end{align}
with
\begin{align}
  \pdv{{Z_{kj}}^{-1}}{\hat{D}^\mr{p}_{mn}} &= - {Z_{ik}}^{-1} \pdv{Z_{ir}}{\hat{D}_{mn}} Z_{rj}^{-1} \\
  \pdv{Z_{ij}}{\hat{D}^{\mr{p}}_{mn}} &= -\frac{\dt}{2} \delta_{im} \delta_{jn} \\
  \pdv{Y_{ij}}{\hat{D}^\mr{p}_{mn}} &= \frac{\dt}{2} \delta_{im} \delta_{jn}
\end{align}
The other derivatives read
\begin{align}
  \pdv{\hat{D}^\mr{p}_{ij}}{\Sigma^\mr{sym}_{mn}} &= \frac{1}{\eta} \left[\hat{N}^\mr{p}_{mn} \hat{N}^\mr{p}_{ij}    + \bar{\Sigma} \pdv{\hat{N}^\mr{p}_{ij}}{\Sigma^{\mr{sym}}_{mn}} \right]\\
  \pdv{\Sigma^{\mr{sym}}_{ij}}{F^\mr{e}_{mn}}
    &= \frac{1}{2} \left( 
  \pdv{\Sigma_{ij}}{F^\mr{e}_{mn}} +
  \pdv{\Sigma_{ji}}{F^\mr{e}_{mn}}
\right) \\
  \pdv{F^{\mr{e}}_{rs}}{F^\mr{p}_{mn}} &= -F_{rk} \left(F^\mr{p}_{km}\right)^{-1} \left(F^\mr{p}_{ns}\right)^{-1}  \label{eq:dFe-dFp} \\
\pdv{\hat{D}^\mr{p}_{ij}}{\hat{a}_{q}} &= \frac{1}{\eta} \left[ \pdv{\bar{\Sigma}}{\hat{a}_{q}} \hat{N}^\mr{p}_{ij}    + \bar{\Sigma} \pdv{\hat{N}^\mr{p}_{ij}}{\hat{a}_{q}} \right]\\
\pdv{\hat{a}_q}{F^\mr{p}_{mn}} &= \frac{1}{\norm{\hat{a}}}\left[ \delta_{qm}  - \frac{1}{\norm{\hat{a}}^2} \hat{A}_{qm} \right] a^0_{n} \label{eq:da-dF}
\end{align}
with
\begin{align}
  \pdv{\Sigma_{ij}}{F^\mr{e}_{mn}} &= \Ifourth_{kmin}\,\sigma_{kl} \left(F^\mr{e}_{jl}\right)^{-1} + F^\mr{e}_{ki}\pdv{\sigma_{kl}}{F^\mr{e}_{mn}} \left(F^\mr{e}_{jl}\right)^{-1}- F^\mr{e}_{ki}\,\sigma_{kl} \left(F^{\mr{e}}_{jm}\right)^\mr{-1} \left(F^\mr{e}_{nl}\right)^\mr{-1} \label{eq:dM-dFe} \\
  \pdv{\hat{N}^p_{ij}}{\Sigma^\mr{sym}_{kl}} &= \frac{1}{\bar{\Sigma}} \left(  \pdv{^2 \hat{I}_1}{\Sigma^\mr{sym}_{kl} \partial \Sigma^\mr{sym}_{ij}} + \alpha_2 \pdv{^2 \hat{I}_2}{\Sigma^\mr{sym}_{kl} \partial \Sigma^\mr{sym}_{ij}} - \hat{N}^p_{kl} \hat{N}^p_{ij}\right) \label{eq:dNp-dMsym} \\
  \pdv{\bar{\Sigma}}{\hat{a}_m} &= \frac{1}{\bar{\Sigma}} \left(\pdv{\hat{I}_1}{\hat{a}_m} + \alpha_2 \pdv{\hat{I}_2}{\hat{a}_m} \right) \label{eq:dMsym-da} \\
  \pdv{\hat{N}^{\mr{p}}_{ij}}{\hat{a}_m} &= \frac{1}{\bar{\Sigma}} \left(\pdv{^2 \hat{I}_1}{\hat{a}_m  \partial \Sigma^\mr{sym}_{ij}} + \alpha_2 \pdv{^2 \hat{I}_2}{\hat{a}_m \partial \Sigma^\mr{sym}_{ij}} - \hat{N}^{\mr{p}}_{ij} \pdv{\bar{\Sigma}}{\hat{a}_m} \right) \label{eq:dNp-da}
\end{align}
The derivative of the Cauchy stress with respect to the elastic deformation in \Cref{eq:dM-dFe} is given as
\begin{align}
 \pdv{\sigma_{ij}}{F^{\mr{e}}_{kl}} &= \pdv{\sigma^{\mr{iso}}_{ij}}{F^{\mr{e}}_{kl}} + \pdv{\sigma^{\mr{trn}}_{ij}}{F^{\mr{e}}_{kl}} \label{eq:dsigma-dFe} \\  
  \pdv{\sigma^{\mr{iso}}_{ij}}{F^{\mr{e}}_{kl}} &= \frac{\mu}{J^{\mr{e}}} \left( \pdv{B^{\mr{e}}_{ij}}{F^{\mr{e}}_{kl}} - (B^{\mr{e}}_{ij} - \delta_{ij}) (F^{\mr{e}}_{kl})^{-\mr{T}} \right) + \lambda J^{\mr{e}} \delta_{ij} (F^{\mr{e}}_{kl})^{-\mr{T}} \\
\pdv{\sigma^{\mr{trn}}_{ij}}{F^{\mr{e}}_{kl}} &= \frac{1}{J^{\mr{e}}} \left( \Phi^{1}_{ijkl} + \Phi^{2}_{ijkl} + \Phi^{3}_{ijkl}   \right) - \frac{1}{J^{\mr{e}}} \sigma^{\mr{trn}}_{ij} (F^{\mr{e}}_{kl})^{-\mr{T}}\\
\Phi^1_{ijkl} &= 2\beta (\xi_2 - 1) \pdv{B^{\mr{e}}_{ij}}{F^{\mr{e}}_{kl}} \\
\Phi^2_{ijkl} &= 4 \beta A_{ij} F^{\mr{e}}_{kl} \\
\Phi^3_{ijkl} &= -\alpha \left( \pdv{B^{\mr{e}}_{im}}{F^{\mr{e}}_{kl}} A_{mj} + \pdv{B^{\mr{e}}_{jm}}{F^{\mr{e}}_{kl}} A_{im} \right) \\
\pdv{B^{\mr{e}}_{ij}}{F^{\mr{e}}_{kl}} &= F^{\mr{e}}_{jl} \delta_{ik} + F^{\mr{e}}_{il} \delta_{jk} 
\end{align}
The other terms in \Cref{eq:dNp-dMsym,eq:dMsym-da,eq:dNp-da} can be expanded as
\begin{align}
  \pdv{\hat{I}_1}{\hat{a}_m} &= \Sigma^{\mr{pind}}_{rs} \pdv{\Sigma^{\mr{pind}}_{rs}}{\hat{a}_m} - \pdv{\hat{I}_2}{\hat{a}_m} \\
  \pdv{\hat{I}_2}{\hat{a}_m} &= \Sigma^{\mr{pind}}_{mj} \Sigma^{\mr{pind}}_{jk} \hat{a}_k +  \hat{a}_q \Sigma^{\mr{pind}}_{qp} \Sigma^{\mr{pind}}_{pm} +  \pdv{\hat{I}_2}{\Sigma^{\mr{pind}}_{rs}} \pdv{\Sigma^\mr{pind}_{rs}}{\hat{a}_m}\\
  \pdv{^2 \hat{I}_1}{\Sigma^\mr{sym}_{rs} \partial \Sigma^\mr{sym}_{kl}} &= \left(\hat{\Ppind}_{ijrs} - \pdv{^2\hat{I}_2}{\Sigma^\mr{sym}_{rs} \partial \Sigma^{\mr{pind}}_{ij}} \right) \hat{\Ppind}_{ijkl}  \\
  \pdv{^2 \hat{I}_2}{\Sigma^\mr{sym}_{rs} \partial \Sigma^\mr{sym}_{kl}} &=  \pdv{^2\hat{I}_2}{\Sigma^\mr{sym}_{rs} \partial \Sigma^{\mr{pind}}_{ij}} \hat{\Ppind}_{ijkl}  \\
  \pdv{^2 \hat{I}_1}{\hat{a}_m \partial \Sigma^{\mr{sym}}_{kl}} &= \pdv{\hat{I}_1}{\Sigma^{\mr{pind}}_{ij}} \pdv{\hat{\Ppind}_{ijkl}}{\hat{a}_m} + \left(\pdv{\Sigma^{\mr{pind}}_{ij}}{\hat{a}_m} - \pdv{^2 \hat{I}_2}{\hat{a}_m  \partial \Sigma^{\mr{pind}}_{ij}} \right) \hat{\Ppind}_{ijkl} \\
  \pdv{^2 \hat{I}_2}{\hat{a}_m  \partial \Sigma^{\mr{sym}}_{kl}}  &= {\pdv{\hat{I}_2}{\Sigma^{\mr{pind}}_{ij}}} \pdv{\hat{\Ppind}_{ijkl}}{\hat{a}_m} + \pdv{^2 \hat{I}_2 }{\hat{a}_m \partial \Sigma^{\mr{pind}}_{ij}}  \hat{\Ppind}_{ijkl}
\end{align}
where
\begin{align}
  \pdv{\Sigma^{\mr{pind}}_{rs}}{\hat{a}_m} &= \pdv{\Sigma^{\mr{pind}}_{rs}}{{\hat{\Ppind}}_{ijkl}}  \pdv{{\hat{\Ppind}}_{ijkl}}{\hat{a}_{m}} \\ 
  \pdv{{\hat{\Ppind}}_{ijkl}}{\hat{a}_{m}} &= \pdv{{\hat{\Ppind}}_{ijkl}}{\hat{A}_{rs}} \pdv{\hat{A}_{rs}}{\hat{a}_m} \\ 
  \pdv{^2 \hat{I}_2 }{\hat{a}_m \partial \Sigma^{\mr{pind}}_{ij}} &=  \left( \pdv{\hat{A}_{ir}}{\hat{a}_m} \Sigma^{\mr{pind}}_{rj} + \hat{A}_{ir} \pdv{\Sigma^{\mr{pind}}_{rj}}{\hat{a}_m} + \pdv{\Sigma^{\mr{pind}}_{ir}}{\hat{a}_{m}} \hat{A}_{rj} + \Sigma^{\mr{pind}}_{ir} \pdv{\hat{A}_{rj}}{\hat{a}_m} \right)  
\end{align}
The remaining derivatives can be computed at each \emph{internal} iteration
\begin{align}
  \pdv{^2\hat{I}_2}{\Sigma^\mr{sym}_{rs} \partial \Sigma^{\mr{pind}}_{ij}} &= \left(\hat{A}_{im} \hat{\Ppind}_{mjrs} + \hat{\Ppind}_{imrs} \hat{A}_{mj}\right)  \\
  \pdv{\hat{I}_1}{\Sigma^{\mr{pind}}_{rs}} &= 2 \Sigma^{\mr{pind}}_{rs} \\
  \pdv{\hat{I}_2}{\Sigma^{\mr{pind}}_{rs}} &= \hat{A}_{rk} \, \Sigma^{\mr{pind}}_{ks} + \Sigma^{\mr{pind}}_{rj}  \, \hat{A}_{js}  \\
  \pdv{\Sigma^{\mr{pind}}_{rs}}{{\hat{\Ppind}}_{ijkl}} &= \delta_{ir} \delta_{js} \Sigma_{kl}  \\
  \pdv{{\hat{\Ppind}}_{ijkl}}{\hat{A}_{rs}} &= -\frac{3}{2} \left( \delta_{ir} \delta_{js} \hat{A}_{kl} + \hat{A}_{ij} \delta_{kr} \delta_{ls} \right)
  + \frac{1}{2}\left( \delta_{kl} \delta_{ir} \delta_{js} + \delta_{ij} \delta_{kr} \delta_{ls} \right) \\
  \pdv{\hat{A}_{rs}}{\hat{a}_m} &= \delta_{rm} \hat{a}_s + \hat{a}_r \delta_{sm} \label{eq:dA-da}
\end{align}

\section{Derivatives external scheme}
\label{sec:derivatives-external}

\noindent $\bullet$ The derivatives in \Cref{eq:internal-residual} are given as
\begin{align}
  \pdv{R_{a_\sigma}}{\sigmaEq} &= - \frac{1}{\sigma_0} \left[ 
    \sinh^{-1}\left(\nicefrac{\bar{\sigma}}{\sigma_0}\right) - \frac{\bar{\sigma}}{\sigma_0}
    \frac{\cosh\left(\nicefrac{\bar{\sigma}}{\sigma_0}\right)}{\sinh^{2}\left(\nicefrac{\bar{\sigma}}{\sigma_0}\right)}
  \right] \exp\left(-\mu \frac{I_3}{\sigma_0}\right) \label{eq:dRa-dsigma} \\
    \pdv{R_{a_\sigma}}{I_3} &=  \frac{\nicefrac{\sigmaEq}{\sigma_0}}{\sinh\left(\nicefrac{\sigmaEq}{\sigma_0}\right)}\exp\left(-\mu \frac{I_3}{\sigma_0}\right) \frac{\mu}{\sigma_0} \label{eq:dRa-dI3} \\
    \pdv{I_3}{\stress} &= \left(\textbf{I}-\bar{\textbf{A}}\right) \label{eq:dI3-dsigma}
  \end{align}
  $\bullet$ Applying the chain rule to the second term on the RHS of \Cref{eq:dFp-da} yields
\begin{align}
  \pdv{ \textbf{R}_{\F_{\mr{p}i}} }{a_\sigma} = -\pdv{\bm{f}_i}{\bm{\Pi}_i}:\pdv{\bm{\Pi}_i}{\hat{\textbf{D}}_{\mr{p}i}} : \pdv{\hat{\textbf{D}}_{\mr{p}i}}{a_\sigma}
  \label{eq:dRa-dashift}
\end{align}
where $\hpdv{\hat{\textbf{D}}_{\mr{p},i}}{a_\sigma} = -\hfrac{\bar{\sigma}}{\left(\eta_{0i} a^2_\sigma\right)} \hat{\textbf{N}}_{\mr{p},i} = -\hfrac{1}{a_\sigma} \hat{\textbf{D}}_{\mr{p},i}$
 The expressions for $\hpdv{\bm{f}_i}{\bm{\Pi}_i}$ and $\hpdv{\bm{\Pi}_i}{\hat{\textbf{D}}_{\mr{p}i}}$ are given by \Cref{eq:df-dXi,eq:dFincr-dD}, respectively. 

\section{Derivatives for consistent tangent modulus}
\label{sec:derivatives-tangent}

\noindent $\bullet$ The derivative $\hpdv{\textbf{R}_{\F_{\mr{p}i}}}{\F}$ in \Cref{eq:dFp-dF} reads
\begin{align}
   \pdv{\textbf{R}_{\F_{\mr{p}i}}}{\F} &=
  - \pdv{\bm{f}}{\bm{\Pi}} : \pdv{\bm{\Pi}}{\hat{\textbf{D}}_{\mr{p}i}} : \pdv{\hat{\textbf{D}}_{\mr{p}i}}{\bm{\Sigma}^{\mr{sym}}_i}
  : \pdv{\Mstress^\mr{sym}_i}{\Mstress_i} :\left[ \pdv{\Mstress_i}{\F_{\mr{e}i}}: \pdv{\F_{\mr{e}i}}{\F}+ \pdv{\Mstress_i}{\stress_i} : \pdv{\stress_i}{\bm{a}} \cdot \pdv{\bm{a}}{\F} \right]
  \label{eq:dR-dF}
\end{align}
The derivatives $\hpdv{\bm{f}}{\bm{\Pi}}$,
$\hpdv{\bm{\Pi}}{\hat{\textbf{D}}_{\mr{p}i}}$,
$\hpdv{\Mstress_i}{\F_{\mr{e}i}}$ and $\hpdv{\stress_i}{\bm{a}}$, 
are given by \Cref{eq:df-dXi,eq:dFincr-dD,eq:dM-dFe,eq:dsigmai-da}, respectively. The other terms can be derived by differentiating \Cref{eq:mandel-like,eq:a,eq:deformation-decomposition}. Applying the chain rule to  $\hpdv{\hat{\textbf{D}}_{\mr{p}i}}{\bm{\Sigma}^{\mr{sym}}}$ gives
\begin{align}
\pdv{\hat{\textbf{D}}_{\mr{p}i}}{\bm{\Sigma}^{\mr{sym}}} = 
\frac{1}{\eta_i} \left[ \hat{\textbf{N}}_{\mr{p}i} \otimes \pdv{\bar{\Sigma}_i}{\bm{\Sigma}^{\mr{sym}}_i} + \bar{\Sigma}^i \pdv{\hat{\textbf{N}}_{\mr{p}i}}{\bm{\Sigma}^{\mr{sym}}_i}\right] 
\end{align}
where $\hpdv{\bar{\Sigma}_i}{\Mstress^\mr{sym}_i} = \hat{\textbf{N}}_{\mr{p}i}$ and $\hpdv{\hat{\textbf{N}}_{\mr{p}i}}{\Mstress^\mr{sym}_i}$ is given by \Cref{eq:dNp-dMsym}. 
The derivative $\hpdv{\stress_i}{\bm{a}}$ in \Cref{eq:dR-dF} reads
\begin{align}
  \pdv{\stress_i}{\bm{a}} &= \frac{1}{J_{\mr{e}i}} \left[\bm{\Lambda}_1 +\bm{\Lambda}_2 + \bm{\Lambda}_3+ \bm{\Lambda}_4\right]  
\label{eq:dsigmai-da} \\
  \bm{\Lambda}_1 &= 4\beta_i \left(\textbf{B}_{\mr{e}i} \otimes  \bm{a}\right) \\
  \bm{\Lambda}_2 &= 2\left[\alpha_i + \beta_i  \left( \xi_{1i}-3 \right)  + 2 \gamma \left(\xi_{2i} - 1 \right) \right] \pdv{\textbf{A}}{\bm{a}} \\
  \bm{\Lambda}_3 &= 8\gamma_i \left(\textbf{A} \otimes  \bm{a} \right) \\
  \bm{\Lambda}_4 &= - \alpha_i \left(\textbf{B}_{\mr{e}i} \cdot \pdv{\textbf{A}}{\bm{a}}
    + \pdv{\textbf{A}}{\bm{a}} \cdot \textbf{B}_{\mr{e}i} \right)
\end{align}
where derivative $\hpdv{\textbf{A}}{\bm{a}}$ is given in \Cref{eq:dA-da}, by replacing $\hat{\textbf{A}}$ and $\hat{\bm{a}}$ with $\textbf{A}$ and $\bm{a}$.

\vspace{1cm}

\noindent $\bullet$ The derivative $\hpdv{R_{a_\sigma}}{\F}$ in equation \Cref{eq:dashift-dF} reads
\begin{equation}
  \pdv{R_{a_\sigma}}{\F} = \left[\pdv{R_{a_\sigma}}{\sigmaEq} \pdv{\sigmaEq}{\stress} +  
    \pdv{R_{\sigma}}{I_3}\pdv{I_3}{\stress} \right]
    : \sum^{N}_i \pdv{\stress_i}{\F} 
+
      \left[
    \pdv{R_{\sigma}}{\sigmaEq} \pdv{\sigmaEq}{\bar{\bm{a}}}+
  \pdv{R_{a_\sigma}}{I_3} \pdv{I_3}{\bar{\bm{a}}} \right] 
  \cdot \pdv{\bar{\bm{a}}}{\F} 
  \label{eq:dRashift-dF}
\end{equation}
where the $\hpdv{R_{a_\sigma}}{\sigmaEq}$, $\hpdv{R_{a_\sigma}}{I_3}$, $\hpdv{I_3}{\stress}$ are given by \Cref{eq:dRa-dsigma,eq:dRa-dI3,eq:dI3-dsigma}, respectively. 
Derivative $\hpdv{\bar{\bm{a}}}{\textbf{F}}$ is given by \Cref{eq:da-dF}, replacing $\hat{\bm{a}}$ by $\bm{a}$. The other derivatives read
\begin{align}
\pdv{I_3}{\bar{\bm{a}}} &= \purple{- 2 \stress \cdot \bar{\bm{a}}} \\
  \pdv{\sigmaEq}{\bar{\bm{a}}} &= \frac{1}{\bar{\sigma}} \left(\pdv{I_1}{\bar{\bm{a}}} + \alpha_2 \pdv{I_2}{\bar{\bm{a}}} \right) \label{eq:dI2-da} \\
  \pdv{\sigmaEq}{\stress} &= \frac{1}{\bar{\sigma}} \left(\pdv{I_1}{\stress} + \alpha_2 \pdv{I_2}{\stress} \right)  \label{eq:dI2-dsigma} \\
  \pdv{\stress_i}{\F} &= \pdv{\stress_i}{\F_{\mr{e}i}} : \left[ \pdv{\F_{\mr{e}i}}{\F} + \pdv{\F_{\mr{e}i}}{\F_{\mr{p}i}} : \pdv{\F_{\mr{p}i}}{\F}\right] + \pdv{\stress_i}{\bm{a}} \cdot \pdv{\bm{a}}{\F} 
      \label{eq:dsigmai-dF-consta}
\end{align}
where $\{\hpdv{I_j}{\bar{\bm{a}}}\}_{j=1,2}$ and $\{\hpdv{I_j}{\stress}\}_{j=1,2}$ are derived in \Cref{sec:seq-mode} and can be found by replacing \emph{intermediate} configuration quantities $\Mstress^{\mr{pind}}_i$ and $\hat{\bm{a}}_i$ for each mode $i$ by \emph{total} \emph{current} configuration quantities $\stress$ and $\bm{a}$.
The derivative $\partial \F_{\mr{p}i} / \partial {\F}$ in \Cref{eq:dsigmai-dF-consta} is found by solving the first consistency condition (see \Cref{eq:dFp-dF}).

\section*{Declaration of competing interest}
The authors declare that they have no known competing financial interests or personal relationships that could have appeared to influence the work reported in this paper.

\section*{Acknowledgement}
This research was carried out as part of the project ENLIGHTEN (project number N21010h) in the framework of the Partnership Program of the Materials innovation institute M2i (www.m2i.nl) and the Netherlands Organization for Scientific Research (www.nwo.nl).




\bibliographystyle{elsarticle-num} 
\bibliography{my_library_manual,my_library}





\end{document}